\documentclass[twocolumn]{aastex63}

\usepackage{epsfig,graphicx}
\usepackage{changepage}
\usepackage{hyperref,url}

\graphicspath{{./figures/}{./}}

\begin{document}

\title{The Panchromatic Hubble Andromeda Treasury: Triangulum Extended Region (PHATTER) I. Ultraviolet to Infrared Photometry of 22 Million Stars in M33}

\author[0000-0002-7502-0597]{Benjamin F. Williams}
\affiliation{Department of Astronomy, University of Washington, Box 351580, U.W., Seattle, WA 98195-1580, USA}

\author[0000-0002-0786-7307]{Meredith J. Durbin}
\affiliation{Department of Astronomy, University of Washington, Box 351580, U.W., Seattle, WA 98195-1580, USA}

\author[0000-0002-1264-2006]{Julianne J. Dalcanton}
\affiliation{Department of Astronomy, University of Washington, Box 351580, U.W., Seattle, WA 98195-1580, USA}

\author[0000-0002-1172-0754]{Dustin Lang}
\affiliation{McWilliams Center for Cosmology, Department of Physics, Carnegie Mellon University, 5000 Forbes Ave., Pittsburgh, PA} 

\author[0000-0002-6301-3269]{Leo Girardi}
\affiliation{Padova Astronomical Observatory, Vicolo dell'Osservatorio 5, Padova, Italy}

\author[0000-0003-2599-7524]{Adam Smercina}
\affiliation{Department of Astronomy, University of Washington, Box 351580, U.W., Seattle, WA 98195-1580, USA}

\author[0000-0001-8416-4093]{Andrew Dolphin}
\affiliation{Raytheon, Tucson, AZ 85726, USA}
\affiliation{Steward Observatory, University of Arizona, Tucson, AZ 85726, USA}

\author[0000-0002-6442-6030]{Daniel R. Weisz}
\affiliation{Astronomy Department, University of California, Berkeley, CA 94720, USA}

\author[0000-0003-1680-1884]{Yumi Choi}
\affiliation{Space Telescope Science Institute, 3700 San Martin
  Drive, Baltimore, MD, 21218, USA} 

\author[0000-0002-5564-9873]{Eric F. Bell}
\affiliation{Department of Astronomy, University of 
Michigan, 323 West Hall, 1085 S. University Ave., Ann Arbor, MI, 48105-1107, USA} 
  
\author[0000-0002-5204-2259]{Erik Rosolowsky}
\affiliation{University of Alberta, Department of Physics, 4-183 CCIS, Edmonton AB T6G 2E1, Canada}

\author[0000-0003-0605-8732]{Evan Skillman}
\affiliation{Minnesota Institute for Astrophysics, 116 Church Street SE, Minneapolis, MN 55455}

\author[0000-0001-9605-780X]{Eric W. Koch}
\affiliation{University of Alberta, Department of Physics, 4-183 CCIS, Edmonton AB T6G 2E1, Canada}

\author[0000-0003-0588-7360]{Christine W. Lindberg}
\affiliation{JHU/STScI, 3700 San Martin
  Drive, Baltimore, MD, 21218, USA}

\author[0000-0001-8918-1597]{Lea Hagen}
\affiliation{Space Telescope Science Institute, 3700 San Martin
  Drive, Baltimore, MD, 21218, USA} 

\author[0000-0001-5340-6774]{Karl D.~Gordon}
\affiliation{Space Telescope Science Institute, 3700 San Martin
  Drive, Baltimore, MD, 21218, USA} 

\author[0000-0003-0248-5470]{Anil Seth}
\affiliation{University of Utah, Salt Lake City, UT 84112}

\author[0000-0003-0394-8377]{Karoline Gilbert}
\affiliation{Space Telescope Science Institute, 3700 San Martin
  Drive, Baltimore, MD, 21218, USA} 

\author[0000-0001-8867-4234]{Puragra Guhathakurta}
\affiliation{University of California - Santa Cruz, 1156 High Street, Santa Cruz, CA, 95064} 

\author[0000-0003-3234-7247]{Tod Lauer}
\affiliation{National Optical Astronomy Observatory, PO Box 26732, Tucson, AZ 85726}

\author[0000-0001-7746-5461]{Luciana Bianchi}
\affiliation{JHU, 3400 North Charles St., 473 Bloomberg center for Physics and Astronomy, Baltimore, MD, 21218}


\begin{abstract}

We present panchromatic resolved stellar photometry for 22 million stars in the Local Group dwarf spiral Triangulum (M33), derived from {\it Hubble Space Telescope} (HST) observations with the {\it Advanced Camera for Surveys} (ACS) in the optical (F475W, F814W), and the {\it Wide Field Camera 3} (WFC3) in the near ultraviolet (F275W, F336W) and near-infrared (F110W, F160W) bands.  The large, contiguous survey area covers $\sim$14 square kpc and extends to 3.5~kpc (14~arcmin, or 1.5--2 scale lengths) from the center of M33. \deleted{Compared to the companion observations of M31 from the Panchromatic Hubble Andromeda Treasury (PHAT), the M33 survey area has a 10 times higher star formation rate intensity and present-day metallicities that lie between those in the outer disk of M31 and the Large Magellanic Clouds.} The PHATTER observing strategy and photometry technique closely mimics that of Panchromatic Hubble Andromeda Treasury (PHAT), but with updated photometry techniques that take full advantage of all overlapping pointings (aligned to within $<$5-10 milliarcseconds) and improved treatment of spatially varying PSFs.  The photometry reaches a completeness-limited depth of F475W$\sim$28.5 in the lowest surface density regions observed in M33 \added{and F475W$\sim$26.5 in the most crowded regions found near the center of M33.  We find the young populations trace several relatively tight arms, while the old populations show a clear, looser two-armed structure.} \deleted{but in the dense, high surface brightness central regions, stellar crowding limits the photometry to F475W$\sim$26.5.}  We present extensive analysis of the data quality including artificial star tests to quantify completeness, photometric uncertainties, and flux biases.  This stellar catalog is the largest ever produced for M33, and is publicly available for
download by the community.

\end{abstract}

\section{Introduction}

Resolved stellar photometry has the potential to constrain fundamental processes in astrophysics, including star formation, stellar evolution, feedback into the interstellar medium, galaxy formation and evolution, and chemical enrichment.  The stars themselves are the fossil record of these processes, which leave signatures in the properties of individual stars, their mass distribution, their distribution of colors and magnitudes, and their spatial distribution with respect to other galactic tracers.  

While Gaia is transforming our understanding of the stars and structure of the Milky Way disk \citep{gaia2018}, we still need comparably detailed population studies of other disks to put our Galaxy and its stellar populations in context.  The best targets for such studies are the galaxies in the Local Group, which contains two spirals other than the Milky Way --- M31, a ``green valley" Sb galaxy, and M33, a blue sequence, star-forming \deleted{bulgeless} dwarf spiral. Along with the Milky Way, these two galaxies form our best anchors for baryonic processes in spiral galaxies.  This set of three galaxies spans a large dynamic range, giving ample opportunities for contrasting how astrophysical processes are shaped by other parameters. For example, both M31 and the Galaxy are of similar mass and metallicity \citep{watkins2010,gregersen2015}, but M31 seems to have a much more dramatic recent merger history \citep[e.g.,][]{hammer2018,dsouza2018,kruijssen2019}. 

In contrast to both of these more massive partners, M33 is of lower mass and metallicity, and appears to have a relatively quiescent merger history, as suggested by its inside-out growth \citep{magrini2007,williams2009,beasley2015,mostoghiu2018} and lack of a significant extended stellar halo \citep{mcconnachie2010,mcmonigal2016} or prominent thick disk \citep{wyse2002,vanderkruit2011}.  Thus, M33 probes a different set of physical and chemical evolution properties than the other Local Group disk galaxies, and we can constrain these in exquisite detail through measurements of M33's constituent stars.

Along with M31, previously surveyed with HST as part of the Panchromatic Hubble Andromeda Treasury (PHAT, \citealp{dalcanton2012}), M33 is one of the richest galaxies in the Local Group for obtaining photometric measurements of resolved stars in a spiral galaxy.  It is close enough that we can resolve stars all the way down to the ancient main sequence \citep{williams2009} over much of the disk.
All of its stars are at the same distance and foreground extinction, alleviating issues related to the wide range of distances and extinctions of stars in the Galaxy.  Furthermore, M33 has no \added{well-established} significant bulge component beyond its nuclear cluster \citep{mclean1996, kormendy1993}, \added{and at most a weak pseudobulge or bar \citep{regan1994,minniti1993,stephens2002,corbelli2007}}, meaning that there is \added{little} confusion between disk and bulge populations.  


\added{The metallicity gradient in M33 is well known, and has been measured many times with planetary nebulae and HII regions \citep[e.g.,][]{cioni2008,rosolowsky2008, magrini2009, magrini2010, bresolin2010,toribio2016,lin2017}.  These studies have provided a range of measured central metallicities of $8.4 < 12 + log (O/H) < 8.8$ and a range of slopes of $-0.05 < dex/kpc < -0.02$.  Our representation in the right panel of Figure~\ref{m33_vs_m31_fig} is approximate and extrapolated based on these measurements, and shows how M33 covers the gap in metallicity between the LMC and M31.}

\added{The star formation rate (SFR) density of M33 is significantly higher than in M31, making it an excellent probe of higher intensity star formation environments.  The total SFR from the GALEX far-UV (FUV) and Spitzer 24~$\mu$m observations is $\sim$0.5 M$_{\odot}$ yr$^{-1}$ \citep[e.g.,][]{verley2009}, which is higher than the rate in the entire PHAT survey:   $\sim$0.3 M$_{\odot}$ yr$^{-1}$ in an area about 8 times larger \citep{lewis2015}.  These measurements are consistent with the order of magnitude relative difference in star formation intensity distributions shown in Figure~\ref{m33_vs_m31_fig}.}

M33's value for obtaining knowledge about disk stellar populations is reflected in its rich history of resolved star studies, dating back to the 19th century \citep[e.g,][]{roberts1899}.  Since then, ground-based observations have studied the bright massive stars in great detail, providing estimates of star formation rate and constraints on the evolution of massive stars \citep[e.g.,][and many others]{madore1974,humphreys1980,massey1996}. The bright stars that can be resolved from the ground were finally fully cataloged by \citet{massey2006}, and its extended halo was probed by the Pan-Andromeda Archaeological Survey \citep[PAndAS;][]{mcconnachie2010}.  More recent ground-based work focuses on the variability of these massive stars to further constrain their complex evolutionary stages \citep[e.g.,][and many others]{gordon2016,humphreys2017,smith2020}.

Over the past few decades, past ground-based studies of M33 have been supplemented with {\it HST} imaging, both farther into the ultraviolet \citep[e.g.,][]{chandar1999,hoopes2000} and to much fainter depth in the optical \citep[e.g.,][]{mighell1995,sarajedini2000,barker2007,barker2007b,williams2009}.  These capabilities have provided deep insight into the properties of the youngest and oldest stars and stellar clusters, as well as the formation processes of the M33 disk \citep[e.g.,][and references therein]{vanderkruit2011}.  

M33's stellar population studies benefit from the legacy of surveys across virtually all wavelengths.  Its cold interstellar medium (ISM) has been mapped through 21cm maps of atomic H{\sc i} \citep{deul1987, gratier2010, koch18}, through millimeter maps of molecular gas in the CO($J=1-0$) and CO($J=2-1$) lines \citep[Figure~\ref{footprints_multiwavelength}, see e.g.,][]{heyer2004, gratier2010, engargiola2003, rosolowsky2003, rosolowsky2007,druard2014}, and through extensive studies of dust through Spitzer \citep{hinz2004,mcquinn2007}, Herschel \citep[most notably the HerM33es project; e.g.,][]{kramer2010,xilouris2012}, and long-wavelength facilities like APEX and Planck \citep[e.g.,][]{hermelo2016, depaolis2016, tibbs2018}.  M33 has been mapped with GALEX near and far UV \citep{thilker2005} 
\deleted{ and H$\alpha$}  
\deleted{and kinematics of the stellar populations are being resolved with Keck/DEIMOS (K. Gilbert et al. in preparation).  Finally, there are data that constrain M33's population of high-energy sources, including,}
\added{in} hard X-rays from {\it NuSTAR} \citep{west2018}, softer X-rays from {\it Chandra} \citep{tuellmann2011} and {\it XMM-Newton} \citep{williams2015}, gamma rays from {\it Fermi} \citep{xi2020}, and deep radio continuum \citet{white2019}.  This rich compendium of multi-wavelength data, and its associated catalogs, can serve both to support interpreting M33's resolved HST photometry, and to be interpreted in turn by improved knowledge of M33's stellar content and spatially resolved star formation history.

Recently, the power of wide-area, panchromatic imaging of nearby galaxies has been demonstrated by PHAT  \citep{dalcanton2012,williams2014}, which covered roughly one third of the star forming disk of M31 in 6-bands ranging from the near-UV to the near-IR.  The PHAT survey has produced scientific return spanning a wide range of topics, from star clusters \citep{johnson2015} and the initial mass function (IMF) \citep{weisz2015} to star formation history \citep{lewis2015,williams2017}, calibration of star formation rate indicators \citep{lewis2017}, metal retention \citep{telford2019}, mass-to-light ratios \citep{telford2020}, and dust \citep{dalcanton2015}.  \added{We expect similar scientific return from studies of M33 based on the Panchromatic Hubble Andromeda Treasury: Triangulum Extended Region (PHATTER).} 

In this paper we present first results from an equivalent high-resolution, 6-band survey of M33, so that we may provide a resolved stellar photometry catalog with the same quality, giving the community the ability to probe the same processes in a galaxy with very different physical properties, including lower mass, lower metallicity, and higher star formation intensity.  Once the stellar populations of both galaxies are measured in such exquisite detail, the power of direct comparison will likely lead to even more illuminating results. \added{While this paper mainly presents the photometric catalogs and demonstrates the quality of the measurements, we anticipate a number of future dedicated papers on M33’s stellar populations and their connection to their host.} 

Herein we describe our large {\it HST} survey of M33, PHATTER.  Section 2 describes our observing strategy and data reduction techniques.  Section 3 provides our results, including our final catalog of 6-band panchromatic photometry of all of the stars detected in our observations.  Section 4 then investigates the quality of the photometry in the catalog, including analysis of the luminosity function and artificial star tests.  Finally, Section 5 summarizes the paper.  Throughout, we assume a distance to M33 of 859 kpc ($m-M=24.67$; \citealp{degrijs2017}).

\section{Observations and Data Analysis}


\subsection{Observing Strategy}

The highest impact science we anticipate from the new M33 observations comes from exploring galactic environments that are distinct from other Local Group galaxies. Our observing strategy was therefore designed to make comparisons as straightforward as possible, by reproducing the observing strategy for M31, but targeting regions of M33 with complementary properties. 

As shown in Figure~\ref{m33_vs_m31_fig}, M33 is a lower metallicity galaxy than most of M31 at the present day, and its inner regions nicely bridge the metallicity gap between the LMC and M31's outer regions \added{\citep[similar gradient to e.g.,][]{cioni2008,rosolowsky2008, magrini2009, magrini2010, bresolin2010,toribio2016,lin2017}}.  Those same inner regions of M33 also have a typical star formation rate intensity that is nearly a factor of 10 higher than in the area covered by the PHAT survey in M31, adding considerable leverage to studies of the interaction between stars and the ISM.  We therefore targeted the new M33 observations on these inner regions, where there is also considerable multiwavelength coverage from other observatories, \added{covering, for example, nearly all of the CO($J=1-0$) molecular cloud detections from \citet{rosolowsky2007}, as shown in Figure~\ref{footprints_multiwavelength}}.

\begin{figure}
    \centering
    \includegraphics[width=0.7\textwidth]{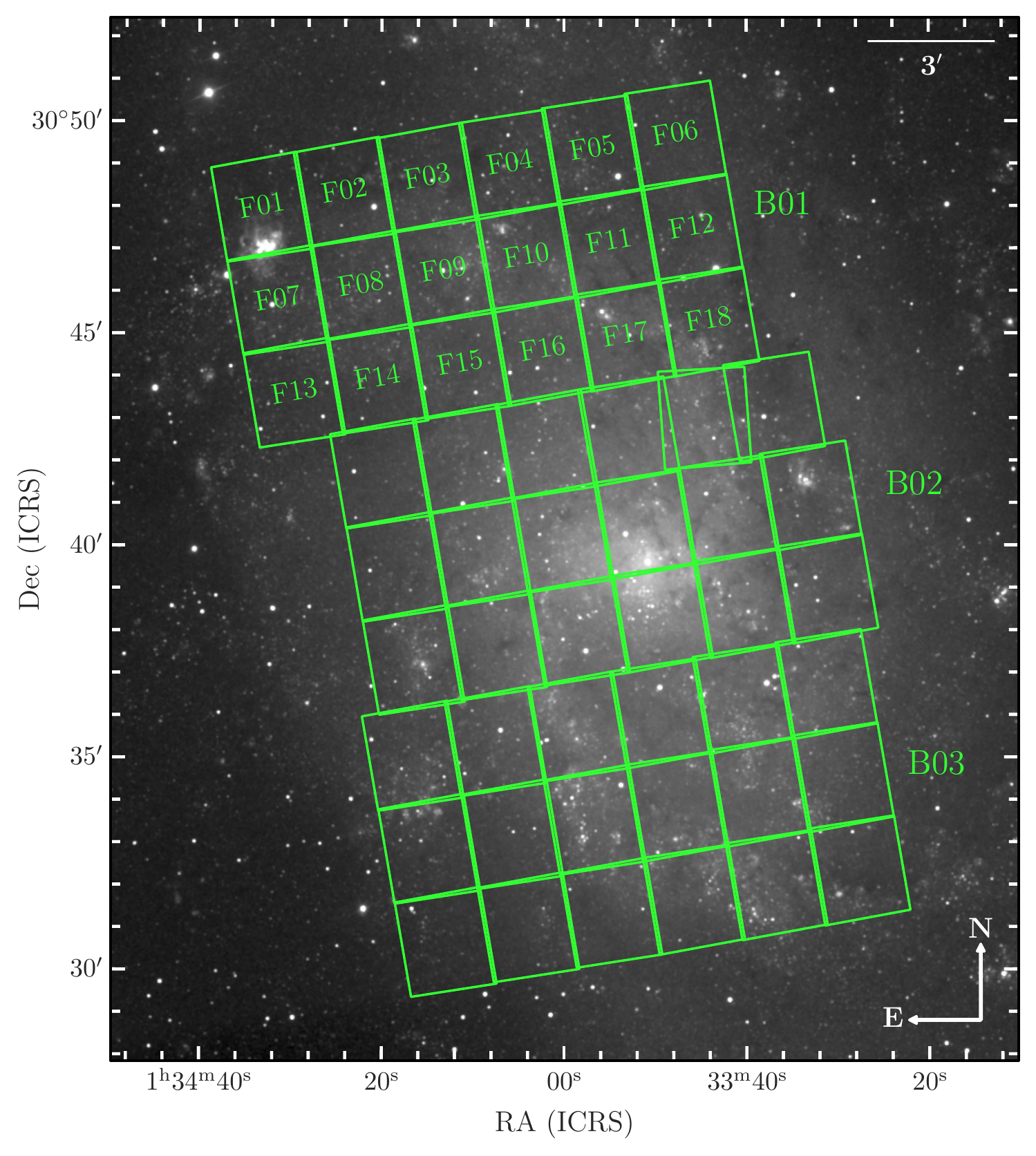} 
    \caption{The WFC3/IR footprints of our M33 survey are plotted on a Sloan Digital Sky Survey (SDSS) image of M33.  Brick 1 is marked by the 6$\times$3 set of footprints surrounding the galaxy center.  Brick 1 is the northernmost 6$\times$3 footprints and Brick 3 is the southernmost set.  \added{The field numbers for each brick are 1$-$18, where 1$-$6 are the top row from left to right, 6$-$12 are the next row down, etc.  The orientation of Field 5 in Brick 2 was shifted due to a lack of available guide stars in the standard orientation, and Field 6 was shifted slightly to compensate.}}
    \label{footprints}
\end{figure}

We build up this survey area using the same PHAT tiling strategy, as described in \citet{dalcanton2012}.  Observations are organized into ``bricks" of 3$\times$6 WFC3/IR footprints (Figure~\ref{footprints}), with observations of each 3$\times$3 half-brick taken $\sim$6 months apart, after the telescope has rotated 180$^\circ$ (see Figure~\ref{footprints}), with \texttt{ORIENT}=55 for one half brick and 235 for the other.  At each pointing, WFC3/UVIS observations \added{(F275W$_{250-300\rm{nm}}$ and F336W$_{310-360\rm{nm}}$ filters)} are taken in one orbit and WFC3/IR observations \added{(F110W$_{900-1400\rm{nm}}$ and F160W$_{1400-1700\rm{nm}}$ filters)} are taken in another, while ACS/WFC operates in parallel observing in \added{F475W$_{400-550\rm{nm}}$ and F814W$_{700-950\rm{nm}}$} covering the adjacent half brick.  When the telescope rotates orientations in $\sim$6 months, the primary WFC3 observations cover the area of the original ACS parallels, and vice versa.  Note that this produces a time difference between the optical and UV$+$IR observations, which may produce unusual colors for time-varying sources. Observations for this program (GO-14610) were taken between February 21, 2017, and February 25, 2018.  The downloaded calibrated images used for photometry were processed under OPUS versions 2016\_2 - 2017\_3b.  For ACS/WFC and WFC3/UVIS, we start with the CTE-corrected \citep{2010PASP..122.1035A, andersen2018}, {\tt flc} image files. For WFC3/IR, we start with {\tt flt} image files.

We chose a 3-brick mosaic to maximize coverage of the high star-formation intensity regions and existing CO detections (Figure~\ref{footprints}).  Brick 1 is the 3$\times$6 array covering the northern portion of the galaxy, Brick 2 covers the center, and Brick 3 is to the south.  Within each Brick, each WFC3/IR pointing area is given a field number, with Field 1 being the upper left on Figure~\ref{footprints} and Field 18 being the lower right.  ACS observations are labeled with the WFC3 field that they overlap.  Of the 54 pointings, one field (Brick 2, Field 5) had no guide stars available in the desired orientation, and was therefore rotated slightly to make observations possible.  This change led to a slight ($\sim$20 arcsec) gap in coverage at the northwest corner of Brick 2 (01:33:30, 30:44:00). In total, the survey area tiled the inner 13.2$\times$19.8 arcmin (3.1$\times$4.6 kpc, projected; 4.3$\times$4.6 kpc, deprojected) of M33, extending to roughly $\sim$1.6 disk scale lengths, assuming a $6^\prime$ scale length \citep{regan1994}.

We adopted an identical exposure sequence (Table~\ref{table_obs}) and dithering strategy as in PHAT, with the only significant change being switching to using UV pre-flash to minimize CTE losses in WFC3/UVIS ({\tt{FLASH}}$=10$ for F336W and $=11$ for F275W).   WFC3/IR exposures were taken with 13 \texttt{MULTIACCUM} non-destructive read samples of the \texttt{STEP100} sequence for a single F110W exposure, three F160W exposures with 9 samples of the \texttt{STEP200} sequence, and one additional F160W exposure with 10 samples of the \texttt{STEP100} sequence. The adopted dithers are designed to produce Nyquist sampled images in F475W, F814W, and F160W, but do not fill in the ACS chip gap. Instead, the ACS chip gaps are filled by overlapping exposures from observations in adjacent fields. The two WFC3/UVIS exposures for each filter are dithered to fill the chip gap, but, have challenging cosmic ray rejection, due to having only 1-2 overlapping images.  The ACS observations also include very short ``guard" exposures in F475W (10 seconds) and F814W (15 seconds) to capture photometry for the brightest stars, which can be saturated in the longer individual exposures.  \deleted{The original PHAT survey paper \citep{dalcanton2012} provides more details.}  A table of the exposures at each position is supplied in Table~\ref{table_obs}.

\begin{figure}
    \centering
    \includegraphics[width=0.9\textwidth]{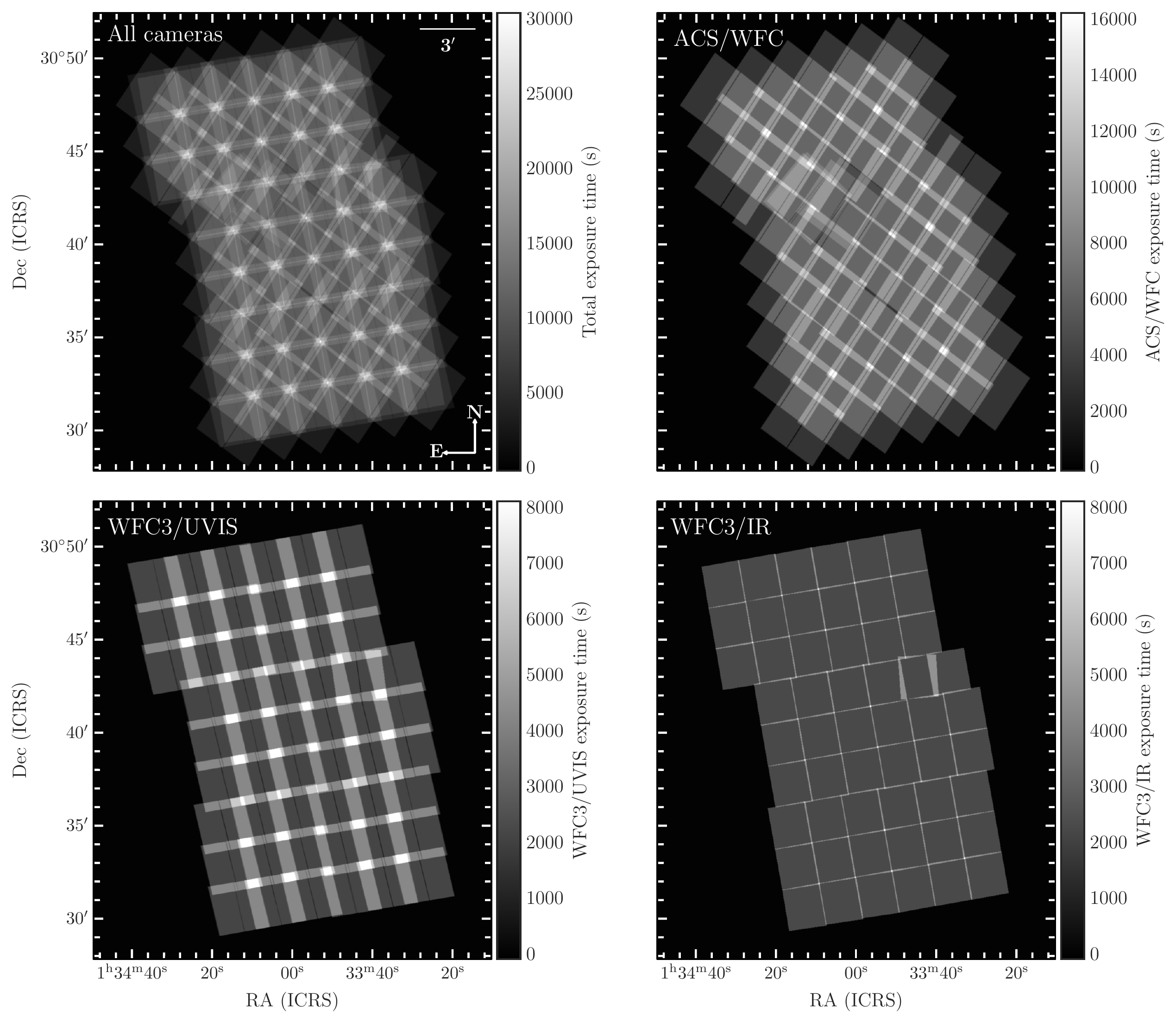} 
    \caption{Top left: The exposure map of the entire survey for all 3 cameras (WFC3/IR, WFC3/UVIS, and ACS/WFC). The grayscale is the amount of total exposure in each location in seconds.
    Top right: Exposure map for ACS/WFC only.
    Bottom left: The same for WFC3/UVIS only.
    Bottom right: The same for WFC3/IR only.
    The WFC3/IR footprints are the same as those shown in Figure~\ref{footprints}.
    }
    \label{expmap}
\end{figure}

The resulting map of exposure times in all cameras is shown in Figure~\ref{expmap}. Notable features are the slightly larger WFC3/UVIS fields of view, which lead to large rectangular overlaps between adjacent fields than the minimally overlapping WFC3/IR fields, and the diagonal overlaps of the even larger ACS/WFC exposures. Some of the inconsistencies in the tiling pattern are due to adjustments that ensured coverage with the non-standard Brick 1, Field 5 rotation.  The most highly overlapped regions in F475W and F814W have over 30,000 seconds of total exposure time.  However, because the majority of observations in the optical and IR are crowding-limited, rather than photon limited, the varying exposure times due to the overlapping pointings tend to affect the measured source density in less obvious ways that are often only noticeable at faint magnitudes. 

\subsection{Photometry}

We measured point spread function (PSF) fitting photometry on the location of every star detected in our survey footprint on every exposure that covered the position of the star.  We closely followed the process used for the PHAT survey photometry to simplify comparisons; however, there have been some improvements made to the process from lessons learned by PHAT.  

The first improvement was the use of charge transfer efficiency (CTE) corrected {\tt flc}-type) for photometry. In PHAT, no correction was used for WFC3/UVIS photometry, and ACS/WFC photometry was corrected at the catalog level.  This change was implemented to address systematic uncertainties that appeared to be related to CTE in \citet{williams2014} at the faint end.  In addition, we implemented spatially-varying TinyTim PSFs \citep{krist2011} for all cameras to address the systematic uncertainties that appeared to be related to the PSF in \citet{williams2014}.  \added{All of these changes affect the photometry at the $\lesssim 0.1$ mag level, but may help to mitigate systematics related to position on the detector.}

A high-level overview of the process of measuring the stellar photometry is provided as follows. The first step was the astrometric alignment of all 972 individual exposures with the Gaia catalog.  These images were then combined into mosaic images, which were used for identifying and flagging bad pixels and cosmic ray affected pixels for masking during photometry, as well as for public release images\footnote{\url{https://hubblesite.org/image/4305/gallery}}.  The aligned individual images were processed with the DOLPHOT software package \citep{dolphin2000,dolphin2016} to measure PSF corrections and aperture corrections, which largely correct for variations in telescope focus.  All overlapping individual exposures were stacked in memory to search for all statistically-significant detections using the full survey depth. At each detected centroid, the appropriate PSF was fit to the detection's locations in each of the overlapping exposures, for all filters simultaneously.  DOLPHOT then reported the measured fluxes and corresponding magnitudes in each image, as well as the combined flux and magnitude in each observed band.  Finally, the raw photometry output was processed to flag possible artifacts and generate summary catalogs containing a subset of the many thousands of columns required to describe the complete measurement suite.  

We describe each of these steps in detail below.

\subsubsection{Astrometric Alignment \& Mosaicking}

We aligned all {\tt flc} (ACS/WFC, WFC3/UVIS) and {\tt flt} (WFC3/IR) images to the {\it Gaia} DR2 astrometric solution following the workflow presented by \citet{2017wfc..rept...19B}\footnote{\url{https://github.com/spacetelescope/gaia\_alignment}}. Using this workflow, a reference astrometric catalog was retrieved from the {\it Gaia} archive with {\tt astroquery} \citep{2017ascl.soft08004G, 2019AJ....157...98G}, which was then passed to the {\tt TweakReg} function in the {\tt Drizzlepac} package \citep{2012ascl.soft12011S, 2013ASPC..475...49H, 2015ASPC..495..281A}, which finds centroids in each image and matches triangular patterns and updates the image headers with the resulting aligned astrometric solution.  The catalogs from which the final alignment solution was derived typically contained several hundred stars per ACS/WFC pointing, and 50-200 stars per WFC3/UVIS or WFC3/IR pointing.

\begin{figure}
    \centering
    \includegraphics[width=0.5\textwidth]{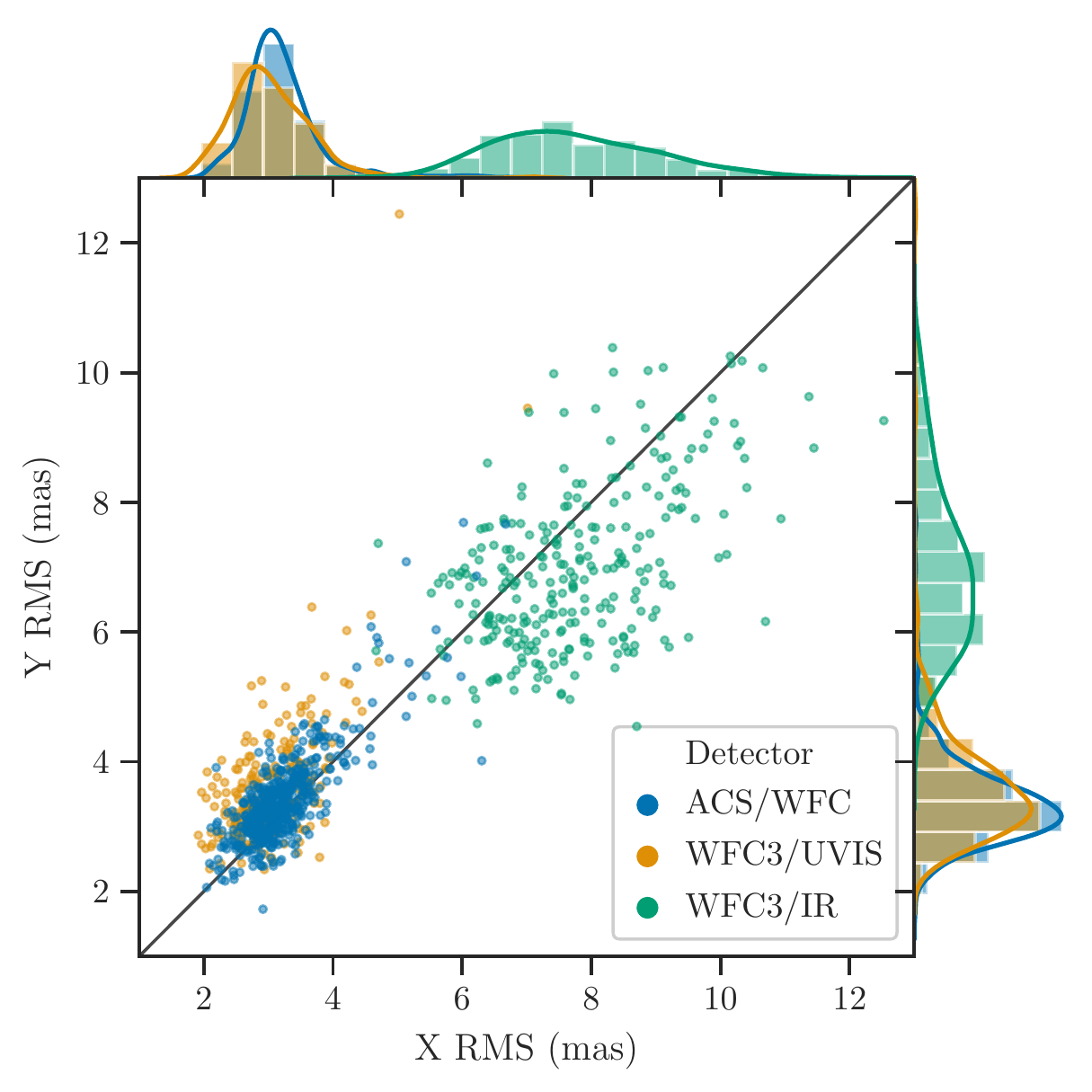}
    \caption{$X$ and $Y$ residual RMS values from {\tt TweakReg} in milliarcseconds. The residual RMS values peak near 3 mas for ACS/WFC and WFC3/UVIS, and near 7 mas for WFC3/IR on both axes.}
    \label{alignment_all}
\end{figure}

The RMS dispersions of the alignment residuals in $X$ and $Y$ are shown for all frames in Figure~\ref{alignment_all}. Typical overall residual dispersions are on the order of 3 mas for ACS/WFC and WFC3/UVIS, and 7 mas for WFC3/IR.



We used the {\tt AstroDrizzle} function of the {\tt Drizzlepac} package \citep{2012ascl.soft12011S, 2013ASPC..475...49H, 2015ASPC..495..281A} to combine the images within each band into a distortion-corrected, high-resolution pixel array (0.035\arcsec/pixel in all bands, combined with a {\tt lanczos3} kernel). This higher resolution array allows the full camera resolution to be recovered from dithered images, which were Nyquist sampled in F475W, F814W, and F160W. 
A {\it minmed} filter flagged statistical outlier pixels on the input exposures for all filters except F110W, for which there is only a single exposure, forcing us to rely on up-the-ramp fitting to flag bad pixels and filter cosmic rays.  These pixels were not considered when generating the combined image, and they can easily be masked in any further analysis using those exposures.  The flagged images were then combined with {\tt astrodrizzle}, weighted by exposure time to produce deep mosaics that take advantage of sub-pixel dithering to improve spatial resolution.

\begin{figure}
    \centering
    \includegraphics[width=0.6\textwidth]{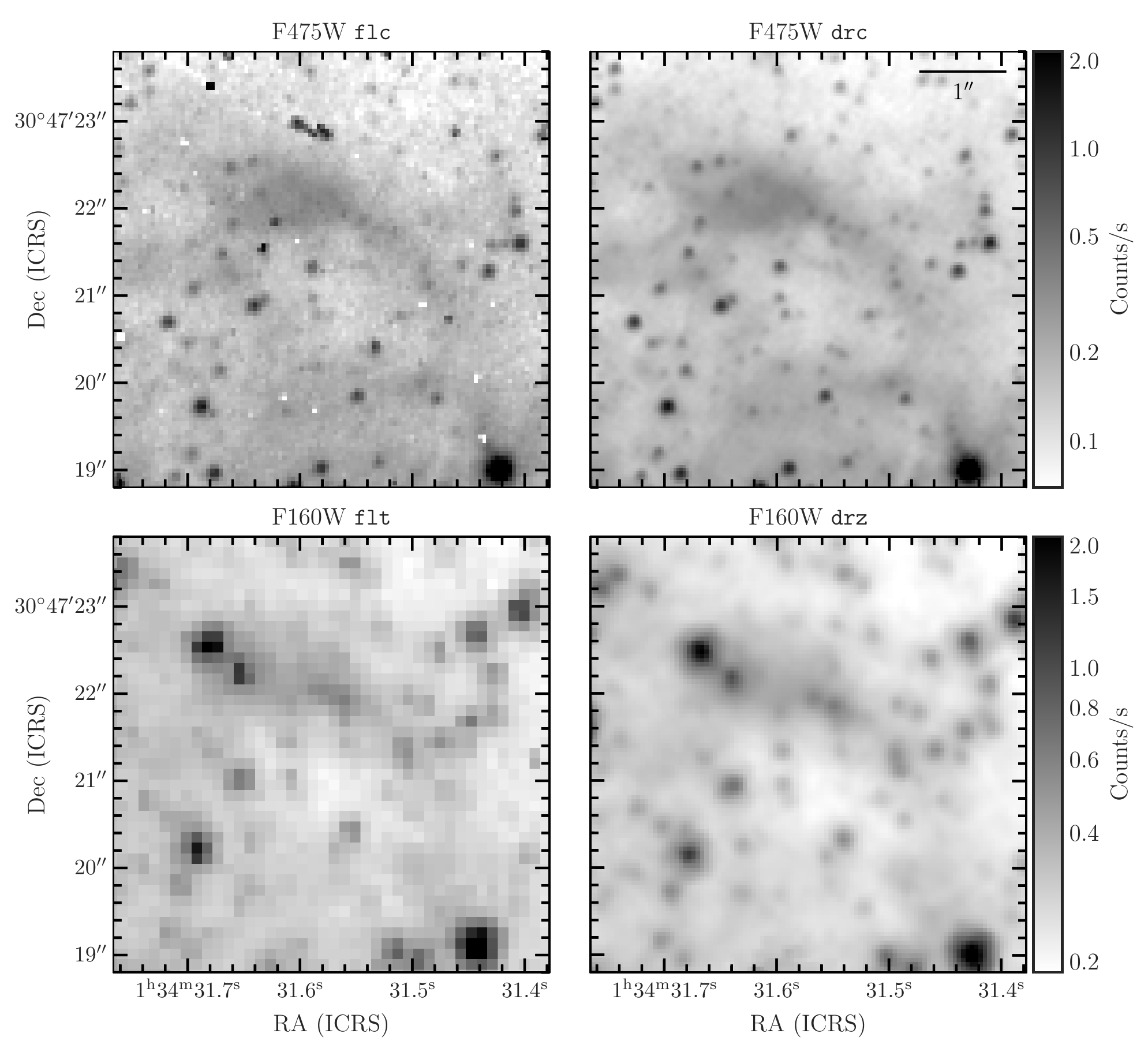}
    \caption{Comparisons of individual (left) and stacked (right) exposures in F475W (top) and F160W (bottom). The individual exposures have not been corrected for geometric distortion, which leads to the slightly different astrometry between the panels on the left.}
    \label{4panel}
\end{figure}

An example of the improvements in depth and resolution is shown in Figure~\ref{4panel}. The final product from the F475W exposures, which is the deepest band with the most sub-pixel dithers, was then applied as the \added{astrometric} reference image for all of the photometry measurements, \added{including the positions of centroids measured in other bands but not detected in the F475W data.}

\subsubsection{Preparing Individual Exposures}

After updating data quality extensions of the individual exposures in the {\tt astrodrizzle} step, we further prepared the individual exposures for photometry with DOLPHOT.  This preparation starts with running the task {\tt acsmask} or {\tt wfc3mask} (depending on camera) on each exposure.  This task masks the flagged pixels in the DQ extensions of each CCD in each exposure and multiplies the image by the appropriate pixel area map to take into account the effects of distortion on the flux measured in each pixel.  We also run this step on the full-depth F475W combined image, which serves as the \added{astrometric} reference image for the final photometry.  DOLPHOT uses this image as the\added{astrometric} reference frame to which all of the individual exposures will be aligned in memory, and from which all of the final star positions will be reported.  As such, it is beneficial to use the deepest and highest spatial resolution image for this purpose.

We then ran the {\tt splitgroups} task to produce separate files for each CCD of each exposure, and then we ran {\tt calcsky} on each of these individual frames to generate maps of the sky level in each exposure.  These sky files, which are simple smoothed versions of the original images, are used by DOLPHOT to find an initial list of statistically significant centroids to align each frame to the \added{astrometric} reference image; in spite of their name, they are not actually used for measuring the true sky level, which instead is measured in a much more sophisticated way described in Section~\ref{dolphot_section}.  We then ran {\tt DOLPHOT} on each individual exposure, to measure the central PSF and aperture corrections of each CCD read, followed by running DOLPHOT's alignment on the full stack of CCD reads to determine and record the parameters that align each individual frame to the \added{astrometric} reference image.  

\subsubsection{Running DOLPHOT on Full Image Stacks}\label{dolphot_section}

With images, alignment parameters, PSF corrections, and aperture corrections for each individual exposure in hand, we could run full-stack photometry on any region of the survey.  We ran these stacks using the DOLPHOT parameters updated from those of the PHAT survey to optimize the resulting catalogs for stellar populations science. The main updates are the removal of catalog-level CTE corrections, because we used the on-image CTE corrections ({\tt{flc}} images), and the use of TinyTim PSFs for all cameras and filters.  Values of all of the adopted DOLPHOT parameters for our reductions are provided in Table~\ref{table_par}.

\begin{figure}
    \centering
    \includegraphics[width=0.8\textwidth]{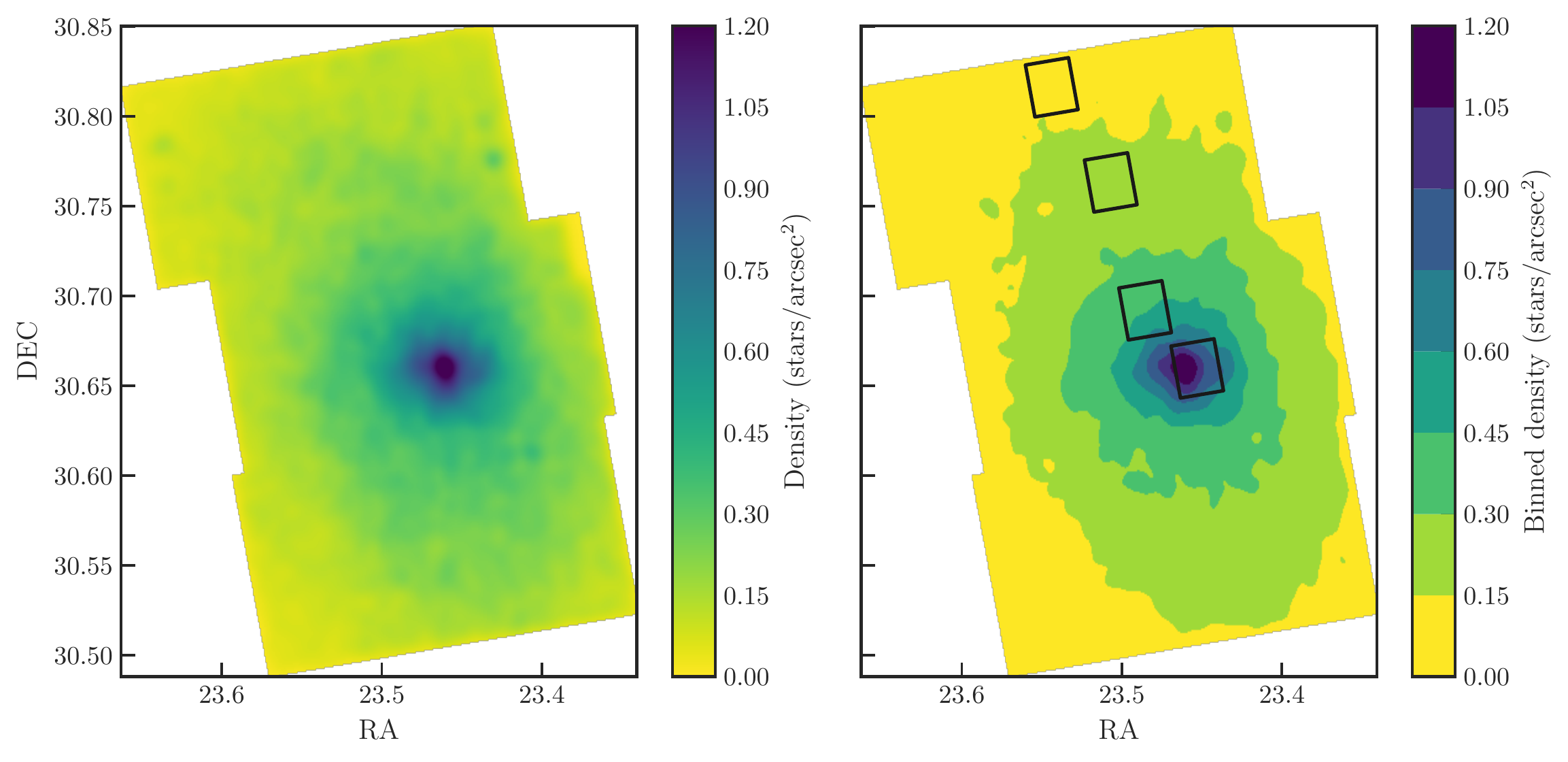}
    \caption{Left: A map of the stellar density of the photometry catalog as determined by star counts per square arcsec with $19.7{<}\mathrm{F160W}{<}20.7$.  Right: Same, with the colormap binned in increments of 0.15 stars/square arcsec. Black boxes mark the areas in which we have artificial star tests (ASTs) for determining the photometric quality (scatter, bias, and completeness) as a function of stellar density.}
    \label{density_map}
\end{figure}

Memory and time limitations prevent us from simply putting the entire set of M33 exposures into DOLPHOT simultaneously.  Instead, we subdivided the data into separate stacks to measure the photometry of different regions of the survey in parallel.  We used DOLPHOT parameters that allow the user to define the region within which it performs photometry to launch multiple photometry processes, each with a different region of the survey including all overlapping individual images.  We made these regions sufficiently small that DOLPHOT could complete the PSF fitting photometry in a reasonable amount of clock time, typically about one week.  We set up 54 separate processes, each covering $\sim$4 square arcminutes of the survey area, overlapping by 100 pixels on a side to avoid introducing edge effects. We then merged the resulting catalogs along the centers of the regions' overlaps to produce one final catalog for the survey.  We then checked for any edge effects from the survey division by plotting the densities of stars.  Such a plot is shown in Figure~\ref{density_map}, \added{which shows our measurement of completeness- and reddening-independent stellar density using the bright end of the red giant branch in our reddest band (number of stars with $19.7{<}\mathrm{F160W}{<}20.7$ per square arcsec).  While this is our adopted standard for measuring stellar density, such maps made without such strict magnitude limits also show no edge effects related to our division used for processing. }


\subsubsection{Flagging and Processing Photometry Output}\label{flagging_section}

DOLPHOT returns a comprehensive table of all of the measurements made on every PSF fit to every image, as well as the combined measurement of every source in every filter.  These measurements include the flux, Vega system magnitude, count-based uncertainty, signal-to-noise ratio, and several measurements of how well the source was fitted by the PSF.  These quality metrics include \texttt{sharpness}, \texttt{roundness}, $\chi$, and \texttt{crowding}.  Full descriptions of these are included in the DOLPHOT documentation\footnote{\url{http://americano.dolphinsim.com/dolphot/}}.  Briefly, the \texttt{sharpness} parameter measures how centrally peaked the source is compared to the PSF, or how much flux is concentrated in its central pixels relative to the outer ones.  High values signify a source with high central concentration, such as a hot pixel or cosmic ray.  Low values indicate that the source is not peaked enough, as expected for blended stars or background galaxies.  The \texttt{roundness} parameter measures how circular the source is (zero is perfectly round), and $\chi$ provides an estimate of the overall goodness of fit to the PSF.  The \texttt{crowding} parameter measures how much the source's photometry is affected by neighboring sources.  The larger the crowding value, the more densely packed the PSF radius is with other sources, and the more likely it is that the reported magnitude has systematic uncertainties due to subtraction of neighbors.

For the PHAT survey, we determined values for the DOLPHOT parameters that tend to indicate good measurements of real stars \citep{williams2014}.  We have adopted these criteria for this catalog as well, and list them here for convenience.  For a complete description of how they were determined, see \citet{williams2014}.  They are different for each camera, as the pixel scale and PSF sampling were different, with the exception of the signal-to-noise ratio, for which we require $\mathrm{S/N}>4$ for all cameras.  For ACS, the other parameters are: \texttt{sharpness}$^2<0.2$ and \texttt{crowding}$\ <2.25$.  For UVIS, they are: \texttt{sharpness}$^2<0.15$; \texttt{crowding}$\ <1.3$, and for WFC3's IR channel they are \texttt{sharpness}$^2<0.15$; \texttt{crowding}$\ <2.25$.  These culling parameters were found to have the best balance of removing a high fraction of sources outside of color-magnitude diagram (CMD) features while keeping a very high fraction of total measurements.  Thus, they lean towards inclusive to avoid over-culling the data, at the expense of allowing a larger fraction of less certain measurements and contaminants.  

\begin{figure}
    \centering
    \includegraphics[width=0.8\textwidth]{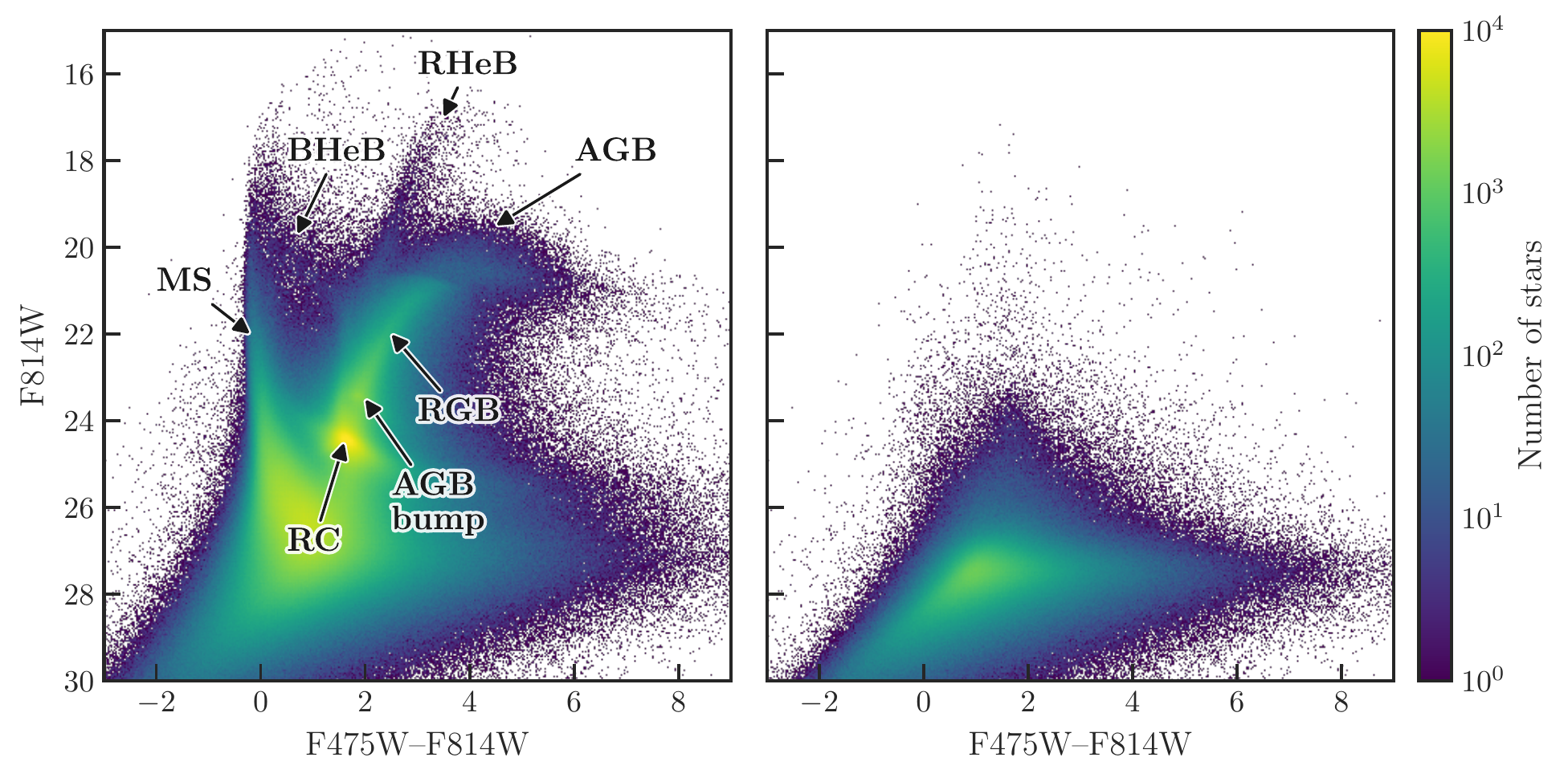}
    \caption{Left: optical (F475W--F814W) \replaced{CMD}{Hess diagram} of all output photometry (phot.fits). Right: same as left, showing measurements that fail our GST quality criteria in both bands. The failing measurements do not tend to mark typical CMD features, suggesting they are not reliable for population work.}
    \label{full_cmd}
\end{figure}

We show the results of the above cuts on the CMD in Figure~\ref{full_cmd}, where the stars that pass the metric make a CMD with well-defined, well-populated features, whereas the rejected stars form a relatively featureless cloud of points. However, there is always some risk in excluding important individual detections that did not produce high-quality PSF fits, such as bright stars in clusters.  Thus, we include in our catalog all measurements, but we add a flag column to each band indicating whether it passes.  This method allows the user to search the full catalog for specific source, but also allows one to easily look at populations without being distracted by artefacts.  For science cases that require a very clean sample, we recommend going to the full catalog and applying more conservative culling criteria than those adopted for our quality columns reported here.

\begin{figure}
    \centering
    \includegraphics[width=0.8\textwidth]{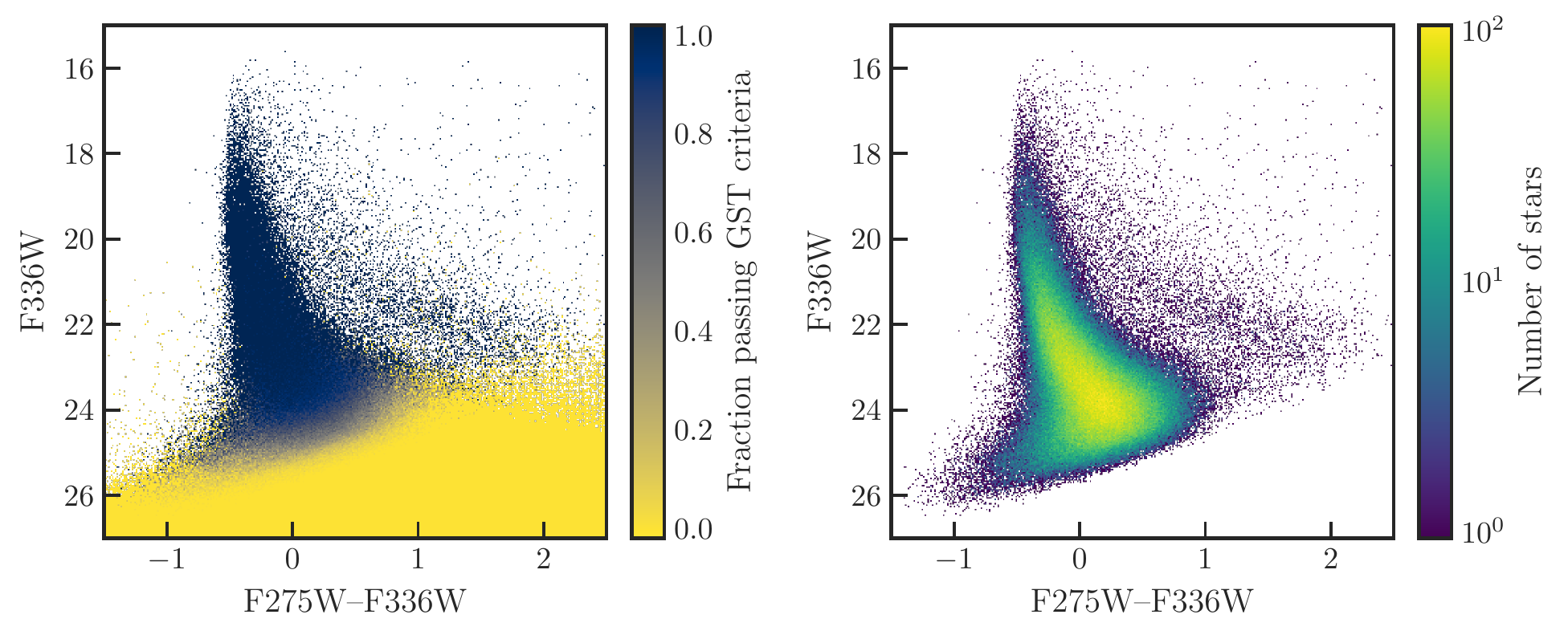}
    \caption{UV color–magnitude diagram for all F275W and F336W measurements in the survey. Left: fraction of measurements flagged as not passing our quality cuts in either of the two bands. Right: CMD produced showing only measurements that pass our quality cuts in both bands. Our quality cuts keep a very high fraction of the stars in the CMD features and a very low fraction of stars outside of these features.}
    \label{uv_cmd}
\end{figure}

\begin{figure}
    \centering
    \includegraphics[width=0.8\textwidth]{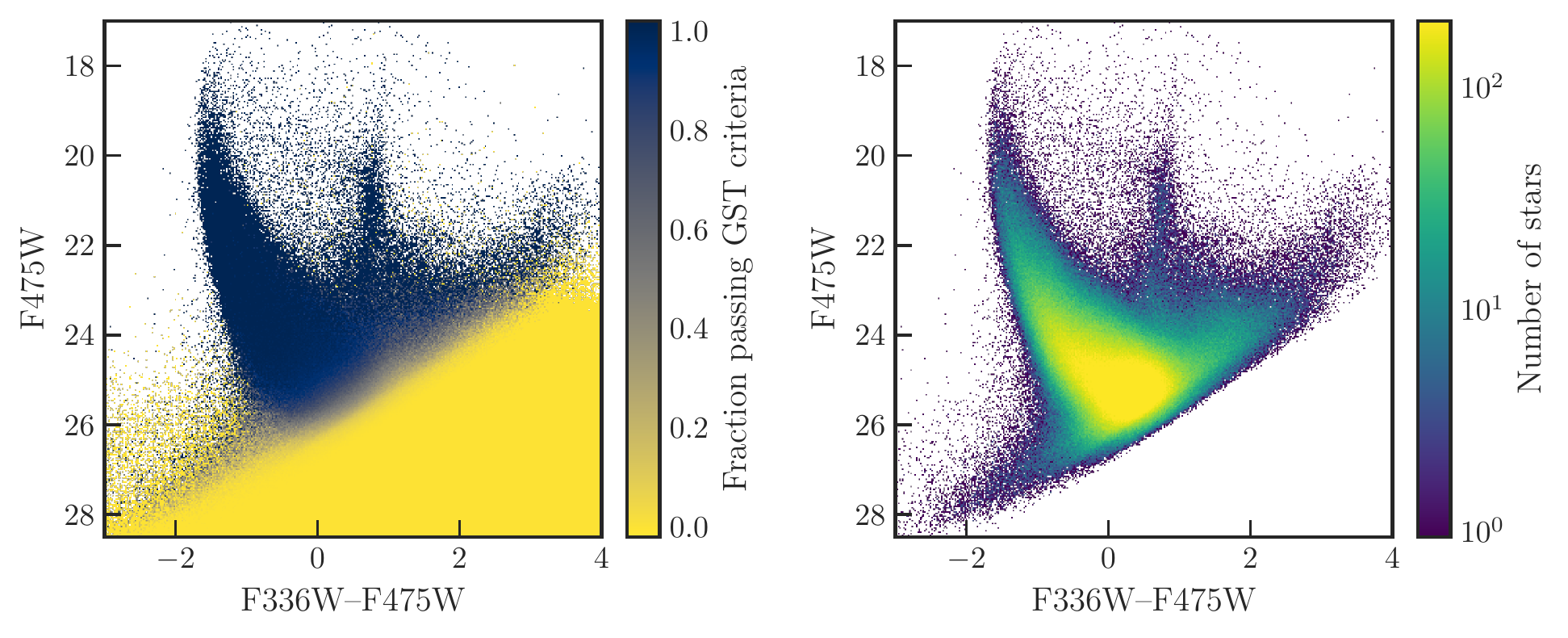}
    \caption{Same as Figure~\ref{uv_cmd}, but showing all F336W and F475W measurements. \added{The vertical feature at F336W-F475W$\sim$0.8 is caused by the Balmer jump at $\sim$3646~$\AA$ which results in stars with a relatively broad range of temperatures ($3.7<log(T_{eff})<3.9$) having a common color in these bands.} }
    \label{F336W_F475W_cmds}
\end{figure}

\begin{figure}
    \centering
    \includegraphics[width=0.8\textwidth]{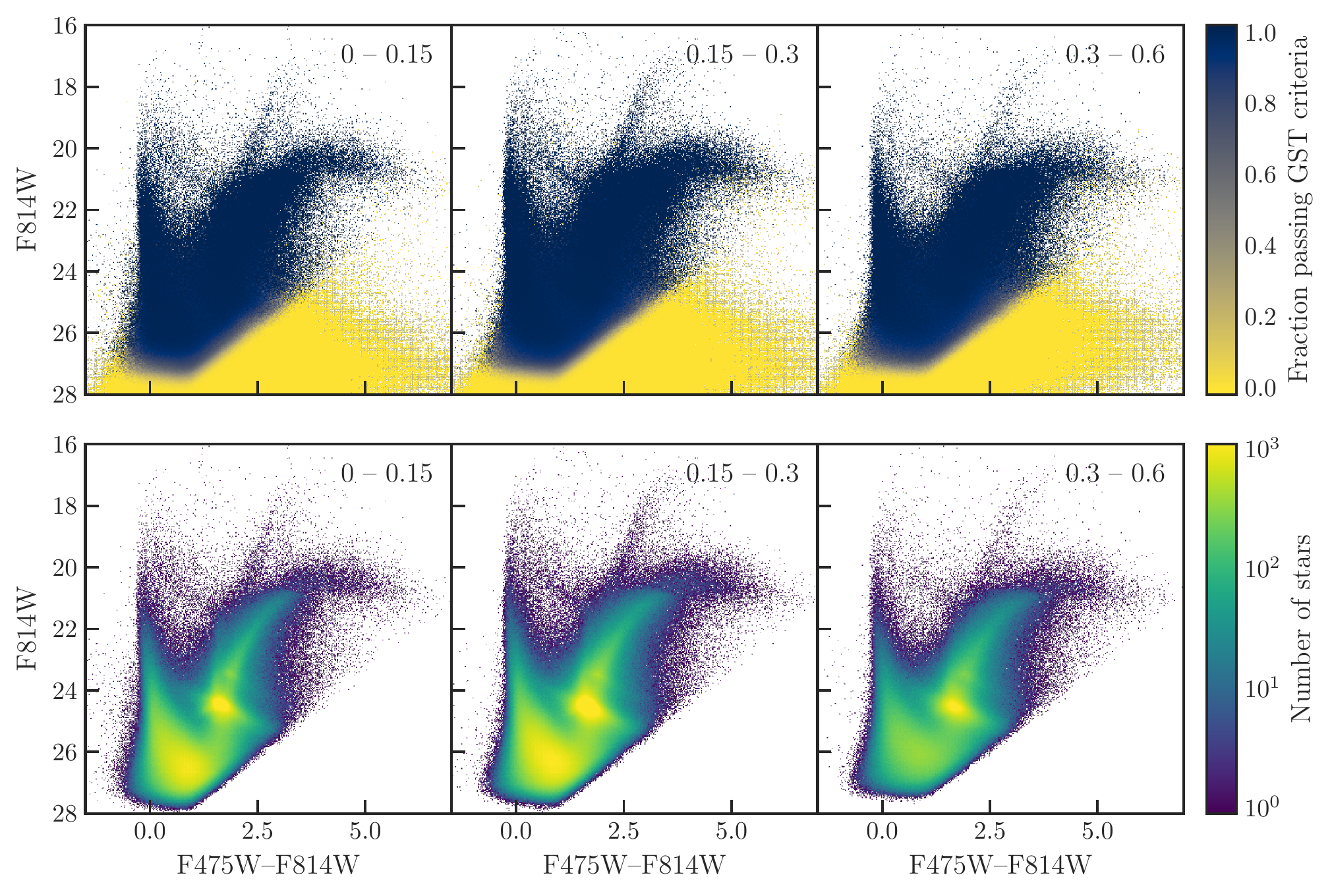}
    \includegraphics[width=0.8\textwidth]{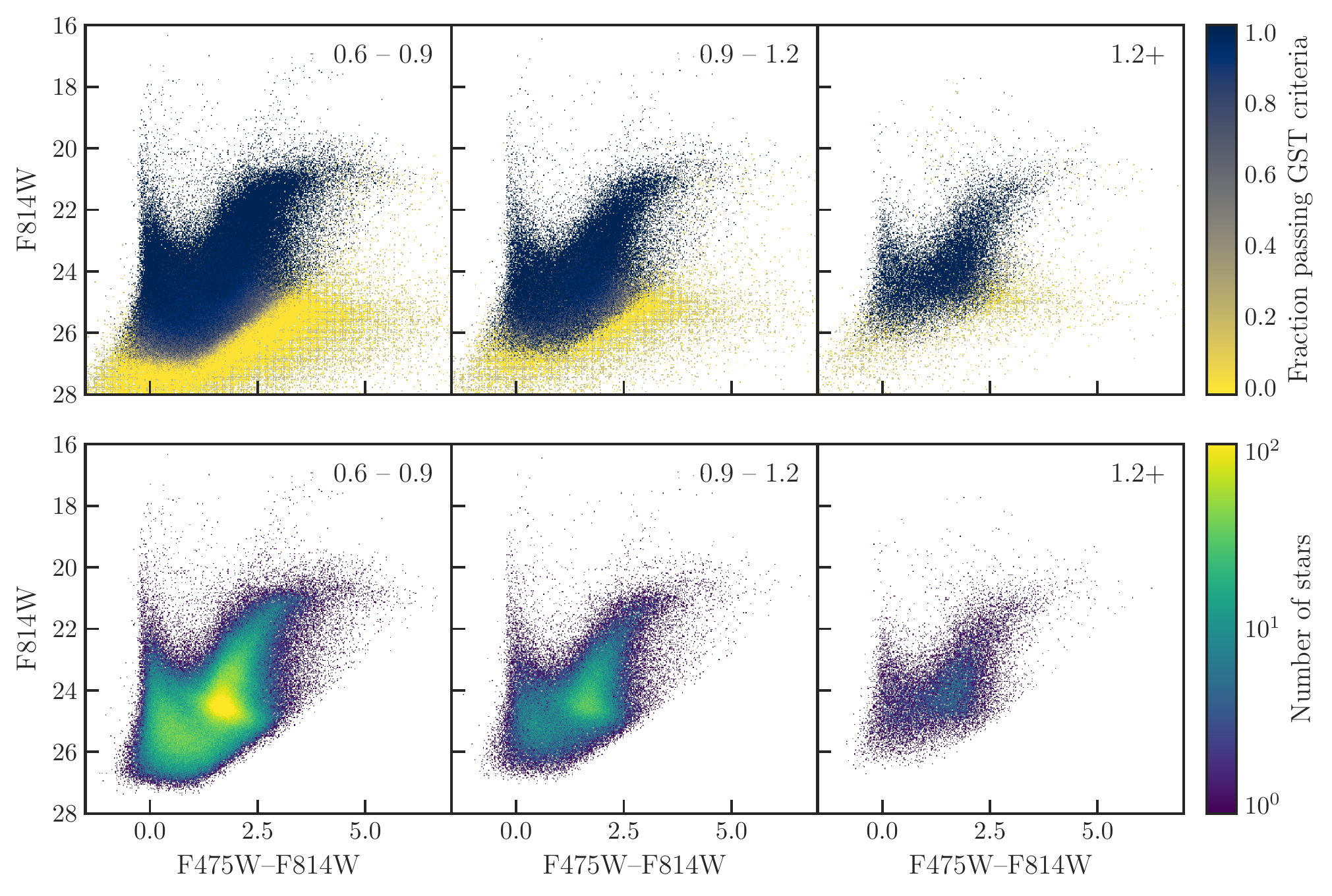}
    \caption{Same as Figure~\ref{uv_cmd}, but for all F475W and F814W measurements.  Here, we split the measurements up by stellar density to show the effects of crowding.  The stellar density range included in each CMD \added{corresponds to the density maps shown in Figure~\ref{density_map} , where density is the number of stars with $19.7{<}\mathrm{F160W}{<}20.7$ per square arcsec, and} is marked in the upper-right corner of each panel in units of stars per arcsec$^{2}$. These bands are strongly affected by crowding, as apparent by the brighter magnitude limit at the higher stellar densities.}
    \label{F475W_F814W_cmds}
\end{figure}

\begin{figure}
    \centering
    \includegraphics[width=0.8\textwidth]{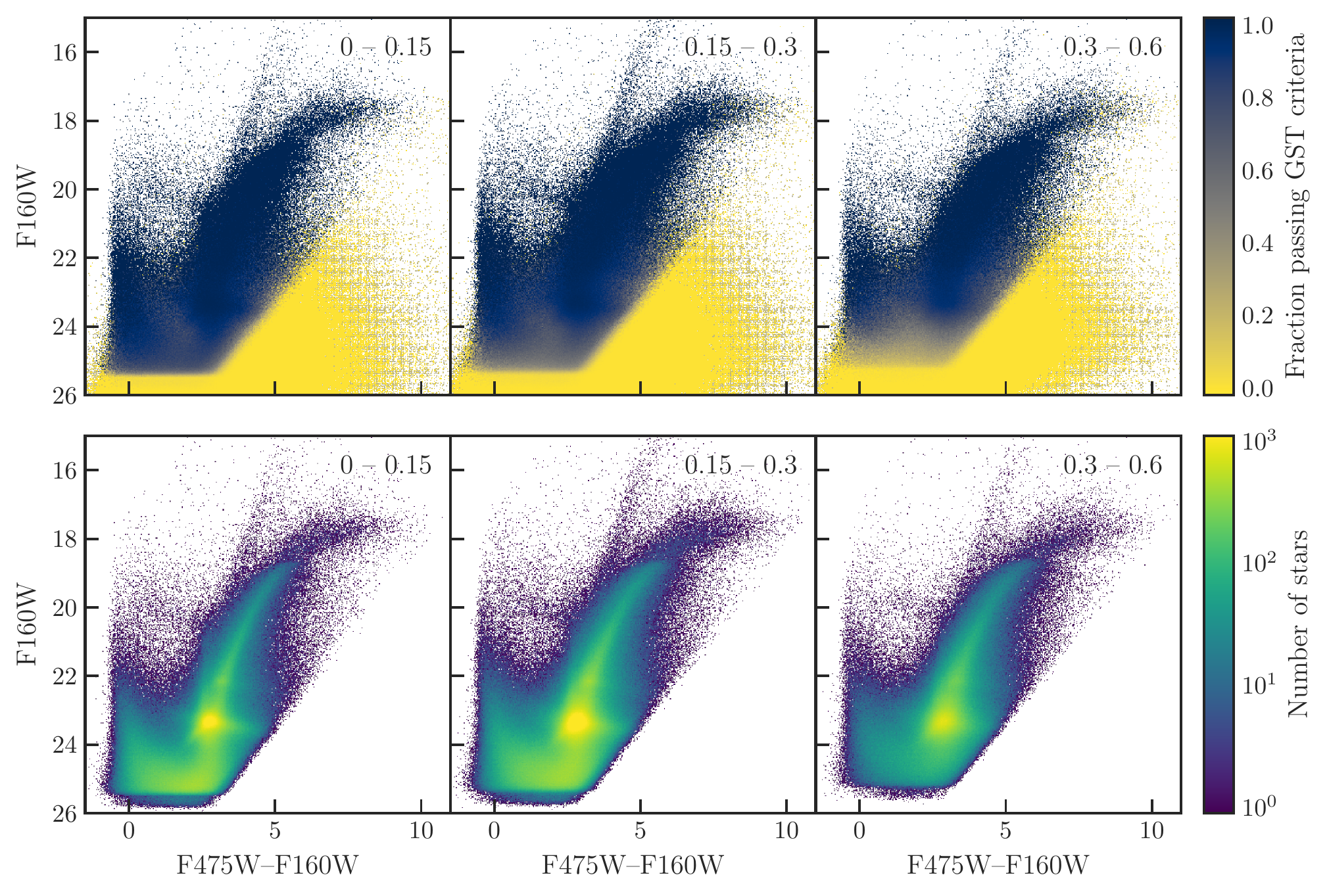}
    \includegraphics[width=0.8\textwidth]{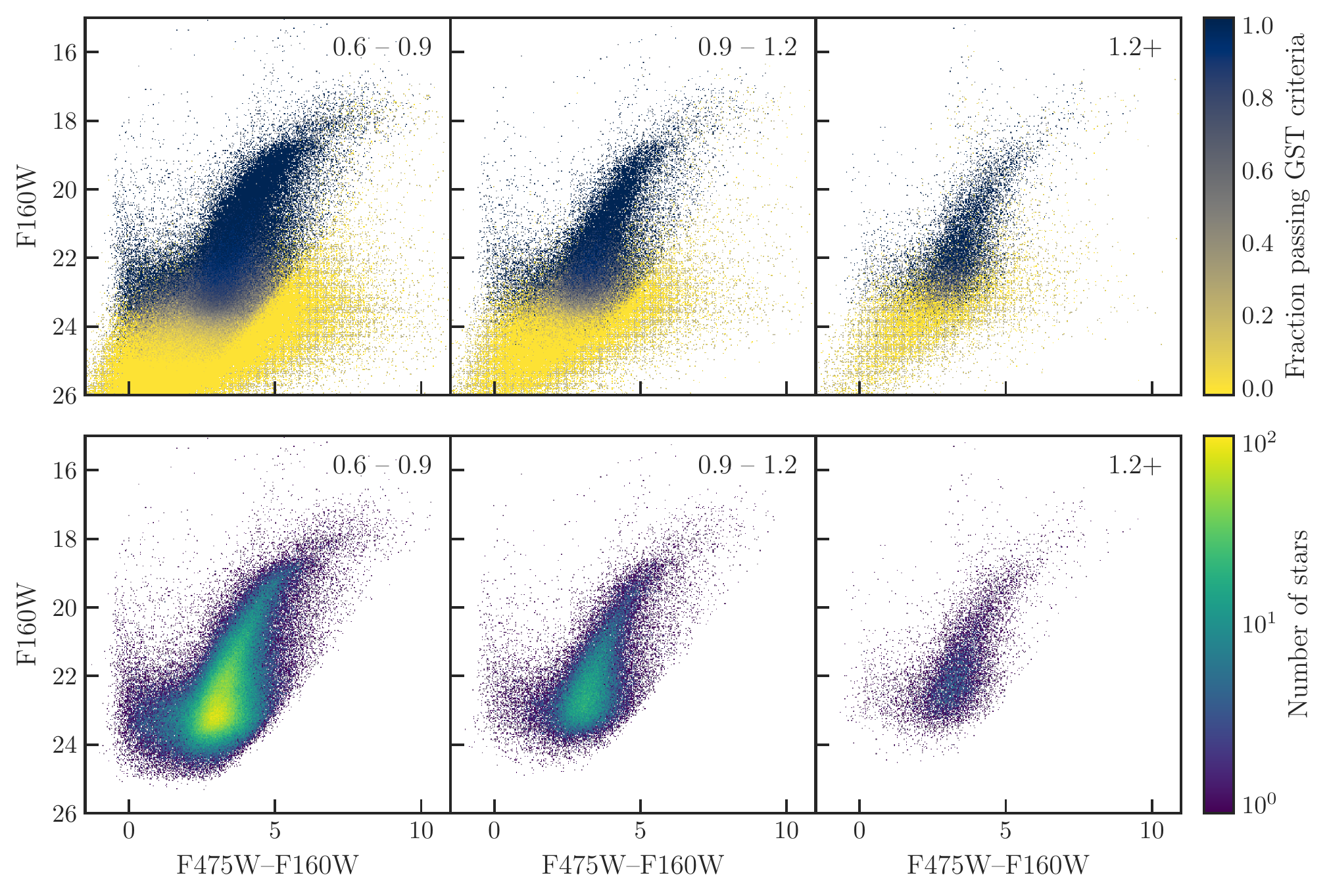}
    \caption{Same as Figure~\ref{F475W_F814W_cmds}, but for the F475W and F160W measurements.}
    \label{F475W_F160W_cmds}
\end{figure}

\begin{figure}
    \centering
    \includegraphics[width=0.8\textwidth]{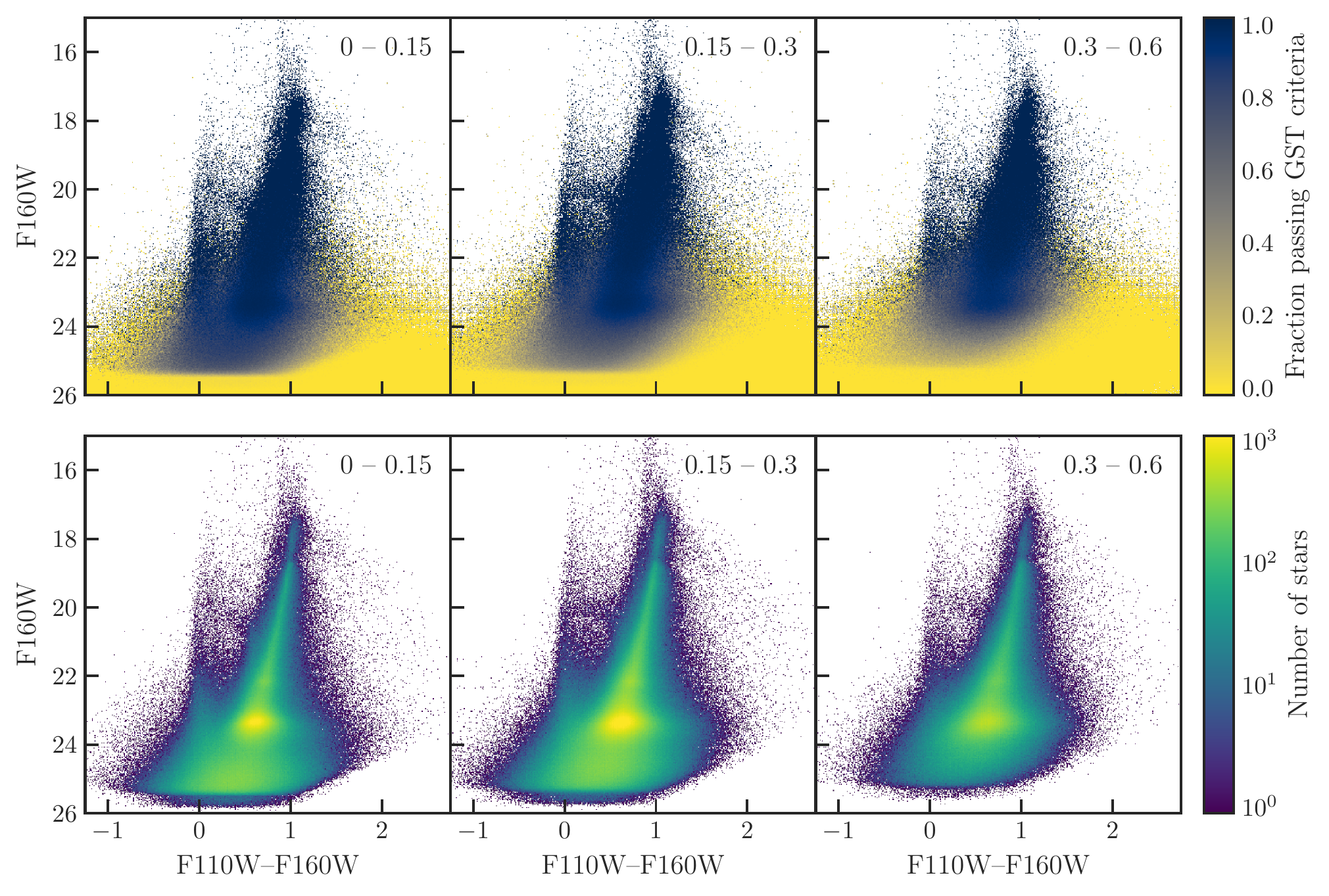}
    \includegraphics[width=0.8\textwidth]{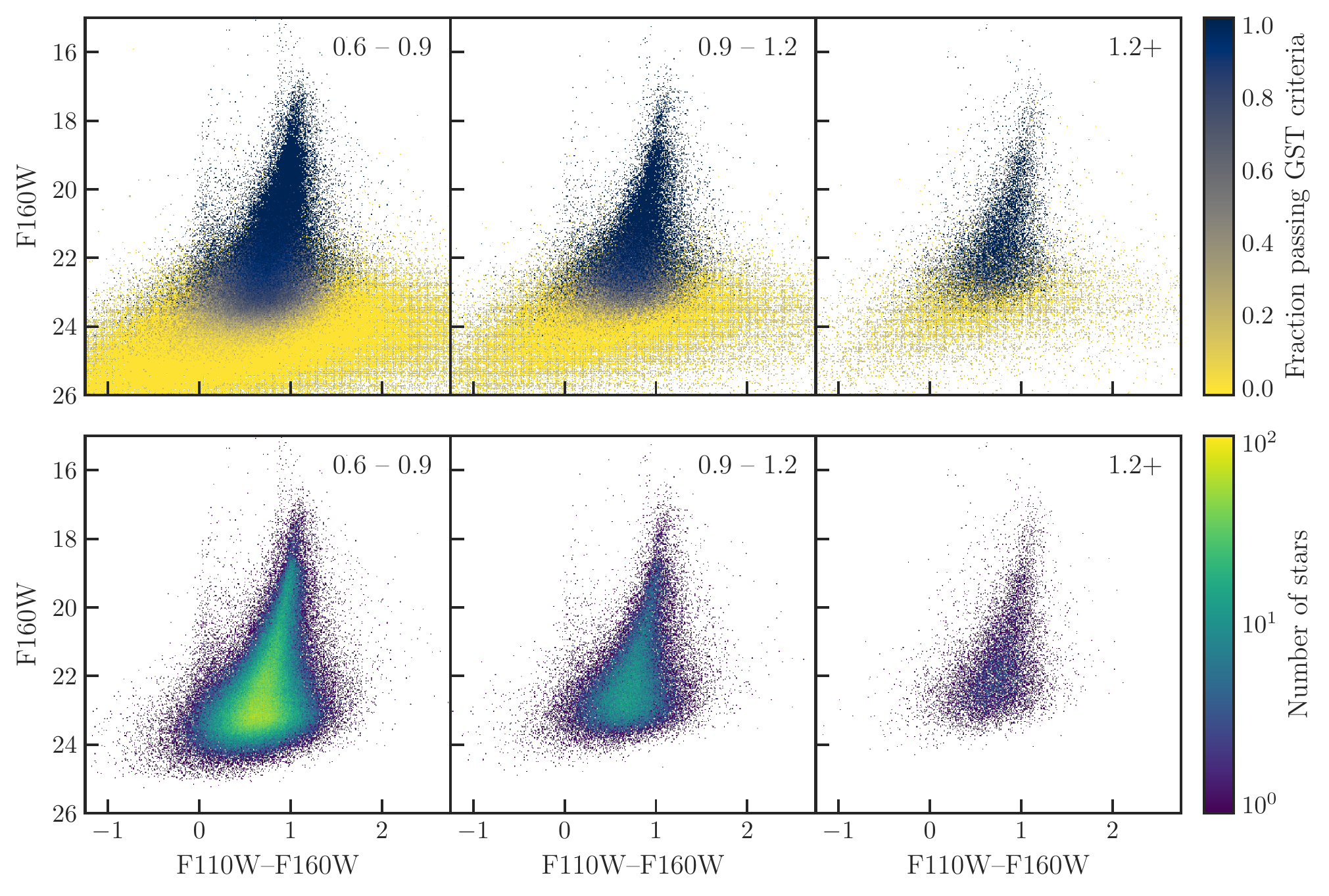}
    \caption{Same as Figure~\ref{F475W_F814W_cmds}, but for the F110W and F160W measurements.}
    \label{ir_cmd}
\end{figure}

The CMDs in all bands for the stars that pass our GST quality checks are shown in Figures~\ref{uv_cmd}-\ref{ir_cmd}.  These figures also show, in the upper panels, the fraction of accepted measurements over the same CMD space.  

In general, the highest impact of any metric on the culling of the data is the signal-to-noise ratio, which culls 100\% of the the measurements fainter than the detection limit in each band. However, in the IR, the quality metrics greatly reduce the amount of scatter in the CMD features at the faint end, as demonstrated by the low fraction of passing measurements up to 2 magnitudes brighter than the detection limit in F160W in the crowded central regions.  This difference is mainly attributable to the lower spatial resolution in the IR, which increases the impact of crowding, making more unreliable measurements that fall outside of the main features of the CMD.

\begin{figure}
    \centering
    \includegraphics[width=\textwidth]{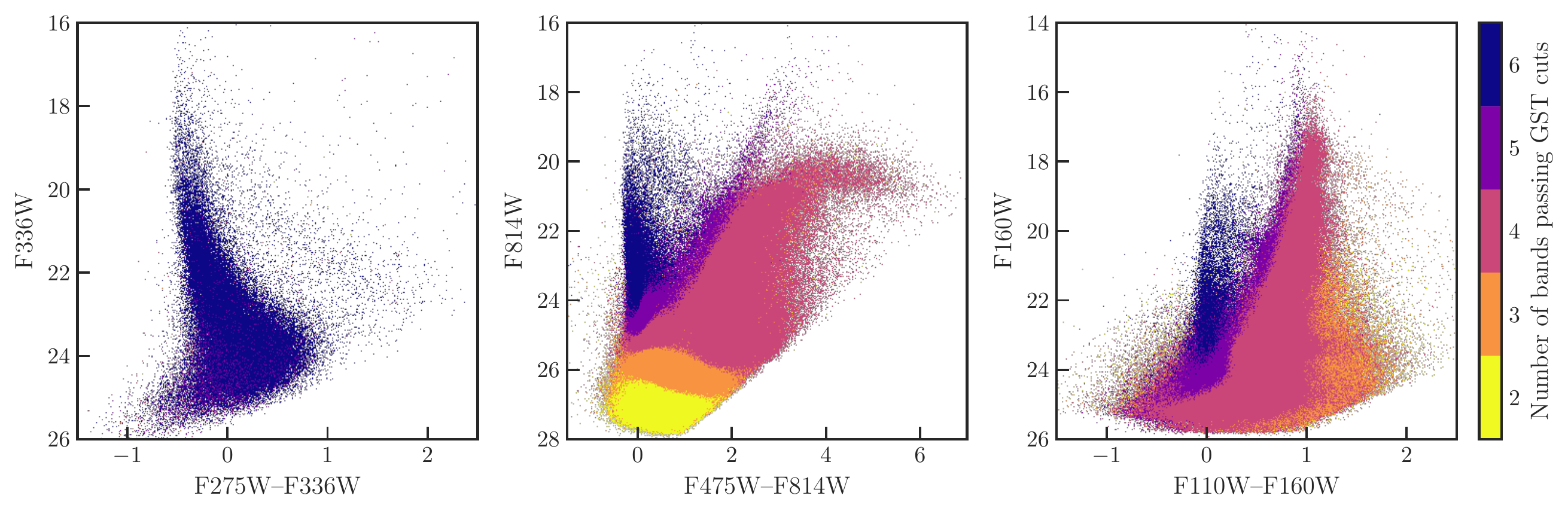}
    \caption{UV (left), optical (center), and IR (right) CMDs of the lowest density bin, with the colorbar showing the mean number of bandpasses in which a star passes the GST criteria. Nearly every detection in the NUV is detected in all 6 bands.  Most RGB and AGB stars are detected in 4 bands, and only the faintest optical stars are limited to 2 bands, highlighting the depth of the ACS data.}
    \label{filter_detections}
\end{figure}

We also show the effects of the depth in each band on our recovery of different features in Figure~\ref{filter_detections}.  Here a representative subsample of stars is plotted on CMDs color-coded by the number of bands in which they were detected.  It is clear from this figure that the UV observations are our shallowest, as nearly every UV detection is also detected in all of the other bands, and no RGB stars are detected in the UV.  On the other hand, nearly every star in the catalog is detected in the optical, and all but the faintest main-sequence stars are detected in the IR.  It is important to keep these depth effects in mind when working with the catalogs to perform analysis on the populations present in M33.

\subsection{Artificial Star Tests}

We quantify the accuracy, precision, and completeness of our photometry through artificial star tests (ASTs), wherein artificial stars with known parameters are injected into the data and then recovered (if possible).   ASTs place stars with realistic spectral energy distributions (SED) at a fixed sky position in each overlapping input image.  We then put those images through the same photometry routine as the original data, and compare the output measurements for the star to the input values.  If the star is not recovered by the photometry routine, that is also recorded.  We repeat this process many thousands of times in many locations in the survey to characterize the quality of our photometry catalogs as a function of survey stellar density, \added{where density is the number of stars with $19.7{<}\mathrm{F160W}{<}20.7$ per square arcsec, which is also a proxy of galactocentric distance, as the stellar density falls off smoothly with radius.}  We describe each step in detail below.

\begin{figure}
    \centering
    \includegraphics[width=\textwidth]{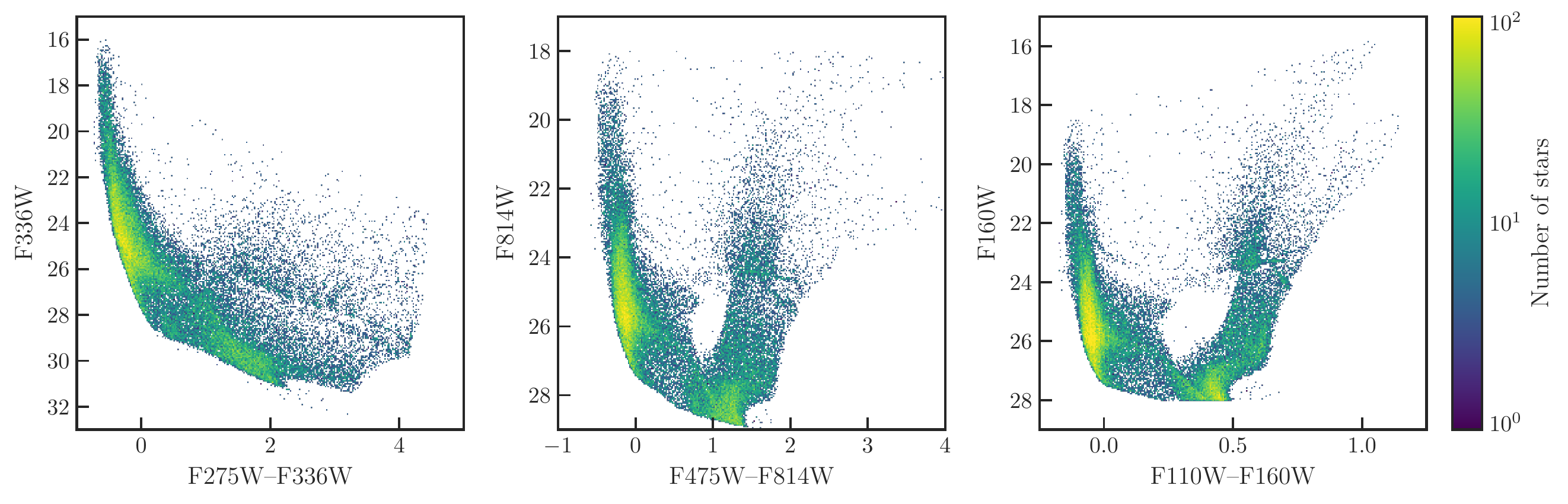}
    \caption{UV-optical-IR CMDs of artificial star inputs.}
    \label{ast_inputs}
\end{figure}

We generated input artificial star magnitudes with MATCH \citep{dolphin2002} using the {\tt fake} utility to produce a simulated 6-band photometric catalog sampled from the MIST model suite \citep{2016ApJ...823..102C}.
We used two age bins, 1 Myr to 1 Gyr and 8 to 16 Gyr, and a metallicity range of $-2 < \mathrm{[Fe/H]} < 0.5$, which together span sufficient color space to be applicable to the majority of our photometry.  We restricted the optical magnitudes to $15 < \mathrm{F475W} < 31$ and $17 < \mathrm{F814W} < 30$, but left the UV and IR magnitudes effectively unconstrained. To ensure sufficient sampling of bright stars, we used a top-heavy IMF.
CMDs of the final AST inputs are shown in Figure~\ref{ast_inputs}. \added{For the purposes of assessing our photometric quality for each band, we performed tests that covered at least one magnitude beyond the full range of magnitudes passing our quality cuts in each band, but we did not cover the full range of detected colors.  Full statistical modeling of the color distributions and stellar spectral energy distributions will require more comprehensive ASTs, but these are sufficient for determining the completeness and precision as a function of magnitude in each of our observed bands.  The results of these tests are supplied in Table~\ref{table_asts}, and described below.}



We select four regions roughly along the major axis that span the full range of stellar densities, as shown in the right panel of Figure~\ref{density_map}. For each region we create input lists of 50,000 artificial stars with random XY locations, for a total of 200,000 ASTs.
We run the stars from the input AST lists through our photometry routine one at a time, such that the ASTs were not able to affect one another.  DOLPHOT's output in AST mode includes the location and flux of each input star, followed by all of the output that is reported for all of the unaltered data.  Quality metrics that were used to flag measurements in the star catalog can then be applied to the AST catalog for consistency.

\begin{figure}
    \centering
    \includegraphics[width=0.6\textwidth]{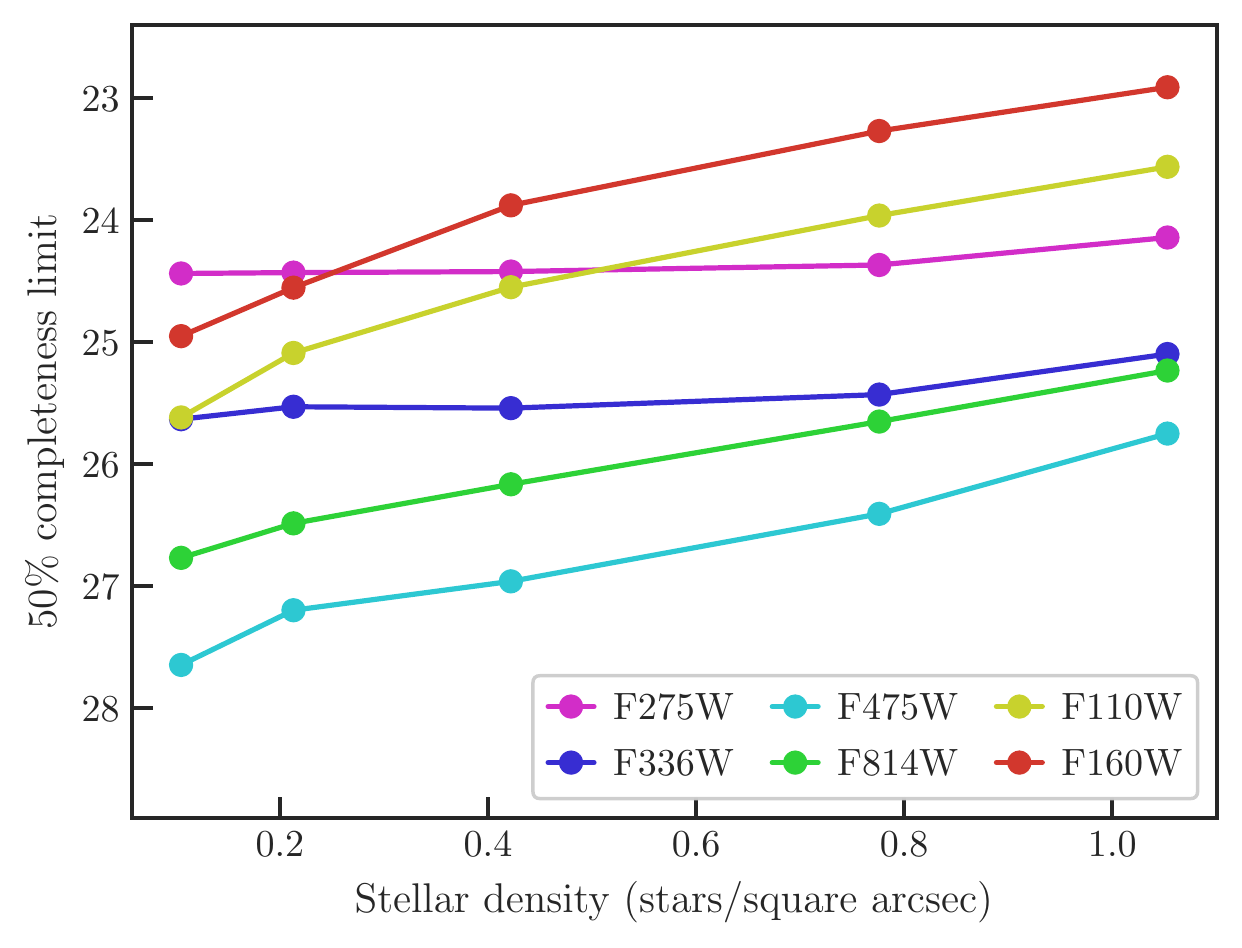}
    \caption{Magnitudes at which we measure 50\% completeness by stellar density in all filters. \added{Stellar density corresponds to the maps in Figure~\ref{density_map}, where density is the number of stars with $19.7{<}\mathrm{F160W}{<}20.7$ per square arcsec; specifically, the right hand panel is binned to complement the data points in this plot.} Completeness limits in the UV are largely consistent over the full density range of the survey, whereas they grow brighter with increasing density in the optical and NIR due to crowding.}
    \label{density_completeness}
\end{figure}

\added{While Table~\ref{table_asts} reports all of the output measured magnitudes, for our quality analysis,} we consider an artificial star to be ``recovered" in a given band if it is within 2 reference frame pixels ($0.07\arcsec$) of the input source position \added{and 1 magnitude of the input magnitude}, and fulfills the GST (``good star") quality requirements for said band discussed in Section~\ref{flagging_section}.
Figure~\ref{density_completeness} and Table~\ref{table_comp} provide the completeness as a function of magnitude as well as the magnitude $m_{50}$ at which 50\% of inserted artificial stars are recovered (the ``50\% completeness limit").  For typical astronomical point sources, this completeness limit is largely set by the number of photons detected from an astronomical source.  However, at high stellar densities, the completeness limit is set by the magnitude at which the surface density of sources (i.e., \# of sources per square arcsec) is so high that they are always blended with brighter sources, rendering the original source undetectable. 
In this ``crowding limited" (rather than ``photon limited") regime, the limiting magnitude is set more by stellar density than by photon counting statistics.  

\begin{figure}
    \centering
    \includegraphics[width=0.45\textwidth]{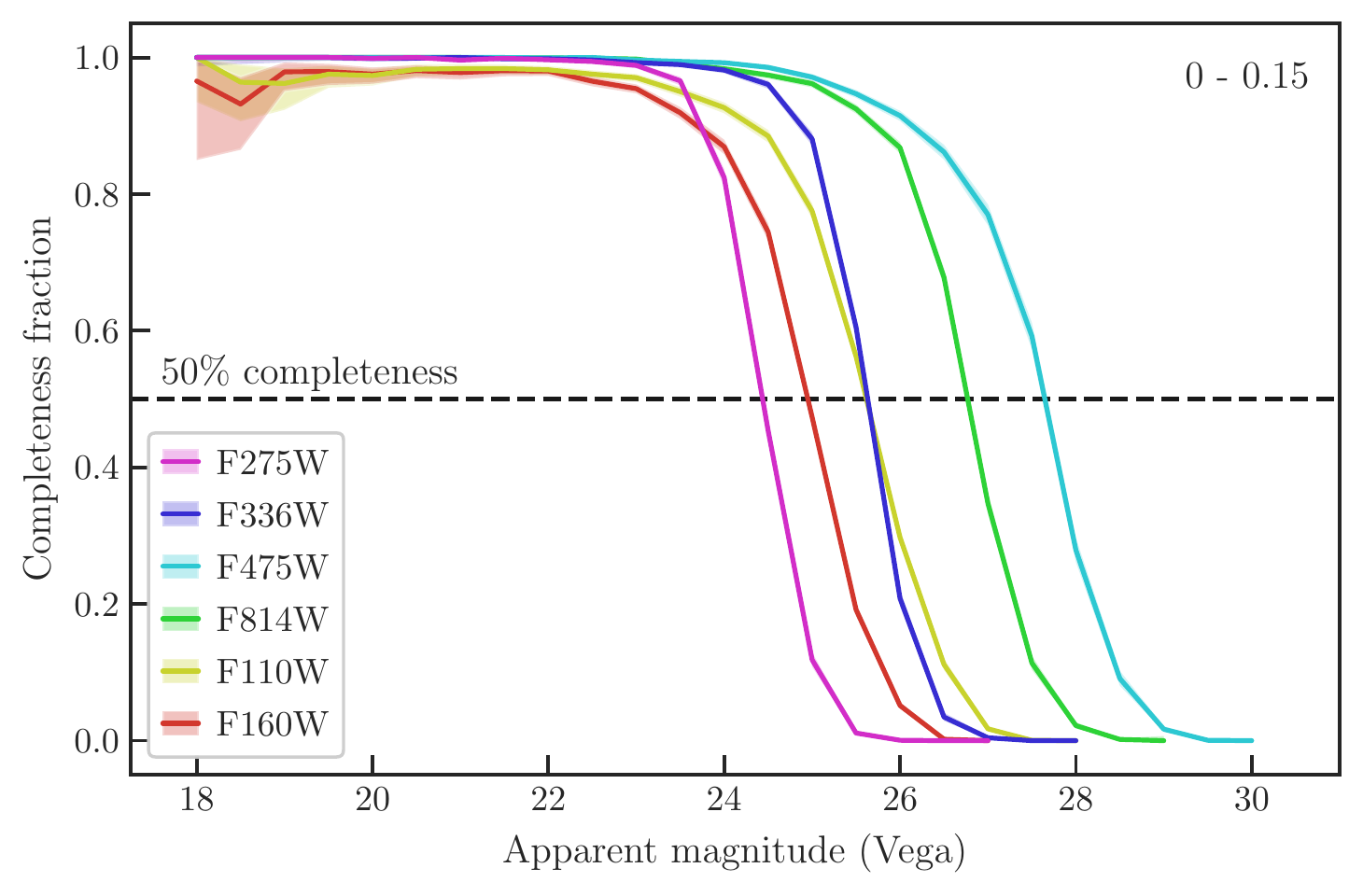}
    \includegraphics[width=0.45\textwidth]{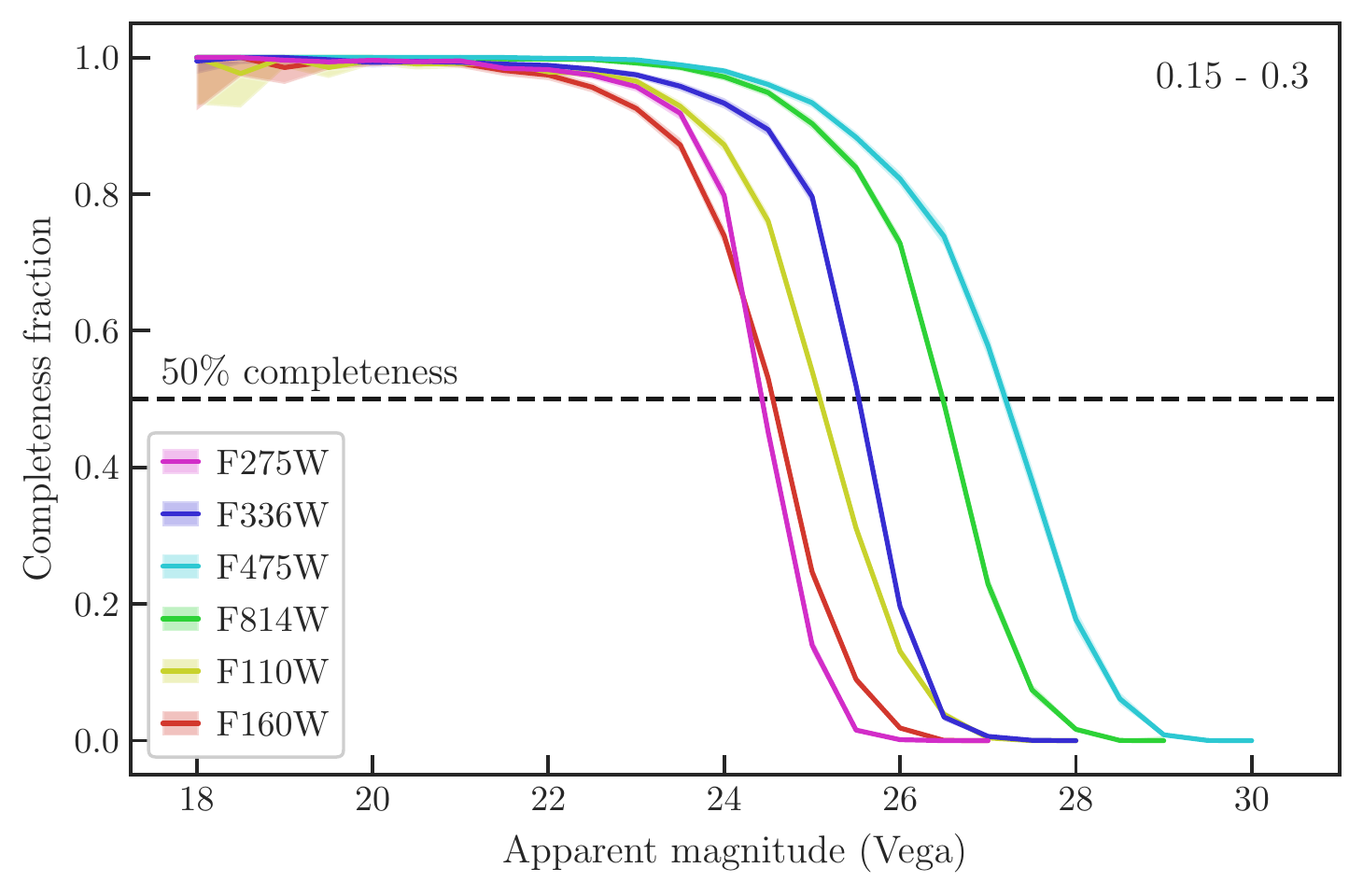}
    \includegraphics[width=0.45\textwidth]{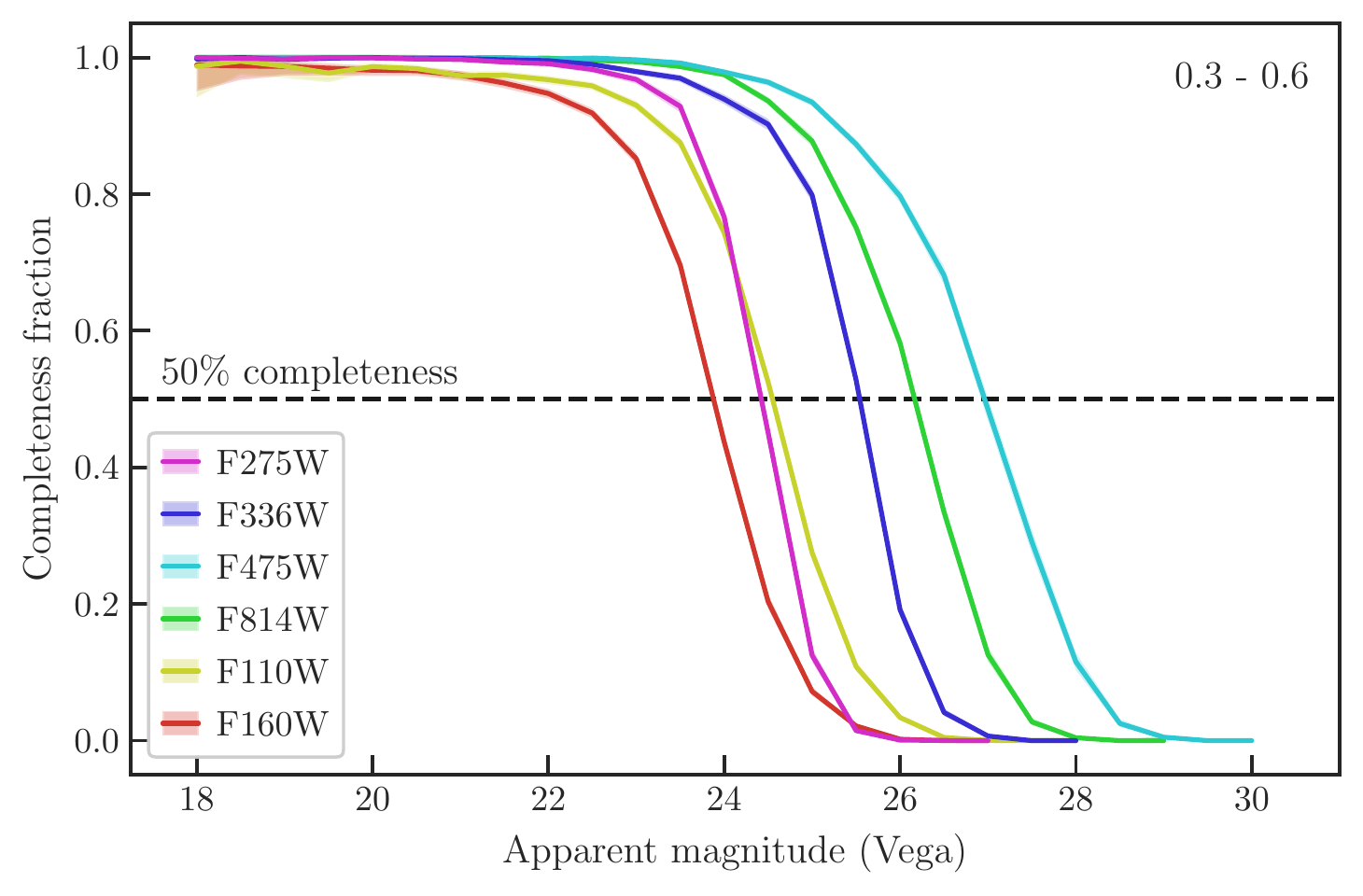}
    \includegraphics[width=0.45\textwidth]{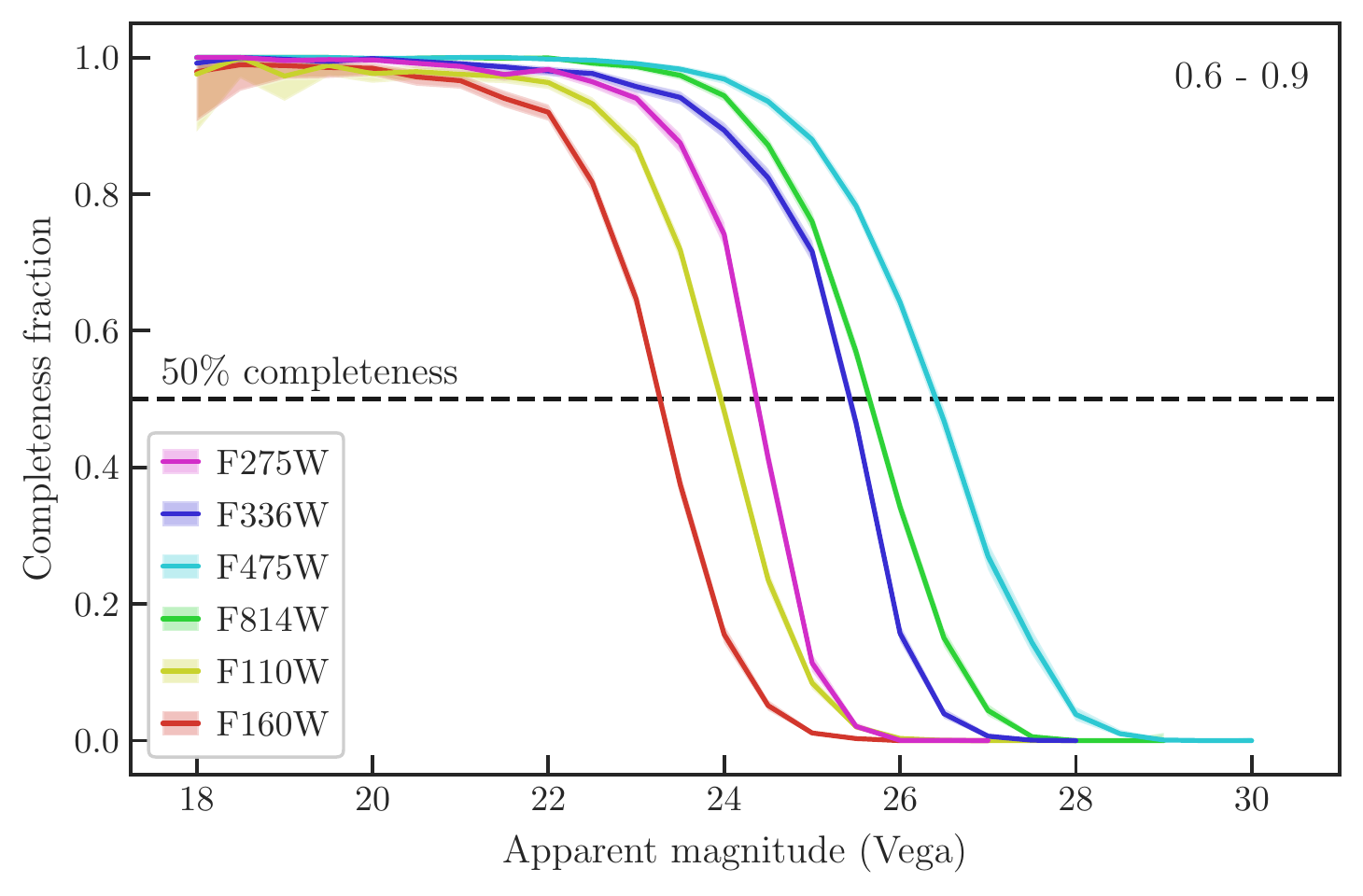}
    \includegraphics[width=0.45\textwidth]{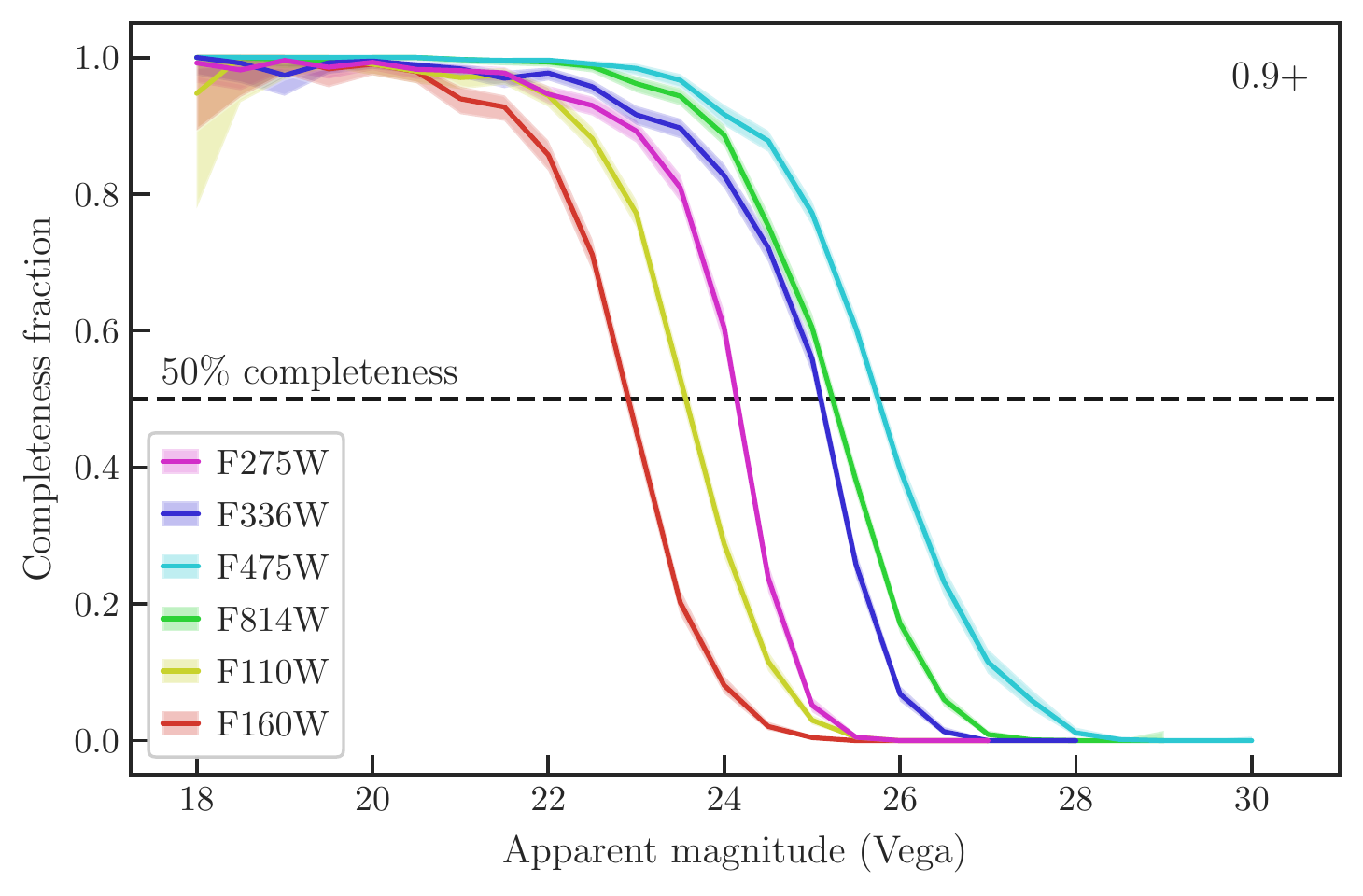}
    \caption{Photometric completeness (fraction of input stars that pass quality cuts) as a function of input magnitude in all filters for five characteristic density bins (labeled in upper right corners,\added{where density is the number of stars with $19.7{<}\mathrm{F160W}{<}20.7$ per square arcsec}). The shaded regions show 95\% confidence using the Jeffreys interval.}
    \label{completeness_functions}
\end{figure}

In both M31 and M33, {\it HST} imaging is crowding limited in the optical and NIR over much of the disk, with the effects being most significant in the NIR where the larger pixel scale and PSF size severely limit detection and reliable measurement of faint stars.  In contrast, the PHAT and PHATTER observations in the UV bands are sufficiently shallow that they do not reach magnitudes where UV-detectable stars are so numerous that they begin to crowd together.  For M33, the UV observations reach F275W$\sim$24.5 relatively independent of stellar density \added{(or galactocentric distance)}, as expected for photon-limited images, whereas for the optical, the depth changes by $\sim$1.3 magnitudes moving from the inner to outer disk, reflecting the role that stellar crowding plays in setting the detection limit.  In addition, the variation in completeness with magnitude (Figure~\ref{completeness_functions}) is qualitatively different in the photon-limited and crowding-limited data.  In the former, the completeness drops from near 100\% to 0\% over a narrow range in magnitude ($\lesssim$1) , whereas in the crowding-limited data, the roll-off in completeness is much more gradual with magnitude ($\gtrsim$2), such that stars begin to be ``hidden" by crowding several magnitudes before the magnitude at which they disappear from the catalog.  The slow roll-off in completeness is the result (in part) of increasing odds that a star will fail the quality cuts with the increasing likelihood of it blending with a star of comparable flux.

As with completeness, photometric uncertainties reflect impacts from both photon-counting uncertainties and crowding.  There are multiple contributors to photometric uncertainty and bias in crowded-field photometry beyond the well-known impact of photon-counting statistics for the source and sky.  These effects include uncertainties and biases from deblending of neighbors and sky estimation, as well as brightward biases from blending with undetected sources (which also increases the chance of detection).  These effects are captured well by artificial star tests, though other systematic effects due to CTE or imperfect PSF models will remain.   These various drivers of uncertainty and bias --- crowding, exposure time, and background --- all vary among filters and cameras, and thus will have different behavior in each.

\begin{figure}
    \centering
    \includegraphics[width=0.8\textwidth]{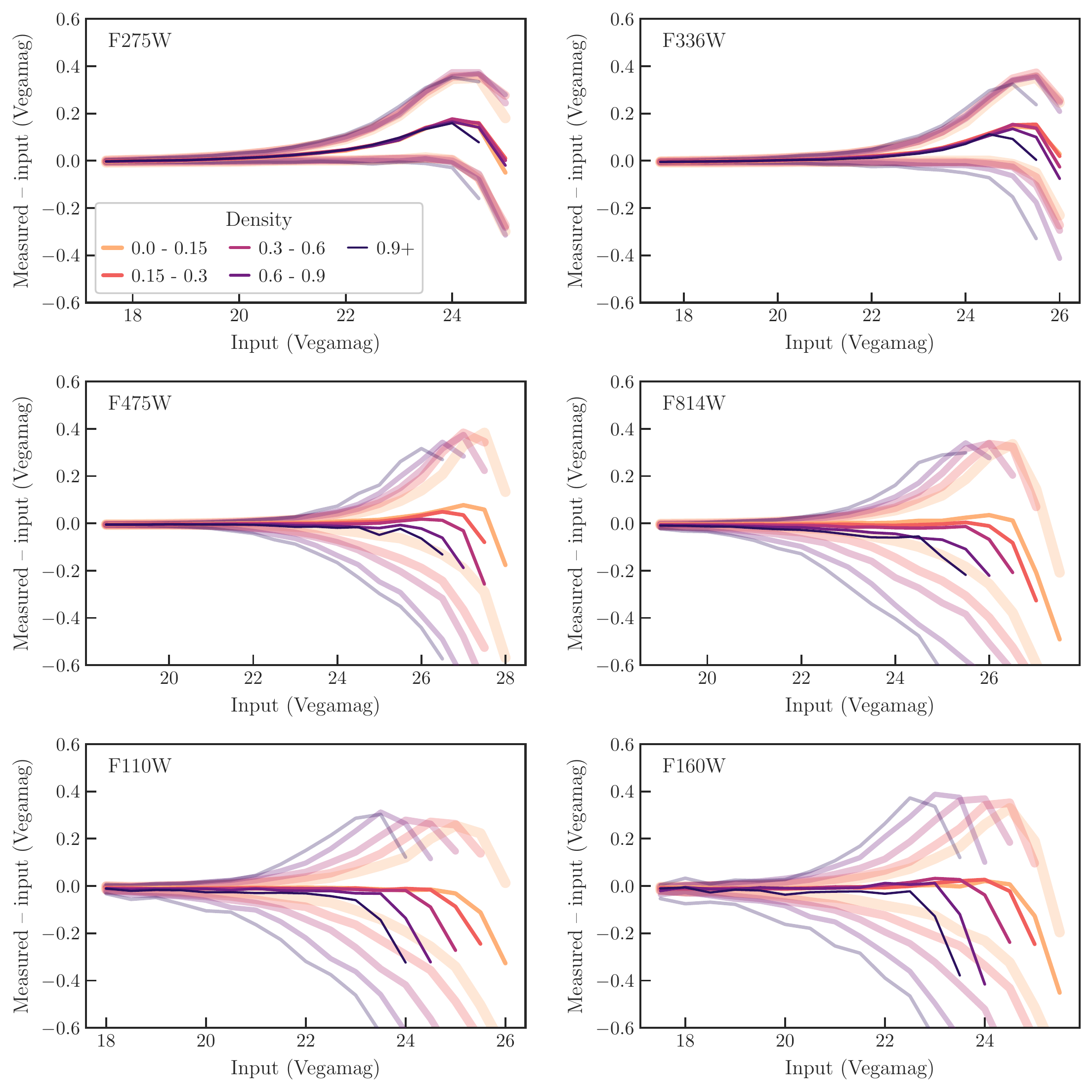}
    \caption{Photometric bias (thin solid lines) and $\pm1\sigma$ uncertainty ranges (thick faded lines) derived from ASTs as a function of input magnitude in each filter for five density bins, \added{, where density is the number of stars with $19.7{<}\mathrm{F160W}{<}20.7$ per square arcsec}. The bias is taken to be the median of the measured minus input AST magnitudes in half-magnitude bins, and the uncertainty bounds are the 16th and 84th percentiles of the same. Darker line colors correspond to higher densities.}
    \label{bias_uncertainty}
\end{figure}

We summarize the AST results for uncertainties and bias in Figure~\ref{bias_uncertainty} and Table~\ref{table_bias}.  Figure~\ref{bias_uncertainty} shows the median difference in magnitude between the recovered and input magnitudes (recovered - input) as a function of input magnitude, and the 16th and 84th percentile ranges for the distribution of differences, shown as solid and transparent lines, respectively, plotted for a range of mean local densities (different color lines, with darker, thinner lines indicating higher stellar densities).  Positive values indicate sources that are recovered at fainter magnitudes than their true magnitudes.  Table~\ref{table_bias} compiles numerical measurements of the bias and uncertainty for different filters and source densities.  The uncertainty is also reported in units of the DOLPHOT-reported photometric uncertainty, which is based entirely on photon-counting uncertainties.  The measured scatter between the true and recovered magnitudes is typically $\sim$20\% larger than the photon-counting uncertainty in the NUV, a factor of $\sim$4 larger in the optical, and a factor of $\sim$5 larger in the NIR.

Figure~\ref{bias_uncertainty} shows that, as expected, both the bias and and the measurement uncertainty increase towards fainter magnitudes, where photon counting and crowding is worse.  The biases are much smaller than the photometric uncertainties (typically by a factor of 2-4) at all but the very faintest limits, where very few sources would be recovered at all.  At a fixed magnitude in the optical or NIR bands, the biases and uncertainties are larger in regions with higher source densities, due to the higher crowding.  In the optical and NIR bands, as sources become intrinsically fainter, their measured fluxes tend to be biased towards brighter magnitudes, due to unresolved, overlapping sources boosting the inserted artificial star above the detection limit.  These effects are somewhat more pronounced in the NIR, most likely due to the camera's larger pixels and longer wavelengths producing lower resolution images (see Figure~\ref{4panel}) and thus larger impacts due to crowding.  No corrections for these biases have been made to the catalog. 

In the UV, the trend of increasing bias and uncertainty for fainter sources is similar to what is seen in the optical and the NIR.  However, the variations with UV magnitude are largely independent of local source density, reflecting the lack of significant crowding except at the very highest density in F336W.  Another notable difference is in the sign of the bias.  Well before completeness begins to decline significantly, the bias begins to become substantial, but has the opposite sign as seen in the optical and NIR, such that measurements seem to be biased significantly faint.  The effect appears most consistent with a slightly high background measurement, since the bias induced from high sky subtraction would be very small for bright sources, and increase for fainter source, as we see in the NUV photometry.  A similar trend was seen in the W14 PHAT photometry study, and the speculation was that perhaps charge transfer efficiency (CTE) effects were causing the sky brightness to be overestimated.  However, in this work we have used pre-flashed, CTE-corrected UVIS images, which should have reduced CTE effects on the sky brightness.  Nonetheless, it is clear that our technique is likely attributing too much flux to sky in the NUV images.  \added{None of these biases have been corrected for in the catalog, although they can be accounted for using our ASTs.}

\section{Results}

Tables~\ref{table_phot} and \ref{table_asts} provide samples of the photometry catalog and AST results from the survey. The catalog included here contains the positions, magnitudes, signal-to-noise ratio, and data quality flag for each detected star. \added{While this table contains all of the stars detected and measured by DOLPHOT, not all of those measurements are likely to be stars in M33 with reliable photometry.  To simplify the use of this version of the catalog, we have supplied a flag column (GST) for each band, which is set to ``T" for sources that pass our quality checks detailed in Section~2.2.4, and to ``F" for sources that do not.  If a source has a T in any band, it is likely to be a real star in M33.  However, it still may have ``F" flags in other bands, which indicates that the star was not reliably measured in that band.  Only sources with ``F" values in all bands are unlikely to be stars in M33.  

Many of the stars in our catalog, as for a large fraction of stars, are likely to be binary systems \citep[e.g.,][]{niu2020}.  At the distance of M33, we are not able to resolve binary stars, so that our photometry contains the light of both members.  However, the impact of binary stars varies by phase of stellar evolution.  Many of the stars in our catalog are red giants, which is a short-lived and very bright phase of stellar evolution in the life of a relatively low-mass star.  The binary companions of almost all of these stars are not in their red giant phase and are faint low mass stars contributing insignificantly to the measured flux. In contrast, the more massive stars that make up the bright main-sequence are more likely to have bright companions of similar mass \citep[e.g.,][]{kobulnicky2007}.  The companions of these stars will affect their photometry by contributing as much as half of the measured flux and should be considered when modeling photometry of upper main-sequence stars in our catalog.}

The comprehensive, and much larger, catalog is available as a high-level science product (HLSP) in the Multimission archive via \dataset[10.17909/t9-ksyp-na40]{\doi{10.17909/t9-ksyp-na40}}.  This comprehensive catalog includes the combined measurements of each star in each band, as well as in each of the individual measurements in all of the survey exposures, along with all of the measurement quality information reported by DOLPHOT (uncertainty, $\chi$, \texttt{sharp}, \texttt{round}, \texttt{crowd}, error flag).  This catalog includes thousands of columns, and is hundreds of GB in size.

The simplified AST results (with limited columns) in Table~\ref{table_asts} are the location, input magnitude, output magnitude, output signal-to-noise, and output quality flag for each artificial star.  The full catalog with all of the columns includes the input counts into each individual exposure, as well as all of the output photometry measurement columns as for the detected stars in the survey.  As such, the HLSP catalog again is much larger and contains thousands of columns for those who would make use of the full AST input and output.

\subsection{Color-magnitude Diagrams}

In Figure~\ref{full_cmd} we plot the entire catalog of detections in the optical bands, and we label the strongest features.  The left panel shows all of the measurements, and the right panel shows the measurements that do not pass our quality metrics. \added{The vast majority of these failed measurements are located at the faint end where spurious and blended measurements are much more likely, while the few brighter ones are most likely contaminants and artifacts.}  This overview CMD shows the high fidelity of the photometry, which produces well-populated and clearly defined features. 
The high-definition of these features, described below, all suggest that a very large fraction of our photometry is reliable.  

On the blue edge, the vertical plume of the upper-main sequence (MS) is narrow and confined to a sharp edge determined by the saturation of color when the effective temperature of stars reaches hotter than $\sim$10$^4$~K.  Slightly to the red of this, is a second, less populated blue plume that is the blue helium burning (BHeB) sequence. This sequence marks the bluest extremity of the loop that characterizes the core-helium-burning phase of stars of intermediate and high masses. Its continuous appearance suggests that M33 has been forming stars at a relatively high intensity for hundreds of Myr.  

The next bright plume to the red (brighter than F814W$\sim$20 and starting at F475W--F814W$\sim$2) is the red helium-burning (RHeB) sequence.  This feature  consists of massive stars in the initial stage of core-helium-burning, with convective envelopes, before the decrease in the central helium content that drives their move towards the blue BHeB. It also contains the stars at the very latest phases of core-helium-burning, that move to the red again as their He-exhausted cores contract and extended convection sets in their envelopes. In theory, this RHeB sequences extends down in the CMD until it merges with the red clump (RC) of low-mass core-helium-burning stars at F814W$\sim$25.  
The width of the color gap between the RHeB and the BHeB is sensitive to the metallicity, as more metal rich stars will be redder during this phase.  

\added{These sequences (MS, BHeB, and RHeB) are all indicative of recent star formation, and their numbers are consistent with the much higher star formation rate density in M33 relative to M31.  For example, our catalog has $\sim$30\% more OB stars ($F336W-F475W<0$, M$_{F475W}<0$) than the entire PHAT catalog ($\sim$300000, vs. $\sim$230000 in PHAT), resulting in a factor of 10 larger density of these young stars in M33 than in M31.  Maps of recent star formation history, similar to those for M31 \citep{lewis2015}, are in preparation.}

\added{The cloud of stars redward of the RHeB consist of asymptotic giant branch (AGB) stars that have expanded to be so large and cool that they become very bright and very red. The brightest among these stars can undergo extreme mass loss and develop dusty circumstellar shells, making them extend to fainter and extremely red optical colors.}

\added{Below the AGB is the familiar red giant branch (RGB) of evolved shell-hydrogen burning low-mass stars, extending from F814W$\sim$21 down through the AGB bump and red clump (RC) to the subgiants at F814W$\sim$26. The AGB bump is comprised of early AGB (EAGB) stars, which are low-mass stars that undergo a pause in evolution when forming their double-shell structure \citep{gallart98}.  Below the AGB bump is the prominent RC of stable core-helium burning low mass stars.  In addition, there is a small tail of stars departing from the RC towards the top-left part of the CMD, which we tentatively identify as blends of RC plus MS stars, either because they are crowded together or are real binaries.  Tails of reddened stars depart from the most prominent features, namely the RC and AGB bump, along the reddening vector, reflecting spreads in the internal extinction that often exceed half a magnitude in the $V$ band.}

\begin{figure}
    \centering
    \includegraphics[width=\textwidth]{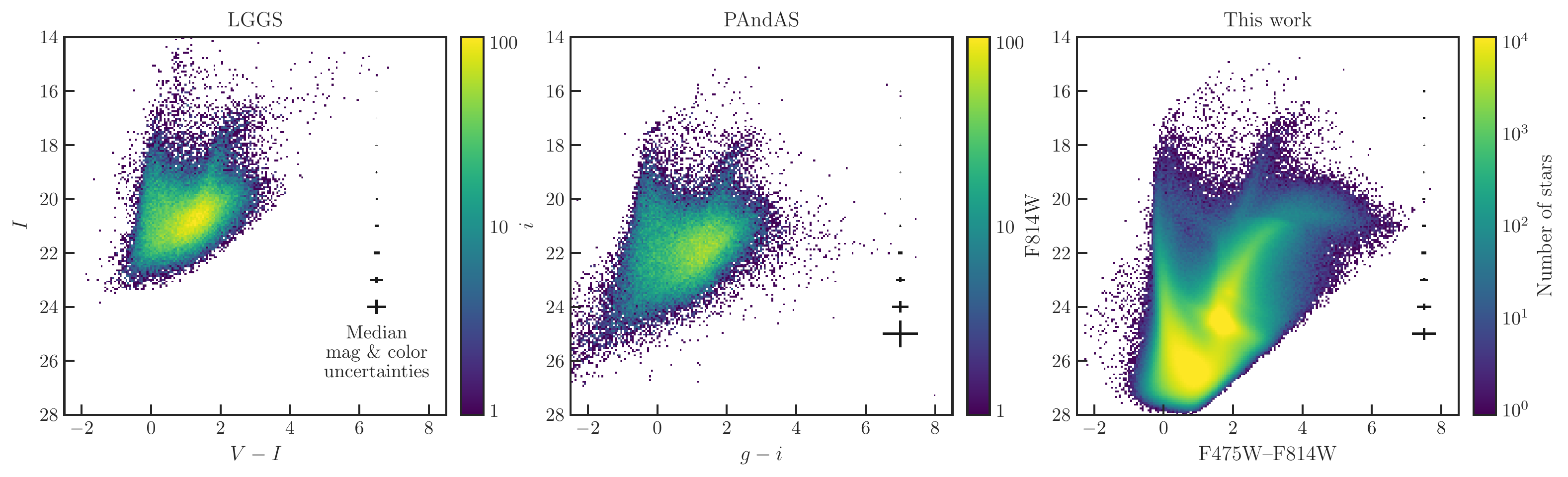}
    \caption{A comparison of optical CMDs of M33 with photometry from 
    LGGS (left, $VI$), PAndAS (center, $gi$), and this work (right, F475W/F814W GST).
    Although the filter systems are not identical, we use a common color and magnitude range on all axes to illustrate the difference in depth that can be achieved with \textit{HST}.
    The LGGS and PAndAS catalogs have been culled to cover approximately the same area as the \textit{HST} survey, but have not been culled on any photometric quality metrics.
    The median color and magnitude uncertainties in 1-mag bins are shown on the right side of each panel (black lines).
    }
    \label{fig:lggs}
\end{figure}

\added{We compare our photometry to ground-based photometry of the same region from LGGS \citep{massey2006} and PAndAS \citep{2018ApJ...868...55M} in Figure~\ref{fig:lggs}. 
Compared to ground-based imaging, our catalog extends 3 to 5 magnitudes deeper, has three orders of magnitude more stars (22 million vs.~$\sim$70,000 in both LGGS and PAndAS), and has qualitatively more and tighter feature.  This comparison clearly demonstrates the improvement in available photometry and the power of HST to resolve stars in the M33 disk.} 

\added{\subsection{Star Clusters}}

\begin{figure}
    \centering
    \includegraphics[width=0.8\textwidth]{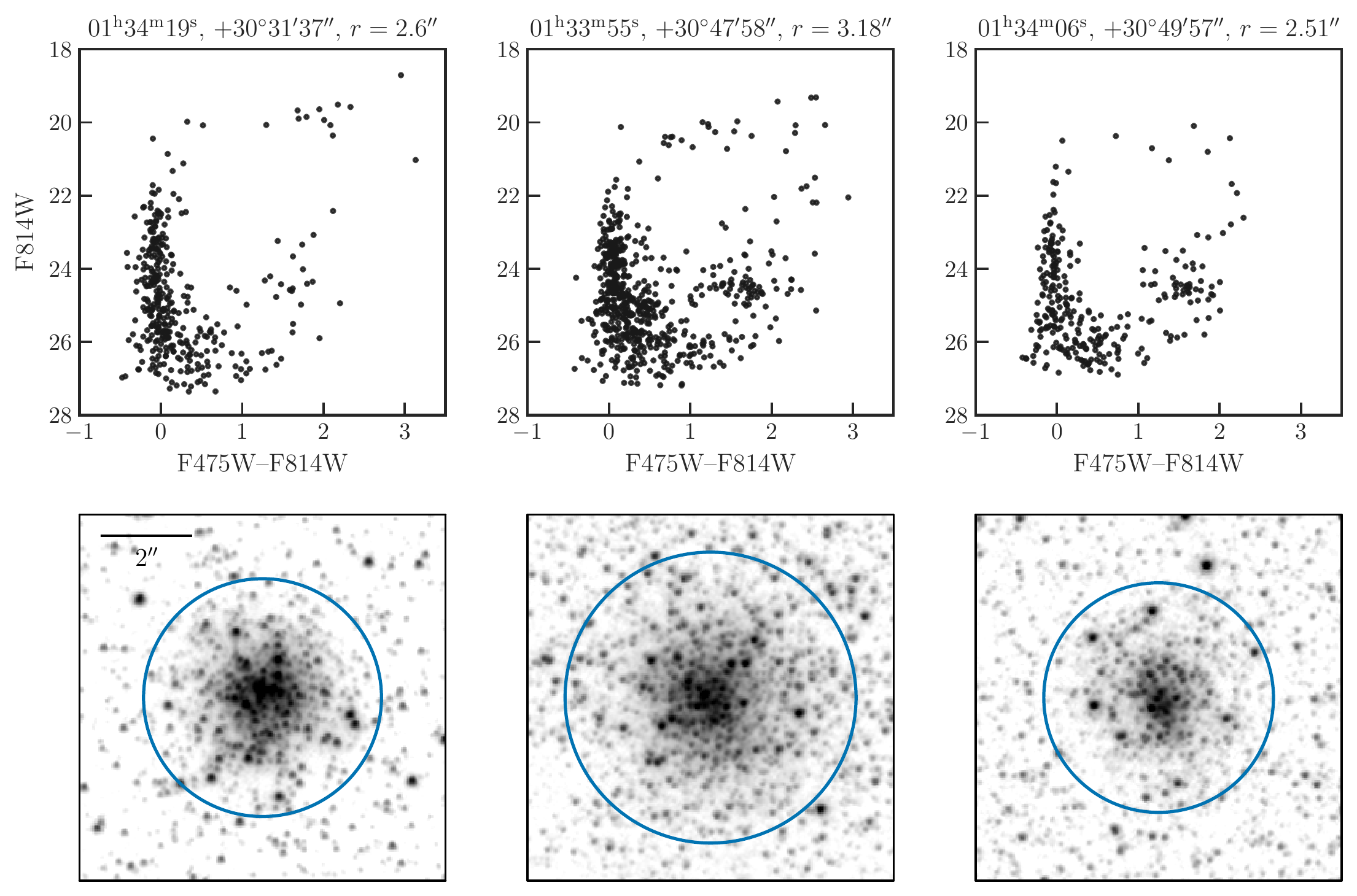}
    \caption{Top row: optical CMDs of three young stellar clusters, labeled with coordinates and radii. Bottom row: corresponding $8\times8\arcsec$ F475W \texttt{drc} cutouts, with clusters encircled at the appropriate radii.}
    \label{cluster_examples}
\end{figure}

\added{The PHATTER data are able to resolve individual stars within star clusters in M33, allowing detailed age and metallicity studies.  We show a few examples of young stellar clusters of a range of masses in Figure~\ref{cluster_examples}, all of which were identified by citizen scientists as part of the Local Group Cluster Search on Zooniverse \citep{2019AAS...23324911J, 2020AAS...23530602W}\footnote{\url{https://www.zooniverse.org/projects/lcjohnso/local-group-cluster-search}}.  The lower panels of this figure show zoomed-in F475W images of the clusters, and the upper panels show optical CMDs of the regions within the clusters' radii.  In all cases, the blue plume of main-sequence stars is visible, showing the young population of cluster stars.  Moreover, the sequence on the CMD is more populated in the more massive, larger clusters, showing the quality of the resolved photometry. While photometry in the clusters does not reach the faintest magnitudes of the survey due to crowding, it is clear that the PHATTER catalog will be a powerful tool for future studies of resolved populations in star clusters in M33.}  

\added{\subsection{Population Maps}}

\begin{figure}
    \centering
    \includegraphics[width=0.8\textwidth]{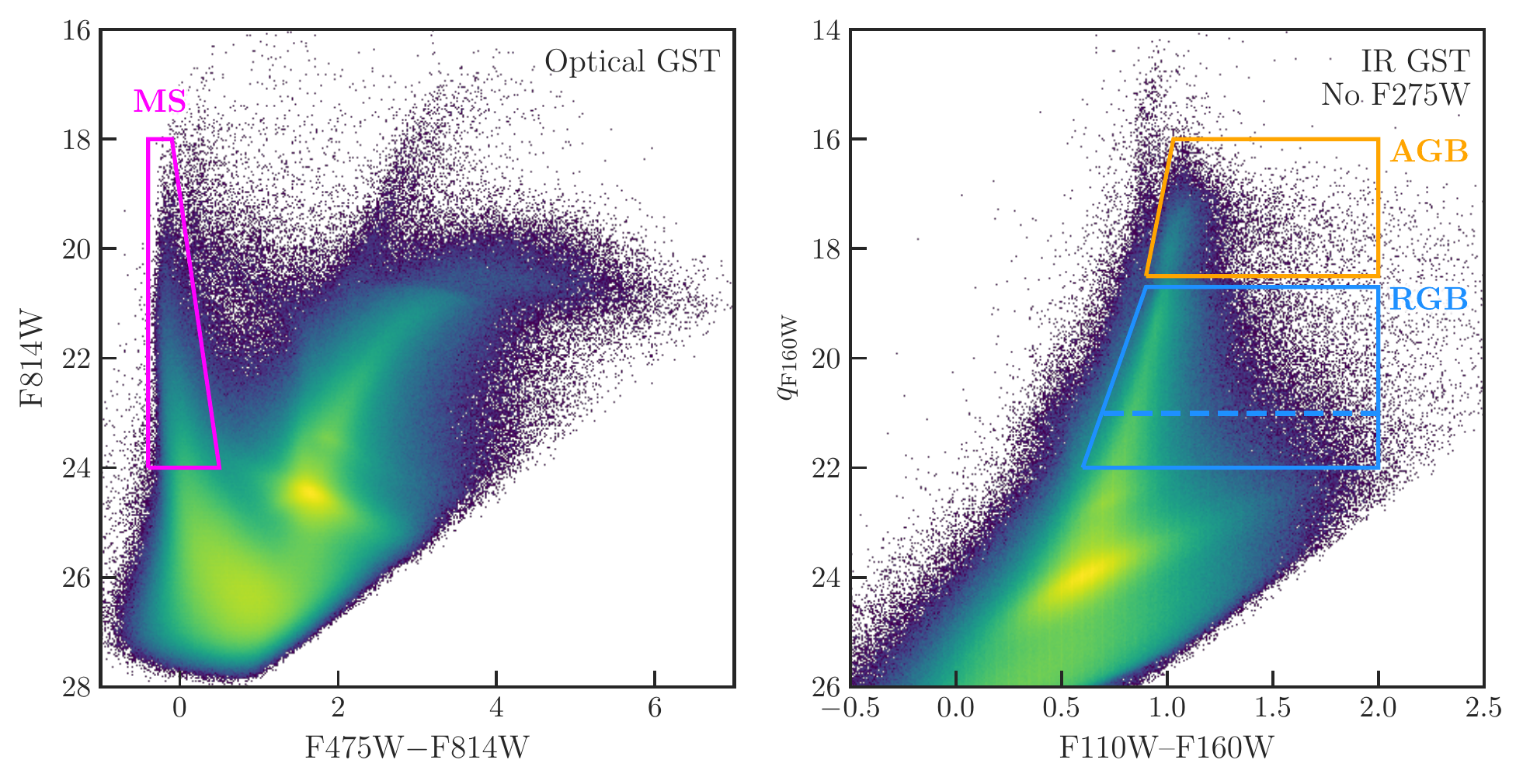} 
    \caption{CMDs showing selection regions for three stellar subpopulations of different characteristic ages. Left: F475W--F814W vs. F814W for stars meeting the GST criteria in the optical. The magenta polygon shows the selection region for young main sequence stars.
    Right: F110W--F160W color vs. reddening-free F160W magnitude ($q_{\mathrm{F160W}}$, as defined in \citealp{dalcanton2015}) for stars that meet the GST criteria in the IR, but do {\em{not}} pass the GST criteria in F275W. The UV constraint helps to eliminate contamination from young BHeB stars in the selection of older populations.
    The orange polygon shows the selection region for asymptotic giant branch stars, and the blue shows the same for the red giant branch.
    The limiting magnitudes of the RGB and MS selection regions roughly correspond to $\gtrsim$80\% completeness in the relevant bands (see Figure~\ref{density_completeness}).
    As completeness in F160W varies substantially with stellar density, RGB stars are selected using two different criteria. Stars located more than $1.2\arcmin$ from the M33 nucleus ($01^\mathrm{h}33^\mathrm{m}51^\mathrm{s}$, $+30^\circ39\arcmin36.72\arcsec$; \citealt{2019ApJ...872...24V} and references therein) are selected with $q_{\rm F160W} < 22$, while stars at radii $< 1.2\arcmin$ are selected with $q_{\rm F160W} < 21$ (dashed line).
    This radial cut roughly corresponds to a stellar surface density cut at 0.6 stars per square arcsec as measured in Fig.~\ref{density_map}.
    }
    \label{cmd_selection}
\end{figure}

\begin{figure}
    \centering
    \includegraphics[width=\textwidth]{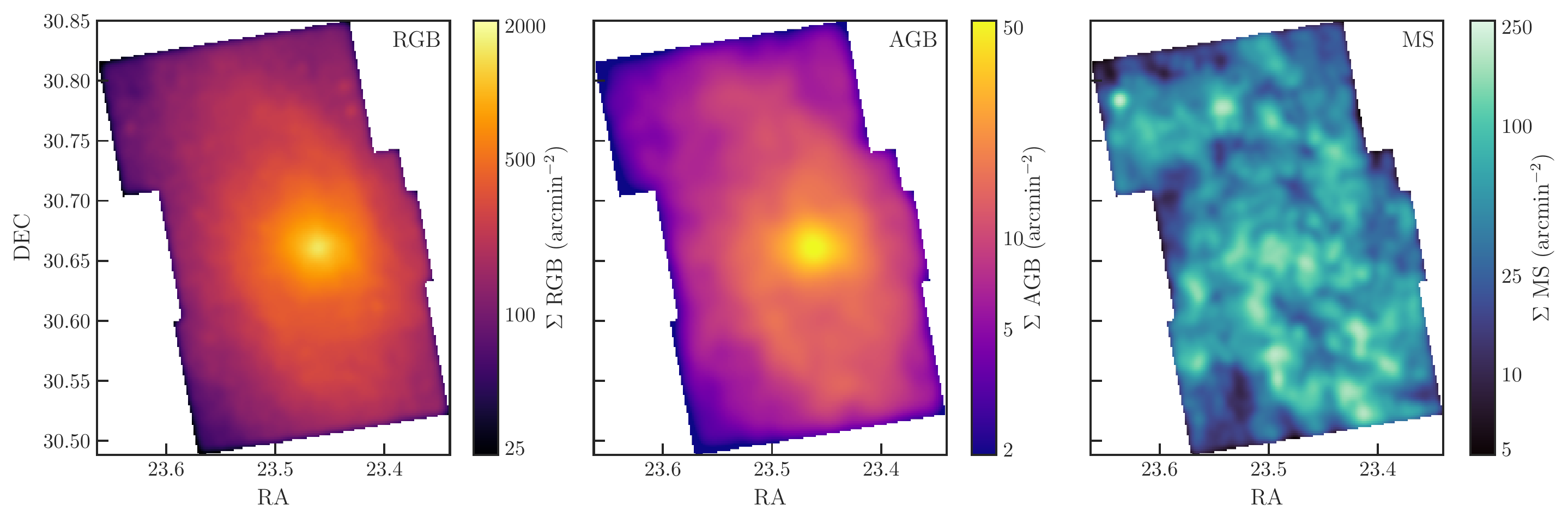} 
    \caption{Stellar density maps of three different subpopulations: old RGB stars (left), intermediate-age AGB stars (center), and young MS stars (right).
    The selection criteria for these subpopulations are shown in Figure~\ref{cmd_selection}.
    For the RGB, star counts in the inner $1.2\arcmin$ have been scaled to correspond to the deeper selection at larger radii.
    The RGB and MS maps have been smoothed with a Gaussian kernel with $\sigma = 0.25\arcmin$, and the AGB with $\sigma = 0.5\arcmin$.
    }
    \label{population_maps}
\end{figure}

\begin{figure}
    \centering
    \includegraphics[width=0.8\textwidth]{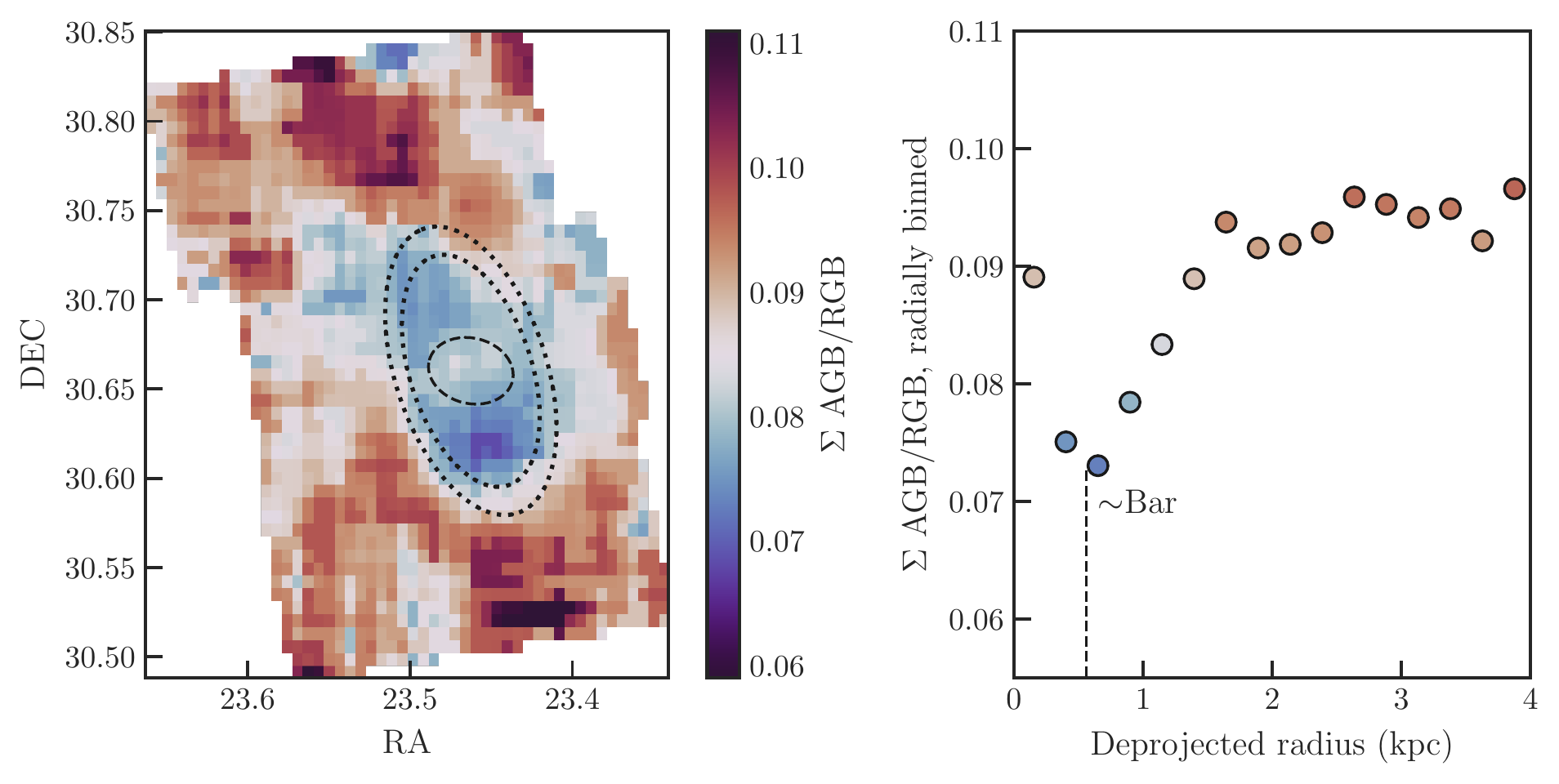}
    \caption{Left: Median-filtered spatial map of the ratio of AGB to upper RGB stars (see Figure~\ref{cmd_selection}). In this case, RGB stars were selected across the entire field with $q_\mathrm{F160W} < 21$ (as opposed to the dual selection shown in Figures~\ref{cmd_selection} and \ref{population_maps}) to eliminate varying completeness as a source of uncertainty in the ratio.
    The large dotted ellipses show one of the radial annuli used for the averaging in the right panel (inclination, position angle, and central coordinates from \citealt{2019ApJ...872...24V} and references therein), while the small dashed ellipse shows the approximate orientation, axis ratio, and maximum scale of M33's weak central bar \citep{corbelli2007}.
    Right: Average AGB/RGB ratio as a function of deprojected distance from the M33 nucleus in kpc. The estimated maximum scale of M33's weak bar is shown for reference ($\sim$559 pc or $2.24\arcmin$). In both the map and radial profile, there is a distinct enhancement of AGB populations in M33's outskirts relative to the center. This supports previous work \citep[e.g.,][]{davidge2003,block2007,verley2009} as evidence of M33's ``inside-out" star-formation history \citep{williams2009,mostoghiu2018}.}
    \label{agb-rgb-ratio}
\end{figure}

\added{Because features on the CMD roughly correspond to distinct populations, we have generated maps of the stellar density within some of the most distinct CMD features.  Figure~\ref{cmd_selection} shows the portions of the CMD used to select MS (young), AGB (intermediate age), and RGB (old) populations.  Figure~\ref{population_maps} shows the density maps of each of these populations.  As might be expected, the young population is highly structured and largely traces the spiral arms and the old population follows a smooth density gradient expected of a disk.  Interestingly, the old and intermediate populations appear to show two faint arms, while the young population appears to trace out more. The structures suggest a long-lived two-armed density structure, along with other, more transient, short-lived arms.

The intermediate age population is not as simple to interpret on its own.  However, in Figure~\ref{agb-rgb-ratio} we show a map of the ratio of AGB to RGB stars, along with a plot of the ratio as a function of distance from the M33 center.  There is a clear radial trend showing a higher fraction of intermediate age stars farther out in the disk.  These maps thus confirm measurements in several previous works \citep[e.g.,][]{davidge2003,block2007,verley2009} and are consistent with other indications of inside-out disk formation in M33 \citep[e.g.,][]{magrini2007,williams2009,beasley2015,mostoghiu2018}, demonstrating the potential for our catalog for future detailed population studies in M33. M33's structure as seen by PHATTER will be explored in detail in a future paper. }

\subsection{Luminosity Functions}

While these initial qualitative evaluations of our photometry are promising, we now move into quantitative tests of the fidelity and consistency of standard CMD features to further assess the robustness and homogeneity of the catalog.  Two features that are very well-suited to such quality checks are the tip of the RGB (TRGB) and the RC.  By comparing the locations of these features in the luminosity function, as a function of position in the survey, we can ensure that any variations are smooth, and thus, most likely related to gradients in the stellar population demographics (e.g., age and metallicity).

\begin{figure}
    \centering
    \includegraphics[width=0.8\textwidth]{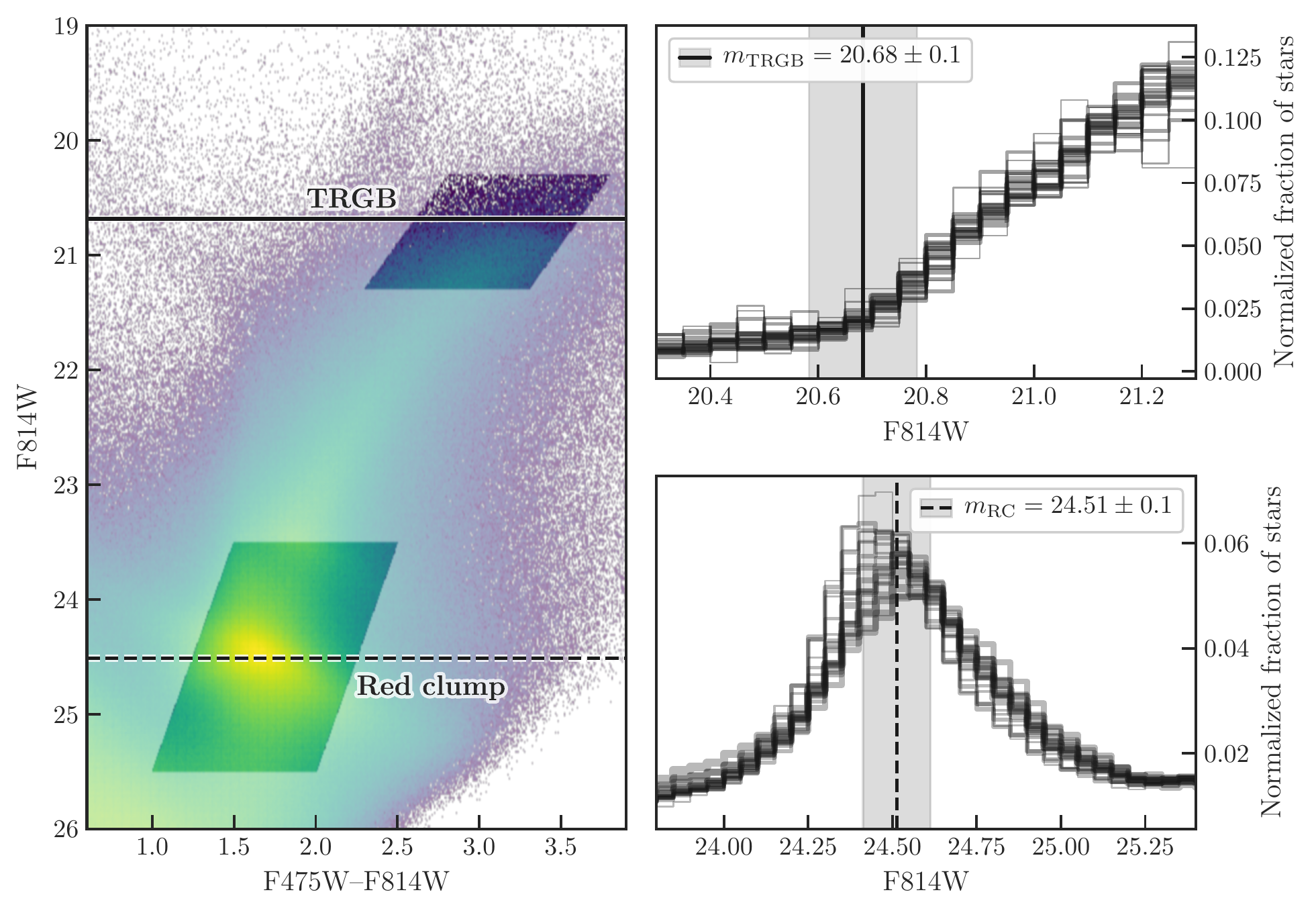}
    \caption{Left: optical CMD showing the selection regions used to measure the F814W magnitude functions for the tip of the red giant branch and red clump features.
    Right: Normalized F814W luminosity functions for the TRGB (top) and red clump (bottom) for 12 $\sim\!6'\times$6$'$ regions of the survey, with lines weighted by the number of stars per sample. We predict apparent TRGB and RC magnitudes using $M^{I}_\mathrm{TRGB} = -4.05$ \citep{2018SSRv..214..113B} and $M_{\rm{RC}}^{I} = -0.22$ \citep{2008A&A...488..935G}, with a distance modulus of $m-M=24.67$ \citep{degrijs2017} and foreground extinction $A_{\mathrm{F814W}}=0.063$ \citep{schlafly2011}. Note the consistency of changes in the magnitude distributions with the predicted TRGB and RC across the entire survey.
    }
    \label{f814w_trgb_rc}
\end{figure}

The left panel of Figure~\ref{f814w_trgb_rc} shows the optical color-magnitude selection regions for the TRGB and the RC on a CMD of the entire survey.
The upper right panel shows the F814W luminosity function, normalized by the total number of stars sampled, for stars in the color range $2.5<$F475W--F814W$<3.5$ at F814W = 20.7 and 20.3$<$F814W$<$21.3 at several locations in the survey, 
which should be dominated by the metal-poor RGB that has a TRGB absolute magnitude of F814W$\sim{-}4.05\pm0.1$ \citep{2018SSRv..214..113B}.
We see that the function steepens at the TRGB, and that the TRGB at this color remains consistent to within $\sim$0.1 mag over the survey, showing that the amount of systematic uncertainty over large areas in our catalog is small.  Furthermore, this TRGB magnitude is within the uncertainties of that expected for a foreground $A_{\rm{F814W}}$ of 0.063 \citep{schlafly2011} and a distance modulus of $24.67\pm0.07$ \citep{degrijs2017}, suggesting that our absolute photometric calibration is also accurate. 

In the lower right panel of Figure~\ref{f814w_trgb_rc}, we show the F814W luminosity function for stars in the color range 1$<$F475W-F814W$<$2 at F814W=25 and 23.5$<$F814W$<$25.5 at several locations to check the position of the RC.  This can be compared to $M_{\rm{RC}}^{I} = -0.22 \pm 0.03$ from \citet{2008A&A...488..935G}, which converts to $M_{\rm{RC}}^{I} = 24.51$ at the distance and extinction of M33.  The magnitude of the peak of the RC remains consistent to within 0.1 mag (24.51$\pm$0.10), confirming that even at much fainter fluxes, the systematic uncertainties over large areas are small.  \added{There is a hint of a small bias to brighter magnitudes in some regions, as evidenced by the slightly brighter peak in some samples. Some bias is expected given the results of the ASTs at this magnitude, as shown in Figure~\ref{bias_uncertainty}. However, some of the variation could also be an effect of the stellar population gradient. A future detailed study of the feature will be required to definitively determine the origin of this variation.}


\replaced{\subsection{Foreground Contamination}}{\subsection{Foreground and Background Contamination}}

Our catalog has a small amount of contamination from Milky Way foreground stars and from background galaxies.  

\begin{figure}
    \centering
    \includegraphics[width=\textwidth]{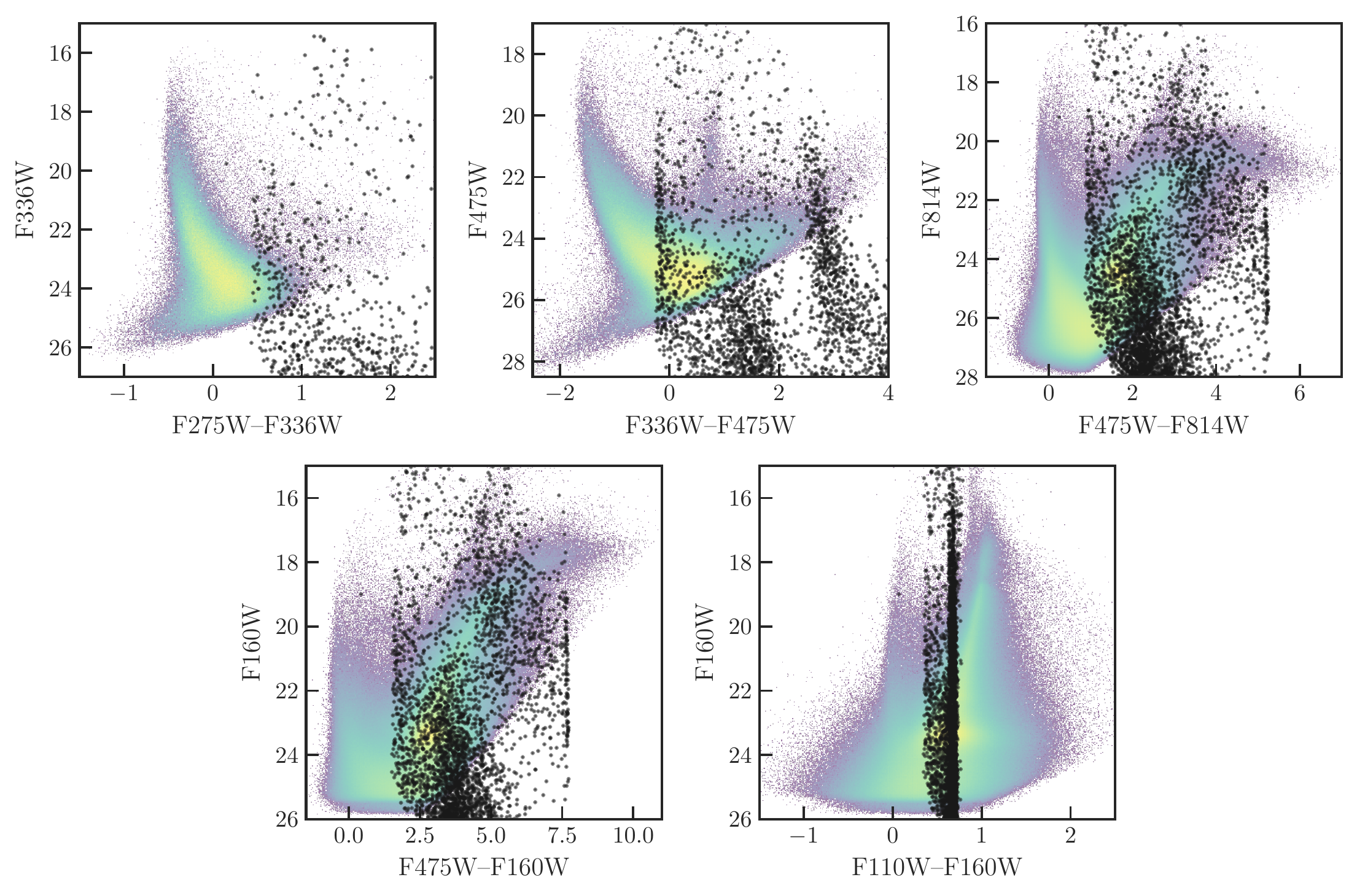}
    \caption{CMDs for the $\sim$5000 foreground stars (black scatter points) predicted by the Trilegal Galactic model for this region of the sky at the depth of our survey, overlaid on the GST CMDs.  While the densities of points are not comparable because the GST CMDs are 2-D histograms, the locations of the foreground stars relative to M33 CMD features is easier to see on the overlay.}
    \label{foreground}
\end{figure}

To estimate the severity of the foreground contamination, we produced model Galactic populations using the {\tt Trilegal} software package \citep{girardi2005}. The model suggests $\sim$3400 foreground stars in our survey footprint with F160W$<$26 with $\sim$2200 of these having F475W$<$28. Thus, our catalog of 22 million stars contains only $<$0.02\% foreground contamination. \added{This fraction increases to $<$0.1\% for the 5 million stars that pass our cuts in at least 2 bands.} However, in certain areas of color-magnitude space, it is important to be able to identify foreground features so that they are not confused with M33 populations.  To aid in this recognition, Figure~\ref{foreground} provides CMDs of the foreground model on the same axes as our survey CMDs.  These plots show the locations of features associated with the foreground populations.  Mostly the foreground occupies the space between the BHeB and the RHeB, along with slightly contaminating the RGB and AGB.  

The only highly visible foreground feature is the narrow bright plume of stars at $\mathrm{F110W}-\mathrm{F160W} = 0.7$ that is the shared color of virtually all of the foreground main sequence stars in the IR.  Interestingly, M33 has a well-populated RHeB feature that is vertical at $\mathrm{F110W}-\mathrm{F160W} = 0.9$.  We have verified that the majority of stars in this feature are the same as those in the bright feature at $\mathrm{F475W}-\mathrm{F160W} \sim 5$, which has no significant foreground equivalent.  Thus, not only are the RHeB stars separated from the foreground in IR color, the foreground contamination in our catalog appears to be less than expected. 

\added{To estimate the severity of the background contamination, we looked for color outliers that were blue in F475W-F814W compared to their F814W-F160W colors, which is similar to the combination used for star-galaxy separation in previous works, such as \citet{robin2007}.  We found outliers from the stellar locus fell at $F475W-F814W<2.5$ and $F814W-F160W>3$.  Only 7183 sources passing our quality criteria in the IR bands have such colors, which corresponds to approximately 0.16\% background contamination when taking the IR quality flags into account.  The contamination drops to 1073 sources, and 0.025\%, for sources passing our quality criteria in the IR and optical bands}.

\section{Conclusions}

We have produced a catalog of resolved stellar photometry for 22 million stars in the field of M33 from 54 {\it HST} pointings covering the inner 3.1$\times$4.6 kpc in 6 bands, including F275W, F336W, F475W, F814W, F110W, and F160W.  The astrometry of this catalog is aligned to the {\it Gaia} DR2 astrometric solution to $\sim$5 milliarcsec.  This catalog reaches $m_{\mathrm{F275W}}=24.5$,  $m_{\mathrm{F336W}}=25$, $m_{\mathrm{F475W}}=28.5$, $m_{\mathrm{F814W}}=27.5$,  $m_{\mathrm{F110W}}=26$, and $m_{\mathrm{F160W}}=25$ with a signal-to-noise limit of 4.  Crowding causes the limiting magnitude to be brighter in the redder bands closer to the center of M33.  This photometry will be studied in great detail by many future studies, such as the history of star formation in M33, the M33 star cluster population, the initial mass function of star clusters in M33, feedback between the stars and interstellar medium in M33, the dust content of M33, and many more.

We have performed many quality checks of the photometry, including ensuring that the tip of the red giant branch is consistent with previous distance measurements of M33, as well as running suites of artificial star tests, where stars of known SEDs are put into the data and the analysis routine was rerun to assess the precision and completeness with which stars are recovered.

A simplified version of our results catalogs are provided here, which will likely provide all of the information required for many science use cases; however, the exhaustive and complete output from our photometry measurements are available from the multimission archive HLSP.

\begin{figure}
    \centering
    \includegraphics[width=0.4\textwidth]{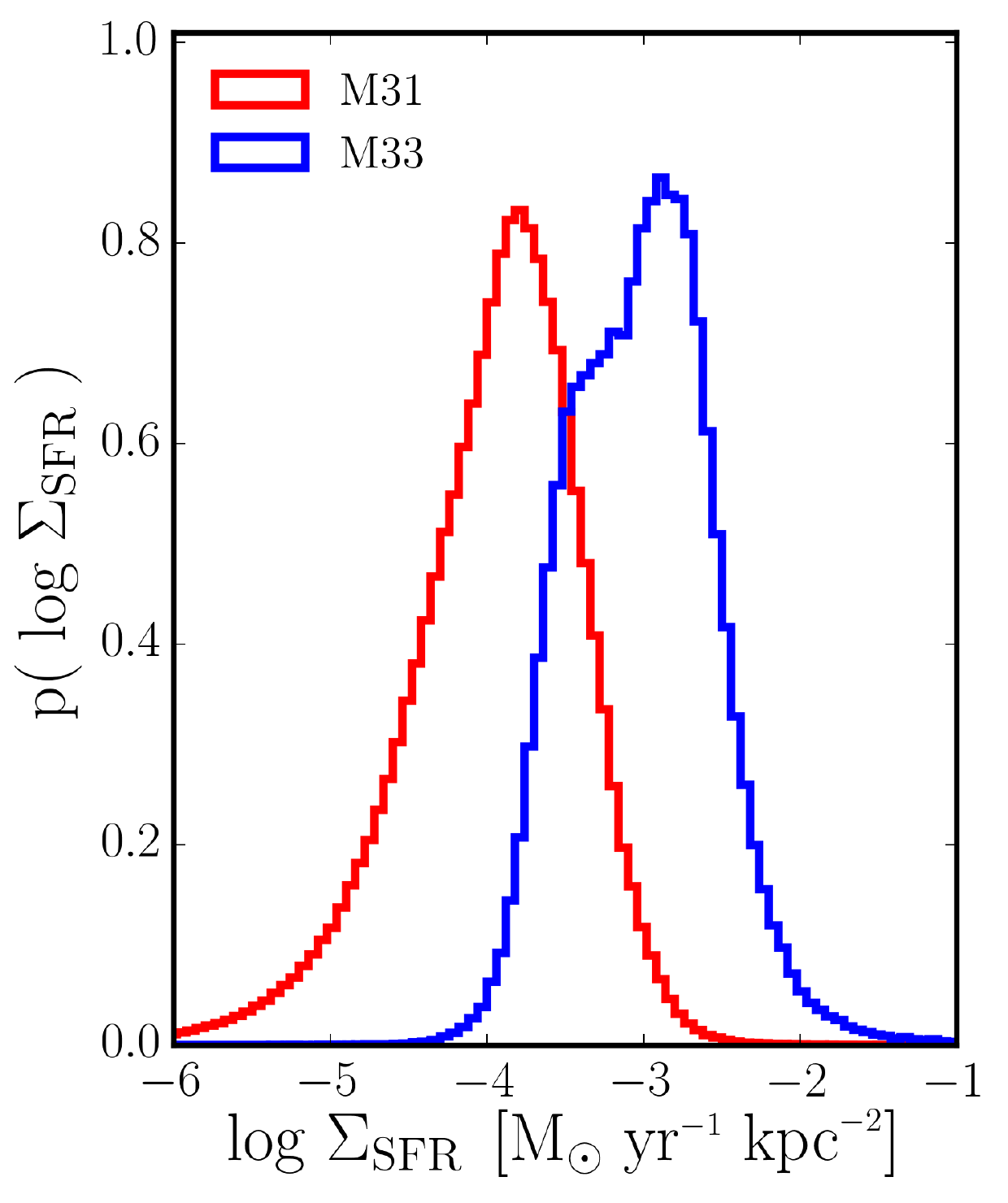}
    \includegraphics[width=0.4\textwidth]{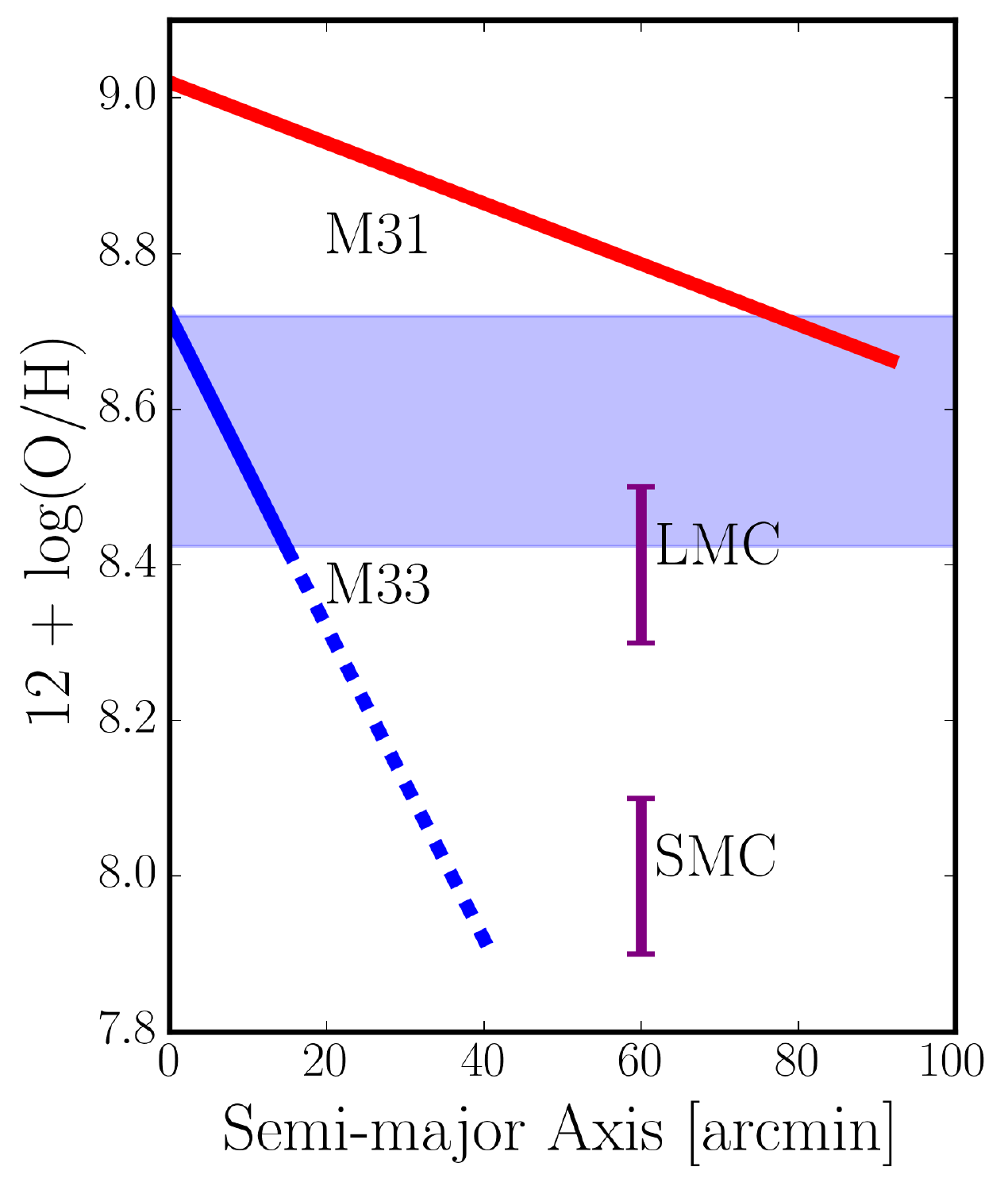}
    \caption{(Left) The relative distribution of star formation rate per unit area (derived from GALEX FUV+24$\mu$m images) for the M33 (blue) and M31 (red) survey areas. The new M33 observations have a significantly higher average SFR intensity than M31.  (Right) The approximate present day metallicity gradient of M33 (blue) and M31 (red), and the range of metallicities in the Magellanic Clouds \citep{gregersen2015,kwitter1981,choudhury2015,choudhury2018}. The solid line shows the extent of the new HST observations. The shaded region shows that the metallicity covered by M33 spans the gap in metallicity between the Large Magellanic Cloud and the M31's outer disk.}
    \label{m33_vs_m31_fig}
\end{figure}

\begin{figure}
    \centering
    \includegraphics[width=0.7\textwidth]{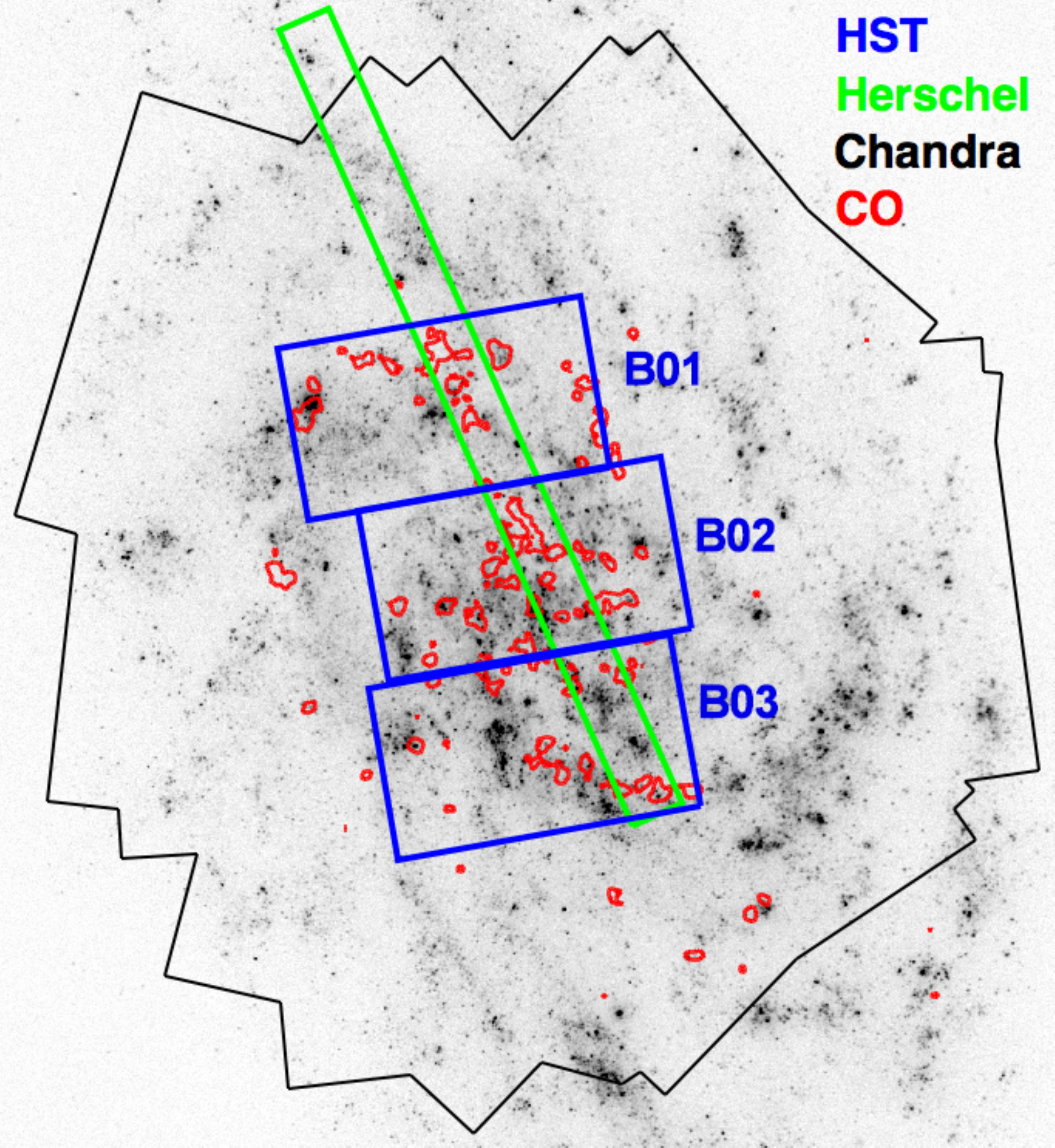} 
    \caption{Locations of the three M33 HST ``bricks" (blue) compared to the FUV (background greyscale image, tracing unobscured star formation), Chandra (black outline, allowing detection of X-ray point sources), CO observations \added{to show coverage} (red contours), and Herschel FIR spectroscopy from the HerM33s survey (green) \citep{rosolowsky2007,kramer2010,xilouris2012,mookerjea2016}}
    \label{footprints_multiwavelength}
\end{figure}

The code used to generate the tables and figures in this paper (with the exception of figures~\ref{m33_vs_m31_fig} and \ref{footprints_multiwavelength}) is available at \url{https://github.com/meredith-durbin/m33_survey_plots}.

\acknowledgments
Support for this work was provided by NASA through grant \#GO-14610 from the Space Telescope Science Institute, which is operated by AURA, Inc., under NASA contract NAS 5-26555.

This research has made use of ``Aladin sky atlas" developed at CDS, Strasbourg Observatory, France, and of the NASA/IPAC Extragalactic Database (NED), which is operated by the Jet Propulsion Laboratory, California Institute of Technology, under contract with the National Aeronautics and Space Administration.

\vspace{5mm}
\facilities{HST(ACS/WFC), HST(WFC3/IR), HST(WFC3/UVIS)}

\software{
          Astropy \citep{2013A&A...558A..33A, 2018AJ....156..123A},
          Astroquery \citep{2017ascl.soft08004G, 2019AJ....157...98G},
          Dask \citep{rocklin2015, dask},
          DOLPHOT \citep{2000PASP..112.1383D, 2016ascl.soft08013D},
          Drizzlepac \citep{2012ascl.soft12011S, 2013ASPC..475...49H, 2015ASPC..495..281A},
          Matplotlib \citep{2007CSE.....9...90H},
          NumPy \citep{numpy, harrisArray2020},
          Pandas \citep{pandas, mckinney2011},
          Seaborn \citep{seabornv090},
          SciPy \citep{scipy},
          Scikit-learn \citep{sklearn},
          Vaex \citep{2018ascl.soft10004B, 2018A&A...618A..13B}
          }


\begin{thebibliography}{}
\expandafter\ifx\csname natexlab\endcsname\relax\def\natexlab#1{#1}\fi
\providecommand{\url}[1]{\href{#1}{#1}}
\providecommand{\dodoi}[1]{doi:~\href{http://doi.org/#1}{\nolinkurl{#1}}}
\providecommand{\doeprint}[1]{\href{http://ascl.net/#1}{\nolinkurl{http://ascl.net/#1}}}
\providecommand{\doarXiv}[1]{\href{https://arxiv.org/abs/#1}{\nolinkurl{https://arxiv.org/abs/#1}}}

\bibitem[{{Anderson} \& {Bedin}(2010)}]{2010PASP..122.1035A}
{Anderson}, J., \& {Bedin}, L.~R. 2010, \pasp, 122, 1035,
  \dodoi{10.1086/656399}

\bibitem[{{Anderson} \& {Ryon}(2018)}]{andersen2018}
{Anderson}, J., \& {Ryon}, J.~E. 2018, {Improving the Pixel-Based
  CTE-correction Model for ACS/WFC}, Instrument Science Report ACS 2018-04

\bibitem[{{Astropy Collaboration} {et~al.}(2013){Astropy Collaboration},
  Robitaille, Tollerud, Greenfield, Droettboom, Bray, Aldcroft, Davis,
  Ginsburg, {Price-Whelan}, Kerzendorf, Conley, Crighton, Barbary, Muna,
  Ferguson, Grollier, Parikh, Nair, G{\"u}nther, Deil, Woillez, Conseil,
  Kramer, Turner, Singer, Fox, Weaver, Zabalza, Edwards, Azalee~Bostroem,
  Burke, Casey, Crawford, Dencheva, Ely, Jenness, Labrie, Lim, Pierfederici,
  Pontzen, Ptak, Refsdal, Servillat, \& Streicher}]{2013A&A...558A..33A}
{Astropy Collaboration}, Robitaille, T.~P., Tollerud, E.~J., {et~al.} 2013,
  Astronomy \& Astrophysics, 558, A33, \dodoi{10.1051/0004-6361/201322068}

\bibitem[{{Astropy Collaboration} {et~al.}(2018){Astropy Collaboration},
  {Price-Whelan}, Sip{\H o}cz, G{\"u}nther, Lim, Crawford, Conseil, Shupe,
  Craig, Dencheva, Ginsburg, {Vand erPlas}, Bradley, {P{\'e}rez-Su{\'a}rez},
  {de Val-Borro}, Aldcroft, Cruz, Robitaille, Tollerud, Ardelean, Babej, Bach,
  Bachetti, Bakanov, Bamford, Barentsen, Barmby, Baumbach, Berry, Biscani,
  Boquien, Bostroem, Bouma, Brammer, Bray, Breytenbach, Buddelmeijer, Burke,
  Calderone, Cano~Rodr{\'i}guez, Cara, Cardoso, Cheedella, Copin, Corrales,
  Crichton, D'Avella, Deil, Depagne, Dietrich, Donath, Droettboom, Earl, Erben,
  Fabbro, Ferreira, Finethy, Fox, Garrison, Gibbons, Goldstein, Gommers, Greco,
  Greenfield, Groener, Grollier, Hagen, Hirst, Homeier, Horton, Hosseinzadeh,
  Hu, Hunkeler, Ivezi{\'c}, Jain, Jenness, Kanarek, Kendrew, Kern, Kerzendorf,
  Khvalko, King, Kirkby, Kulkarni, Kumar, Lee, Lenz, Littlefair, Ma, Macleod,
  Mastropietro, McCully, Montagnac, Morris, Mueller, Mumford, Muna, Murphy,
  Nelson, Nguyen, Ninan, N{\"o}the, Ogaz, Oh, Parejko, Parley, Pascual, Patil,
  Patil, Plunkett, Prochaska, Rastogi, Reddy~Janga, Sabater, Sakurikar,
  Seifert, Sherbert, {Sherwood-Taylor}, Shih, Sick, Silbiger, Singanamalla,
  Singer, Sladen, Sooley, Sornarajah, Streicher, Teuben, Thomas, Tremblay,
  Turner, Terr{\'o}n, {van Kerkwijk}, {de la Vega}, Watkins, Weaver, Whitmore,
  Woillez, Zabalza, \& {Astropy Contributors}}]{2018AJ....156..123A}
{Astropy Collaboration}, {Price-Whelan}, A.~M., Sip{\H o}cz, B.~M., {et~al.}
  2018, The Astronomical Journal, 156, 123, \dodoi{10.3847/1538-3881/aabc4f}

\bibitem[{Avila {et~al.}(2015)Avila, Hack, Cara, Borncamp, Mack, Smith, \&
  Ubeda}]{2015ASPC..495..281A}
Avila, R.~J., Hack, W., Cara, M., {et~al.} 2015, in Astronomical Society of the
  Pacific Conference Series, Vol. 495, Astronomical Data Analysis Software an
  Systems {{XXIV}} ({{ADASS XXIV}}), ed. A.~R. Taylor \& E.~Rosolowsky
  ({Astronomical Society of the Pacific}), 281.
\newblock \doarXiv{1411.5605}

\bibitem[{{Bajaj}(2017)}]{2017wfc..rept...19B}
{Bajaj}, V. 2017, {Aligning HST Images to Gaia: A Faster Mosaicking Workflow},
  Space Telescope WFC3 Instrument Science Report

\bibitem[{{Barker} {et~al.}(2007{\natexlab{a}}){Barker}, {Sarajedini},
  {Geisler}, {Harding}, \& {Schommer}}]{barker2007}
{Barker}, M.~K., {Sarajedini}, A., {Geisler}, D., {Harding}, P., \& {Schommer},
  R. 2007{\natexlab{a}}, \aj, 133, 1138, \dodoi{10.1086/511186}

\bibitem[{{Barker} {et~al.}(2007{\natexlab{b}}){Barker}, {Sarajedini},
  {Geisler}, {Harding}, \& {Schommer}}]{barker2007b}
---. 2007{\natexlab{b}}, \aj, 133, 1125, \dodoi{10.1086/511185}

\bibitem[{{Beasley} {et~al.}(2015){Beasley}, {San Roman}, {Gallart},
  {Sarajedini}, \& {Aparicio}}]{beasley2015}
{Beasley}, M.~A., {San Roman}, I., {Gallart}, C., {Sarajedini}, A., \&
  {Aparicio}, A. 2015, \mnras, 451, 3400, \dodoi{10.1093/mnras/stv943}

\bibitem[{{Beaton} {et~al.}(2018){Beaton}, {Bono}, {Braga}, {Dall'Ora},
  {Fiorentino}, {Jang}, {Mart{\'\i}nez-V{\'a}zquez}, {Matsunaga}, {Monelli},
  {Neeley}, \& {Salaris}}]{2018SSRv..214..113B}
{Beaton}, R.~L., {Bono}, G., {Braga}, V.~F., {et~al.} 2018, \ssr, 214, 113,
  \dodoi{10.1007/s11214-018-0542-1}

\bibitem[{{Block} {et~al.}(2007){Block}, {Combes}, {Puerari}, {Freeman},
  {Stockton}, {Canalizo}, {Jarrett}, {Groess}, {Worthey}, {Gehrz}, {Woodward},
  {Polomski}, \& {Fazio}}]{block2007}
{Block}, D.~L., {Combes}, F., {Puerari}, I., {et~al.} 2007, \aap, 471, 467,
  \dodoi{10.1051/0004-6361:20065908}

\bibitem[{Breddels \& Veljanoski(2018{\natexlab{a}})}]{2018ascl.soft10004B}
Breddels, M.~A., \& Veljanoski, J. 2018{\natexlab{a}}, {{VaeX}}:
  {{Visualization}} and {{eXploration}} of {{Out}}-of-{{Core DataFrames}},
  3.0.0.
\newblock \doeprint{1810.004}

\bibitem[{Breddels \& Veljanoski(2018{\natexlab{b}})}]{2018A&A...618A..13B}
---. 2018{\natexlab{b}}, Astronomy \& Astrophysics, 618, A13,
  \dodoi{10.1051/0004-6361/201732493}

\bibitem[{{Bresolin} {et~al.}(2010){Bresolin}, {Stasi{\'n}ska}, {V{\'\i}lchez},
  {Simon}, \& {Rosolowsky}}]{bresolin2010}
{Bresolin}, F., {Stasi{\'n}ska}, G., {V{\'\i}lchez}, J.~M., {Simon}, J.~D., \&
  {Rosolowsky}, E. 2010, \mnras, 404, 1679,
  \dodoi{10.1111/j.1365-2966.2010.16409.x}

\bibitem[{{Chandar} {et~al.}(1999){Chandar}, {Bianchi}, {Ford}, \&
  {Salasnich}}]{chandar1999}
{Chandar}, R., {Bianchi}, L., {Ford}, H.~C., \& {Salasnich}, B. 1999, \pasp,
  111, 794, \dodoi{10.1086/316393}

\bibitem[{{Choi} {et~al.}(2016){Choi}, {Dotter}, {Conroy}, {Cantiello},
  {Paxton}, \& {Johnson}}]{2016ApJ...823..102C}
{Choi}, J., {Dotter}, A., {Conroy}, C., {et~al.} 2016, \apj, 823, 102,
  \dodoi{10.3847/0004-637X/823/2/102}

\bibitem[{Choudhury {et~al.}(2015)Choudhury, Subramaniam, \&
  Cole}]{choudhury2015}
Choudhury, S., Subramaniam, A., \& Cole, A.~A. 2015, Monthly Notices of the
  Royal Astronomical Society, 455, 1855, \dodoi{10.1093/mnras/stv2414}

\bibitem[{Choudhury {et~al.}(2018)Choudhury, Subramaniam, Cole, \&
  Sohn}]{choudhury2018}
Choudhury, S., Subramaniam, A., Cole, A.~A., \& Sohn, Y.-J. 2018, Monthly
  Notices of the Royal Astronomical Society, 475, 4279,
  \dodoi{10.1093/mnras/sty087}

\bibitem[{{Cioni} {et~al.}(2008){Cioni}, {Irwin}, {Ferguson}, {McConnachie},
  {Conn}, {Huxor}, {Ibata}, {Lewis}, \& {Tanvir}}]{cioni2008}
{Cioni}, M. R.~L., {Irwin}, M., {Ferguson}, A.~M.~N., {et~al.} 2008, \aap, 487,
  131, \dodoi{10.1051/0004-6361:200809366}

\bibitem[{{Corbelli} \& {Walterbos}(2007)}]{corbelli2007}
{Corbelli}, E., \& {Walterbos}, R. A.~M. 2007, \apj, 669, 315,
  \dodoi{10.1086/521618}

\bibitem[{{Dalcanton} {et~al.}(2012){Dalcanton}, {Williams}, {Lang}, {Lauer},
  {Kalirai}, {Seth}, {Dolphin}, {Rosenfield}, {Weisz}, {Bell}, {Bianchi},
  {Boyer}, {Caldwell}, {Dong}, {Dorman}, {Gilbert}, {Girardi}, {Gogarten},
  {Gordon}, {Guhathakurta}, {Hodge}, {Holtzman}, {Johnson}, {Larsen}, {Lewis},
  {Melbourne}, {Olsen}, {Rix}, {Rosema}, {Saha}, {Sarajedini}, {Skillman}, \&
  {Stanek}}]{dalcanton2012}
{Dalcanton}, J.~J., {Williams}, B.~F., {Lang}, D., {et~al.} 2012, \apjs, 200,
  18, \dodoi{10.1088/0067-0049/200/2/18}

\bibitem[{{Dalcanton} {et~al.}(2015){Dalcanton}, {Fouesneau}, {Hogg}, {Lang},
  {Leroy}, {Gordon}, {Sand strom}, {Weisz}, {Williams}, {Bell}, {Dong},
  {Gilbert}, {Gouliermis}, {Guhathakurta}, {Lauer}, {Schruba}, {Seth}, \&
  {Skillman}}]{dalcanton2015}
{Dalcanton}, J.~J., {Fouesneau}, M., {Hogg}, D.~W., {et~al.} 2015, \apj, 814,
  3, \dodoi{10.1088/0004-637X/814/1/3}

\bibitem[{{Dask Development Team}(2016)}]{dask}
{Dask Development Team}. 2016, Dask: {{Library}} for Dynamic Task Scheduling,
  1.0.0.
\newblock \url{https://dask.org}

\bibitem[{{Davidge}(2003)}]{davidge2003}
{Davidge}, T.~J. 2003, \aj, 125, 3046, \dodoi{10.1086/375303}

\bibitem[{{de Grijs} {et~al.}(2017){de Grijs}, {Courbin},
  {Mart{\'\i}nez-V{\'a}zquez}, {Monelli}, {Oguri}, \& {Suyu}}]{degrijs2017}
{de Grijs}, R., {Courbin}, F., {Mart{\'\i}nez-V{\'a}zquez}, C.~E., {et~al.}
  2017, \ssr, 212, 1743, \dodoi{10.1007/s11214-017-0395-z}

\bibitem[{{De Paolis} {et~al.}(2016){De Paolis}, {Gurzadyan}, {Nucita},
  {Chemin}, {Qadir}, {Kashin}, {Khachatryan}, {Sargsyan}, {Yegorian},
  {Ingrosso}, {Jetzer}, \& {Vetrugno}}]{depaolis2016}
{De Paolis}, F., {Gurzadyan}, V.~G., {Nucita}, A.~A., {et~al.} 2016, \aap, 593,
  A57, \dodoi{10.1051/0004-6361/201628780}

\bibitem[{{Deul} \& {van der Hulst}(1987)}]{deul1987}
{Deul}, E.~R., \& {van der Hulst}, J.~M. 1987, \aaps, 67, 509

\bibitem[{{Dolphin}(2016)}]{dolphin2016}
{Dolphin}, A. 2016, {DOLPHOT: Stellar photometry}.
\newblock \doeprint{1608.013}

\bibitem[{Dolphin(2016)}]{2016ascl.soft08013D}
Dolphin, A. 2016, {{DOLPHOT}}: {{Stellar}} Photometry, 2.0.
\newblock \doeprint{1608.013}

\bibitem[{{Dolphin}(2000)}]{dolphin2000}
{Dolphin}, A.~E. 2000, \pasp, 112, 1383

\bibitem[{Dolphin(2000)}]{2000PASP..112.1383D}
Dolphin, A.~E. 2000, Publications of the Astronomical Society of the Pacific,
  112, 1383, \dodoi{10.1086/316630}

\bibitem[{{Dolphin}(2002)}]{dolphin2002}
{Dolphin}, A.~E. 2002, \mnras, 332, 91,
  \dodoi{10.1046/j.1365-8711.2002.05271.x}

\bibitem[{{Druard} {et~al.}(2014){Druard}, {Braine}, {Schuster}, {Schneider},
  {Gratier}, {Bontemps}, {Boquien}, {Combes}, {Corbelli}, {Henkel}, {Herpin},
  {Kramer}, {van der Tak}, \& {van der Werf}}]{druard2014}
{Druard}, C., {Braine}, J., {Schuster}, K.~F., {et~al.} 2014, \aap, 567, A118,
  \dodoi{10.1051/0004-6361/201423682}

\bibitem[{{D'Souza} \& {Bell}(2018)}]{dsouza2018}
{D'Souza}, R., \& {Bell}, E.~F. 2018, Nature Astronomy, 2, 737,
  \dodoi{10.1038/s41550-018-0533-x}

\bibitem[{{Engargiola} {et~al.}(2003){Engargiola}, {Plambeck}, {Rosolowsky}, \&
  {Blitz}}]{engargiola2003}
{Engargiola}, G., {Plambeck}, R.~L., {Rosolowsky}, E., \& {Blitz}, L. 2003,
  \apjs, 149, 343, \dodoi{10.1086/379165}

\bibitem[{{Gaia Collaboration} {et~al.}(2018){Gaia Collaboration}, {Brown},
  {Vallenari}, {Prusti}, {de Bruijne}, {Babusiaux}, {Bailer-Jones}, {Biermann},
  {Evans}, {Eyer}, {Jansen}, {Jordi}, {Klioner}, {Lammers}, {Lindegren},
  {Luri}, {Mignard}, {Panem}, {Pourbaix}, {Randich}, {Sartoretti}, {Siddiqui},
  {Soubiran}, {van Leeuwen}, {Walton}, {Arenou}, {Bastian}, {Cropper},
  {Drimmel}, {Katz}, {Lattanzi}, {Bakker}, {Cacciari}, {Casta{\~n}eda},
  {Chaoul}, {Cheek}, {De Angeli}, {Fabricius}, {Guerra}, {Holl}, {Masana},
  {Messineo}, {Mowlavi}, {Nienartowicz}, {Panuzzo}, {Portell}, {Riello},
  {Seabroke}, {Tanga}, {Th{\'e}venin}, {Gracia-Abril}, {Comoretto},
  {Garcia-Reinaldos}, {Teyssier}, {Altmann}, {Andrae}, {Audard},
  {Bellas-Velidis}, {Benson}, {Berthier}, {Blomme}, {Burgess}, {Busso},
  {Carry}, {Cellino}, {Clementini}, {Clotet}, {Creevey}, {Davidson}, {De
  Ridder}, {Delchambre}, {Dell'Oro}, {Ducourant},
  {Fern{\'a}ndez-Hern{\'a}ndez}, {Fouesneau}, {Fr{\'e}mat}, {Galluccio},
  {Garc{\'\i}a-Torres}, {Gonz{\'a}lez-N{\'u}{\~n}ez}, {Gonz{\'a}lez-Vidal},
  {Gosset}, {Guy}, {Halbwachs}, {Hambly}, {Harrison}, {Hern{\'a}ndez},
  {Hestroffer}, {Hodgkin}, {Hutton}, {Jasniewicz}, {Jean-Antoine-Piccolo},
  {Jordan}, {Korn}, {Krone-Martins}, {Lanzafame}, {Lebzelter}, {L{\"o}ffler},
  {Manteiga}, {Marrese}, {Mart{\'\i}n-Fleitas}, {Moitinho}, {Mora}, {Muinonen},
  {Osinde}, {Pancino}, {Pauwels}, {Petit}, {Recio-Blanco}, {Richards},
  {Rimoldini}, {Robin}, {Sarro}, {Siopis}, {Smith}, {Sozzetti}, {S{\"u}veges},
  {Torra}, {van Reeven}, {Abbas}, {Abreu Aramburu}, {Accart}, {Aerts},
  {Altavilla}, {{\'A}lvarez}, {Alvarez}, {Alves}, {Anderson}, {Andrei},
  {Anglada Varela}, {Antiche}, {Antoja}, {Arcay}, {Astraatmadja}, {Bach},
  {Baker}, {Balaguer-N{\'u}{\~n}ez}, {Balm}, {Barache}, {Barata}, {Barbato},
  {Barblan}, {Barklem}, {Barrado}, {Barros}, {Barstow}, {Bartholom{\'e}
  Mu{\~n}oz}, {Bassilana}, {Becciani}, {Bellazzini}, {Berihuete}, {Bertone},
  {Bianchi}, {Bienaym{\'e}}, {Blanco-Cuaresma}, {Boch}, {Boeche}, {Bombrun},
  {Borrachero}, {Bossini}, {Bouquillon}, {Bourda}, {Bragaglia}, {Bramante},
  {Breddels}, {Bressan}, {Brouillet}, {Br{\"u}semeister}, {Brugaletta},
  {Bucciarelli}, {Burlacu}, {Busonero}, {Butkevich}, {Buzzi}, {Caffau},
  {Cancelliere}, {Cannizzaro}, {Cantat-Gaudin}, {Carballo}, {Carlucci},
  {Carrasco}, {Casamiquela}, {Castellani}, {Castro-Ginard}, {Charlot},
  {Chemin}, {Chiavassa}, {Cocozza}, {Costigan}, {Cowell}, {Crifo}, {Crosta},
  {Crowley}, {Cuypers}, {Dafonte}, {Damerdji}, {Dapergolas}, {David}, {David},
  {de Laverny}, {De Luise}, {De March}, {de Martino}, {de Souza}, {de Torres},
  {Debosscher}, {del Pozo}, {Delbo}, {Delgado}, {Delgado}, {Di Matteo},
  {Diakite}, {Diener}, {Distefano}, {Dolding}, {Drazinos}, {Dur{\'a}n},
  {Edvardsson}, {Enke}, {Eriksson}, {Esquej}, {Eynard Bontemps}, {Fabre},
  {Fabrizio}, {Faigler}, {Falc{\~a}o}, {Farr{\`a}s Casas}, {Federici},
  {Fedorets}, {Fernique}, {Figueras}, {Filippi}, {Findeisen}, {Fonti},
  {Fraile}, {Fraser}, {Fr{\'e}zouls}, {Gai}, {Galleti}, {Garabato},
  {Garc{\'\i}a-Sedano}, {Garofalo}, {Garralda}, {Gavel}, {Gavras}, {Gerssen},
  {Geyer}, {Giacobbe}, {Gilmore}, {Girona}, {Giuffrida}, {Glass}, {Gomes},
  {Granvik}, {Gueguen}, {Guerrier}, {Guiraud}, {Guti{\'e}rrez-S{\'a}nchez},
  {Haigron}, {Hatzidimitriou}, {Hauser}, {Haywood}, {Heiter}, {Helmi}, {Heu},
  {Hilger}, {Hobbs}, {Hofmann}, {Holland}, {Huckle}, {Hypki}, {Icardi},
  {Jan{\ss}en}, {Jevardat de Fombelle}, {Jonker}, {Juh{\'a}sz}, {Julbe},
  {Karampelas}, {Kewley}, {Klar}, {Kochoska}, {Kohley}, {Kolenberg},
  {Kontizas}, {Kontizas}, {Koposov}, {Kordopatis}, {Kostrzewa-Rutkowska},
  {Koubsky}, {Lambert}, {Lanza}, {Lasne}, {Lavigne}, {Le Fustec}, {Le
  Poncin-Lafitte}, {Lebreton}, {Leccia}, {Leclerc}, {Lecoeur-Taibi},
  {Lenhardt}, {Leroux}, {Liao}, {Licata}, {Lindstr{\o}m}, {Lister}, {Livanou},
  {Lobel}, {L{\'o}pez}, {Managau}, {Mann}, {Mantelet}, {Marchal}, {Marchant},
  {Marconi}, {Marinoni}, {Marschalk{\'o}}, {Marshall}, {Martino}, {Marton},
  {Mary}, {Massari}, {Matijevi{\v{c}}}, {Mazeh}, {McMillan}, {Messina},
  {Michalik}, {Millar}, {Molina}, {Molinaro}, {Moln{\'a}r}, {Montegriffo},
  {Mor}, {Morbidelli}, {Morel}, {Morris}, {Mulone}, {Muraveva}, {Musella},
  {Nelemans}, {Nicastro}, {Noval}, {O'Mullane}, {Ord{\'e}novic},
  {Ord{\'o}{\~n}ez-Blanco}, {Osborne}, {Pagani}, {Pagano}, {Pailler},
  {Palacin}, {Palaversa}, {Panahi}, {Pawlak}, {Piersimoni}, {Pineau}, {Plachy},
  {Plum}, {Poggio}, {Poujoulet}, {Pr{\v{s}}a}, {Pulone}, {Racero}, {Ragaini},
  {Rambaux}, {Ramos-Lerate}, {Regibo}, {Reyl{\'e}}, {Riclet}, {Ripepi}, {Riva},
  {Rivard}, {Rixon}, {Roegiers}, {Roelens}, {Romero-G{\'o}mez}, {Rowell},
  {Royer}, {Ruiz-Dern}, {Sadowski}, {Sagrist{\`a} Sell{\'e}s}, {Sahlmann},
  {Salgado}, {Salguero}, {Sanna}, {Santana-Ros}, {Sarasso}, {Savietto},
  {Schultheis}, {Sciacca}, {Segol}, {Segovia}, {S{\'e}gransan}, {Shih},
  {Siltala}, {Silva}, {Smart}, {Smith}, {Solano}, {Solitro}, {Sordo}, {Soria
  Nieto}, {Souchay}, {Spagna}, {Spoto}, {Stampa}, {Steele},
  {Steidelm{\"u}ller}, {Stephenson}, {Stoev}, {Suess}, {Surdej}, {Szabados},
  {Szegedi-Elek}, {Tapiador}, {Taris}, {Tauran}, {Taylor}, {Teixeira},
  {Terrett}, {Teyssand ier}, {Thuillot}, {Titarenko}, {Torra Clotet}, {Turon},
  {Ulla}, {Utrilla}, {Uzzi}, {Vaillant}, {Valentini}, {Valette}, {van Elteren},
  {Van Hemelryck}, {van Leeuwen}, {Vaschetto}, {Vecchiato}, {Veljanoski},
  {Viala}, {Vicente}, {Vogt}, {von Essen}, {Voss}, {Votruba}, {Voutsinas},
  {Walmsley}, {Weiler}, {Wertz}, {Wevers}, {Wyrzykowski}, {Yoldas},
  {{\v{Z}}erjal}, {Ziaeepour}, {Zorec}, {Zschocke}, {Zucker}, {Zurbach}, \&
  {Zwitter}}]{gaia2018}
{Gaia Collaboration}, {Brown}, A.~G.~A., {Vallenari}, A., {et~al.} 2018, \aap,
  616, A1, \dodoi{10.1051/0004-6361/201833051}

\bibitem[{{Gallart}(1998)}]{gallart98}
{Gallart}, C. 1998, \apjl, 495, L43, \dodoi{10.1086/311218}

\bibitem[{Ginsburg {et~al.}(2017)Ginsburg, Parikh, Woillez, Groener, Liedtke,
  Sipocz, Robitaille, Deil, Svoboda, Tollerud, Persson,
  {S{\'e}guin-Charbonneau}, Armstrong, Mirocha, Droettboom, Allen, Moolekamp,
  Egeland, Singer, Barbary, Grollier, Shiga, Moritz~G{\"u}nther, Parejko,
  Booker, Rol, {Edward}, Miller, \& Willett}]{2017ascl.soft08004G}
Ginsburg, A., Parikh, M., Woillez, J., {et~al.} 2017, Astroquery: {{Access}} to
  Online Data Resources, 0.3.0.
\newblock \doeprint{1708.004}

\bibitem[{Ginsburg {et~al.}(2019)Ginsburg, Sip{\H o}cz, Brasseur,
  Cowperthwaite, Craig, Deil, Guillochon, Guzman, Liedtke, Lim, Lockhart,
  Mommert, Morris, Norman, Parikh, Persson, Robitaille, Segovia, Singer,
  Tollerud, {de Val-Borro}, Valtchanov, Woillez, \& {The Astroquery
  collaboration, a subset of the astropy collaboration}}]{2019AJ....157...98G}
Ginsburg, A., Sip{\H o}cz, B.~M., Brasseur, C.~E., {et~al.} 2019, The
  Astronomical Journal, 157, 98, \dodoi{10.3847/1538-3881/aafc33}

\bibitem[{{Girardi} {et~al.}(2005){Girardi}, {Groenewegen}, {Hatziminaoglou},
  \& {da Costa}}]{girardi2005}
{Girardi}, L., {Groenewegen}, M.~A.~T., {Hatziminaoglou}, E., \& {da Costa}, L.
  2005, \aap, 436, 895, \dodoi{10.1051/0004-6361:20042352}

\bibitem[{{Gordon} {et~al.}(2016){Gordon}, {Humphreys}, \&
  {Jones}}]{gordon2016}
{Gordon}, M.~S., {Humphreys}, R.~M., \& {Jones}, T.~J. 2016, \apj, 825, 50,
  \dodoi{10.3847/0004-637X/825/1/50}

\bibitem[{{Gratier} {et~al.}(2010){Gratier}, {Braine}, {Rodriguez-Fernandez},
  {Schuster}, {Kramer}, {Xilouris}, {Tabatabaei}, {Henkel}, {Corbelli},
  {Israel}, {van der Werf}, {Calzetti}, {Garcia-Burillo}, {Sievers}, {Combes},
  {Wiklind}, {Brouillet}, {Herpin}, {Bontemps}, {Aalto}, {Koribalski}, {van der
  Tak}, {Wiedner}, {R{\"o}llig}, \& {Mookerjea}}]{gratier2010}
{Gratier}, P., {Braine}, J., {Rodriguez-Fernandez}, N.~J., {et~al.} 2010, \aap,
  522, A3, \dodoi{10.1051/0004-6361/201014441}

\bibitem[{{Gregersen} {et~al.}(2015){Gregersen}, {Seth}, {Williams}, {Lang},
  {Dalcanton}, {Girardi}, {Skillman}, {Bell}, {Dolphin}, {Fouesneau},
  {Guhathakurta}, {Hamren}, {Johnson}, {Kalirai}, {Lewis}, {Monachesi}, \&
  {Olsen}}]{gregersen2015}
{Gregersen}, D., {Seth}, A.~C., {Williams}, B.~F., {et~al.} 2015, \aj, 150,
  189, \dodoi{10.1088/0004-6256/150/6/189}

\bibitem[{{Groenewegen}(2008)}]{2008A&A...488..935G}
{Groenewegen}, M.~A.~T. 2008, \aap, 488, 935,
  \dodoi{10.1051/0004-6361:200810201}

\bibitem[{Hack {et~al.}(2013)Hack, Dencheva, \& Fruchter}]{2013ASPC..475...49H}
Hack, W.~J., Dencheva, N., \& Fruchter, A.~S. 2013, in Astronomical Society of
  the Pacific Conference Series, Vol. 475, Astronomical Data Analysis Software
  and Systems {{XXII}}, ed. D.~N. Friedel ({Astronomical Society of the
  Pacific}), 49

\bibitem[{{Hammer} {et~al.}(2018){Hammer}, {Yang}, {Wang}, {Ibata}, {Flores},
  \& {Puech}}]{hammer2018}
{Hammer}, F., {Yang}, Y.~B., {Wang}, J.~L., {et~al.} 2018, \mnras, 475, 2754,
  \dodoi{10.1093/mnras/stx3343}

\bibitem[{Harris {et~al.}(2020)Harris, Millman, {van der Walt}, Gommers,
  Virtanen, Cournapeau, Wieser, Taylor, Berg, Smith, Kern, Picus, Hoyer, {van
  Kerkwijk}, Brett, Haldane, {del R{\'i}o}, Wiebe, Peterson,
  {G{\'e}rard-Marchant}, Sheppard, Reddy, Weckesser, Abbasi, Gohlke, \&
  Oliphant}]{harrisArray2020}
Harris, C.~R., Millman, K.~J., {van der Walt}, S.~J., {et~al.} 2020, Nature,
  585, 357, \dodoi{10.1038/s41586-020-2649-2}

\bibitem[{{Hermelo} {et~al.}(2016){Hermelo}, {Rela{\~n}o}, {Lisenfeld},
  {Verley}, {Kramer}, {Ruiz-Lara}, {Boquien}, {Xilouris}, \&
  {Albrecht}}]{hermelo2016}
{Hermelo}, I., {Rela{\~n}o}, M., {Lisenfeld}, U., {et~al.} 2016, \aap, 590,
  A56, \dodoi{10.1051/0004-6361/201525816}

\bibitem[{{Heyer} {et~al.}(2004){Heyer}, {Corbelli}, {Schneider}, \&
  {Young}}]{heyer2004}
{Heyer}, M.~H., {Corbelli}, E., {Schneider}, S.~E., \& {Young}, J.~S. 2004,
  \apj, 602, 723, \dodoi{10.1086/381196}

\bibitem[{{Hinz} {et~al.}(2004){Hinz}, {Rieke}, {Gordon},
  {P{\'e}rez-Gonz{\'a}lez}, {Engelbracht}, {Alonso-Herrero}, {Morrison},
  {Misselt}, {Hines}, {Gehrz}, {Polomski}, {Woodward}, {Humphreys}, {Regan},
  {Rho}, {Beeman}, \& {Haller}}]{hinz2004}
{Hinz}, J.~L., {Rieke}, G.~H., {Gordon}, K.~D., {et~al.} 2004, \apjs, 154, 259,
  \dodoi{10.1086/422558}

\bibitem[{{Hoopes} \& {Walterbos}(2000)}]{hoopes2000}
{Hoopes}, C.~G., \& {Walterbos}, R. A.~M. 2000, \apj, 541, 597,
  \dodoi{10.1086/309487}

\bibitem[{{Humphreys} {et~al.}(2017){Humphreys}, {Gordon}, {Martin}, {Weis}, \&
  {Hahn}}]{humphreys2017}
{Humphreys}, R.~M., {Gordon}, M.~S., {Martin}, J.~C., {Weis}, K., \& {Hahn}, D.
  2017, \apj, 836, 64, \dodoi{10.3847/1538-4357/aa582e}

\bibitem[{{Humphreys} \& {Sandage}(1980)}]{humphreys1980}
{Humphreys}, R.~M., \& {Sandage}, A. 1980, \apjs, 44, 319,
  \dodoi{10.1086/190696}

\bibitem[{Hunter(2007)}]{2007CSE.....9...90H}
Hunter, J.~D. 2007, Computing in Science \& Engineering, 9, 90,
  \dodoi{10.1109/MCSE.2007.55}

\bibitem[{{Johnson}(2019)}]{2019AAS...23324911J}
{Johnson}, L.~C. 2019, in American Astronomical Society Meeting Abstracts, Vol.
  233, American Astronomical Society Meeting Abstracts \#233, 249.11

\bibitem[{{Johnson} {et~al.}(2015){Johnson}, {Seth}, {Dalcanton}, {Wallace},
  {Simpson}, {Lintott}, {Kapadia}, {Skillman}, {Caldwell}, {Fouesneau},
  {Weisz}, {Williams}, {Beerman}, {Gouliermis}, \& {Sarajedini}}]{johnson2015}
{Johnson}, L.~C., {Seth}, A.~C., {Dalcanton}, J.~J., {et~al.} 2015, \apj, 802,
  127, \dodoi{10.1088/0004-637X/802/2/127}

\bibitem[{Jones {et~al.}(2001)Jones, Oliphant, Peterson, {et~al.}}]{scipy}
Jones, E., Oliphant, T., Peterson, P., {et~al.} 2001, {{SciPy}}: {{Open}}
  Source Scientific Tools for {{Python}}.
\newblock \url{http://www.scipy.org/}

\bibitem[{{Kobulnicky} \& {Fryer}(2007)}]{kobulnicky2007}
{Kobulnicky}, H.~A., \& {Fryer}, C.~L. 2007, \apj, 670, 747,
  \dodoi{10.1086/522073}

\bibitem[{{Koch} {et~al.}(2018){Koch}, {Rosolowsky}, {Lockman}, {Kepley},
  {Leroy}, {Schruba}, {Braine}, {Dalcanton}, {Johnson}, \&
  {Stanimirovi{\'c}}}]{koch18}
{Koch}, E.~W., {Rosolowsky}, E.~W., {Lockman}, F.~J., {et~al.} 2018, \mnras,
  479, 2505, \dodoi{10.1093/mnras/sty1674}

\bibitem[{{Kormendy} \& {McClure}(1993)}]{kormendy1993}
{Kormendy}, J., \& {McClure}, R.~D. 1993, \aj, 105, 1793,
  \dodoi{10.1086/116555}

\bibitem[{{Kramer} {et~al.}(2010){Kramer}, {Buchbender}, {Xilouris}, {Boquien},
  {Braine}, {Calzetti}, {Lord}, {Mookerjea}, {Quintana-Lacaci}, {Rela{\~n}o},
  {Stacey}, {Tabatabaei}, {Verley}, {Aalto}, {Akras}, {Albrecht}, {Anderl},
  {Beck}, {Bertoldi}, {Combes}, {Dumke}, {Garcia-Burillo}, {Gonzalez},
  {Gratier}, {G{\"u}sten}, {Henkel}, {Israel}, {Koribalski}, {Lundgren},
  {Martin-Pintado}, {R{\"o}llig}, {Rosolowsky}, {Schuster}, {Sheth}, {Sievers},
  {Stutzki}, {Tilanus}, {van der Tak}, {van der Werf}, \&
  {Wiedner}}]{kramer2010}
{Kramer}, C., {Buchbender}, C., {Xilouris}, E.~M., {et~al.} 2010, \aap, 518,
  L67, \dodoi{10.1051/0004-6361/201014613}

\bibitem[{{Krist} {et~al.}(2011){Krist}, {Hook}, \& {Stoehr}}]{krist2011}
{Krist}, J.~E., {Hook}, R.~N., \& {Stoehr}, F. 2011, in Society of
  Photo-Optical Instrumentation Engineers (SPIE) Conference Series, Vol. 8127,
  \procspie, 81270J, \dodoi{10.1117/12.892762}

\bibitem[{{Kruijssen} {et~al.}(2019){Kruijssen}, {Pfeffer}, {Reina-Campos},
  {Crain}, \& {Bastian}}]{kruijssen2019}
{Kruijssen}, J.~M.~D., {Pfeffer}, J.~L., {Reina-Campos}, M., {Crain}, R.~A., \&
  {Bastian}, N. 2019, \mnras, 486, 3180, \dodoi{10.1093/mnras/sty1609}

\bibitem[{{Kwitter} \& {Aller}(1981)}]{kwitter1981}
{Kwitter}, K.~B., \& {Aller}, L.~H. 1981, \mnras, 195, 939,
  \dodoi{10.1093/mnras/195.4.939}

\bibitem[{{Lewis} {et~al.}(2015){Lewis}, {Dolphin}, {Dalcanton}, {Weisz},
  {Williams}, {Bell}, {Seth}, {Simones}, {Skillman}, {Choi}, {Fouesneau},
  {Guhathakurta}, {Johnson}, {Kalirai}, {Leroy}, {Monachesi}, {Rix}, \&
  {Schruba}}]{lewis2015}
{Lewis}, A.~R., {Dolphin}, A.~E., {Dalcanton}, J.~J., {et~al.} 2015, \apj, 805,
  183, \dodoi{10.1088/0004-637X/805/2/183}

\bibitem[{{Lewis} {et~al.}(2017){Lewis}, {Simones}, {Johnson}, {Dalcanton},
  {Skillman}, {Weisz}, {Dolphin}, {Williams}, {Bell}, {Fouesneau}, {Kapala},
  {Rosenfield}, \& {Schruba}}]{lewis2017}
{Lewis}, A.~R., {Simones}, J.~E., {Johnson}, B.~D., {et~al.} 2017, \apj, 834,
  70, \dodoi{10.3847/1538-4357/834/1/70}

\bibitem[{{Lin} {et~al.}(2017){Lin}, {Hu}, {Kong}, {Gao}, {Zou}, {Wang},
  {Cheng}, {Fang}, {Lin}, \& {Wang}}]{lin2017}
{Lin}, Z., {Hu}, N., {Kong}, X., {et~al.} 2017, \apj, 842, 97,
  \dodoi{10.3847/1538-4357/aa6f14}

\bibitem[{{Madore} {et~al.}(1974){Madore}, {van den Bergh}, \&
  {Rogstad}}]{madore1974}
{Madore}, B.~F., {van den Bergh}, S., \& {Rogstad}, D.~H. 1974, \apj, 191, 317,
  \dodoi{10.1086/152970}

\bibitem[{{Magrini} {et~al.}(2007){Magrini}, {Corbelli}, \&
  {Galli}}]{magrini2007}
{Magrini}, L., {Corbelli}, E., \& {Galli}, D. 2007, \aap, 470, 843,
  \dodoi{10.1051/0004-6361:20077215}

\bibitem[{{Magrini} {et~al.}(2010){Magrini}, {Stanghellini}, {Corbelli},
  {Galli}, \& {Villaver}}]{magrini2010}
{Magrini}, L., {Stanghellini}, L., {Corbelli}, E., {Galli}, D., \& {Villaver},
  E. 2010, \aap, 512, A63, \dodoi{10.1051/0004-6361/200913564}

\bibitem[{{Magrini} {et~al.}(2009){Magrini}, {Stanghellini}, \&
  {Villaver}}]{magrini2009}
{Magrini}, L., {Stanghellini}, L., \& {Villaver}, E. 2009, \apj, 696, 729,
  \dodoi{10.1088/0004-637X/696/1/729}

\bibitem[{{Massey} {et~al.}(1996){Massey}, {Bianchi}, {Hutchings}, \&
  {Stecher}}]{massey1996}
{Massey}, P., {Bianchi}, L., {Hutchings}, J.~B., \& {Stecher}, T.~P. 1996,
  \apj, 469, 629, \dodoi{10.1086/177811}

\bibitem[{{Massey} {et~al.}(2006){Massey}, {Olsen}, {Hodge}, {Strong},
  {Jacoby}, {Schlingman}, \& {Smith}}]{massey2006}
{Massey}, P., {Olsen}, K.~A.~G., {Hodge}, P.~W., {et~al.} 2006, \aj, 131, 2478,
  \dodoi{10.1086/503256}

\bibitem[{{McConnachie} {et~al.}(2010){McConnachie}, {Ferguson}, {Irwin},
  {Dubinski}, {Widrow}, {Dotter}, {Ibata}, \& {Lewis}}]{mcconnachie2010}
{McConnachie}, A.~W., {Ferguson}, A. M.~N., {Irwin}, M.~J., {et~al.} 2010,
  \apj, 723, 1038, \dodoi{10.1088/0004-637X/723/2/1038}

\bibitem[{{McConnachie} {et~al.}(2018){McConnachie}, {Ibata}, {Martin},
  {Ferguson}, {Collins}, {Gwyn}, {Irwin}, {Lewis}, {Mackey}, {Davidge},
  {Arias}, {Conn}, {C{\^o}t{\'e}}, {Crnojevic}, {Huxor}, {Penarrubia},
  {Spengler}, {Tanvir}, {Valls-Gabaud}, {Babul}, {Barmby}, {Bate}, {Bernard},
  {Chapman}, {Dotter}, {Harris}, {McMonigal}, {Navarro}, {Puzia}, {Rich},
  {Thomas}, \& {Widrow}}]{2018ApJ...868...55M}
{McConnachie}, A.~W., {Ibata}, R., {Martin}, N., {et~al.} 2018, \apj, 868, 55,
  \dodoi{10.3847/1538-4357/aae8e7}

\bibitem[{McKinney(2010)}]{pandas}
McKinney, W. 2010, in Proceedings of the 9th Python in Science Conference, ed.
  S.~{van der Walt} \& {Jarrod Millman}, 51--56

\bibitem[{McKinney(2011)}]{mckinney2011}
McKinney, W. 2011, Python for High Performance and Scientific Computing, 14

\bibitem[{{McLean} \& {Liu}(1996)}]{mclean1996}
{McLean}, I.~S., \& {Liu}, T. 1996, \apj, 456, 499, \dodoi{10.1086/176674}

\bibitem[{{McMonigal} {et~al.}(2016){McMonigal}, {Lewis}, {Brewer}, {Irwin},
  {Martin}, {McConnachie}, {Ibata}, {Ferguson}, {Mackey}, \&
  {Chapman}}]{mcmonigal2016}
{McMonigal}, B., {Lewis}, G.~F., {Brewer}, B.~J., {et~al.} 2016, \mnras, 461,
  4374, \dodoi{10.1093/mnras/stw1657}

\bibitem[{{McQuinn} {et~al.}(2007){McQuinn}, {Woodward}, {Willner}, {Polomski},
  {Gehrz}, {Humphreys}, {van Loon}, {Ashby}, {Eicher}, \&
  {Fazio}}]{mcquinn2007}
{McQuinn}, K.~B.~W., {Woodward}, C.~E., {Willner}, S.~P., {et~al.} 2007, \apj,
  664, 850, \dodoi{10.1086/519068}

\bibitem[{{Mighell} \& {Rich}(1995)}]{mighell1995}
{Mighell}, K.~J., \& {Rich}, R.~M. 1995, \aj, 110, 1649, \dodoi{10.1086/117638}

\bibitem[{{Minniti} {et~al.}(1993){Minniti}, {Olszewski}, \&
  {Rieke}}]{minniti1993}
{Minniti}, D., {Olszewski}, E.~W., \& {Rieke}, M. 1993, \apjl, 410, L79,
  \dodoi{10.1086/186884}

\bibitem[{{Mookerjea} {et~al.}(2016){Mookerjea}, {Israel}, {Kramer}, {Nikola},
  {Braine}, {Ossenkopf}, {R{\"o}llig}, {Henkel}, {van der Werf}, {van der Tak},
  \& {Wiedner}}]{mookerjea2016}
{Mookerjea}, B., {Israel}, F., {Kramer}, C., {et~al.} 2016, \aap, 586, A37,
  \dodoi{10.1051/0004-6361/201527366}

\bibitem[{{Mostoghiu} {et~al.}(2018){Mostoghiu}, {Di Cintio}, {Knebe},
  {Libeskind}, {Minchev}, \& {Brook}}]{mostoghiu2018}
{Mostoghiu}, R., {Di Cintio}, A., {Knebe}, A., {et~al.} 2018, \mnras, 480,
  4455, \dodoi{10.1093/mnras/sty2161}

\bibitem[{{Niu} {et~al.}(2020){Niu}, {Wang}, \& {Fu}}]{niu2020}
{Niu}, H., {Wang}, J., \& {Fu}, J. 2020, \apj, 903, 93,
  \dodoi{10.3847/1538-4357/abb8d6}

\bibitem[{Pedregosa {et~al.}(2011)Pedregosa, Varoquaux, Gramfort, Michel,
  Thirion, Grisel, Blondel, Prettenhofer, Weiss, Dubourg, Vanderplas, Passos,
  Cournapeau, Brucher, Perrot, \& Duchesnay}]{sklearn}
Pedregosa, F., Varoquaux, G., Gramfort, A., {et~al.} 2011, Journal of Machine
  Learning Research, 12, 2825.
\newblock \url{http://dl.acm.org/citation.cfm?id=1953048.2078195}

\bibitem[{{Regan} \& {Vogel}(1994)}]{regan1994}
{Regan}, M.~W., \& {Vogel}, S.~N. 1994, \apj, 434, 536, \dodoi{10.1086/174755}

\bibitem[{{Roberts}(1899)}]{roberts1899}
{Roberts}, I. 1899, {A Selection of Photographs of Stars, Star-Clusters and
  Nebulae, together with Records of Results obtained in the pursuit of
  Celestial Photography (Volume 2)} ({Cambridge University Press})

\bibitem[{Robin {et~al.}(2007)Robin, Rich, Aussel, Capak, Tasca, Jahnke,
  Kakazu, Kneib, Koekemoer, Leauthaud, Lilly, Mobasher, Scoville, Taniguchi, \&
  Thompson}]{robin2007}
Robin, A.~C., Rich, R.~M., Aussel, H., {et~al.} 2007, The Astrophysical Journal
  Supplement Series, 172, 545, \dodoi{10.1086/516600}

\bibitem[{Rocklin(2015)}]{rocklin2015}
Rocklin, M. 2015, in Proceedings of the 14th {{Python}} in {{Science
  Conference}}, ed. K.~Huff \& J.~Bergstra, {Austin, TX}, 126--132,
  \dodoi{10.25080/Majora-7b98e3ed-013}

\bibitem[{{Rosolowsky} {et~al.}(2003){Rosolowsky}, {Engargiola}, {Plambeck}, \&
  {Blitz}}]{rosolowsky2003}
{Rosolowsky}, E., {Engargiola}, G., {Plambeck}, R., \& {Blitz}, L. 2003, \apj,
  599, 258, \dodoi{10.1086/379166}

\bibitem[{{Rosolowsky} {et~al.}(2007){Rosolowsky}, {Keto}, {Matsushita}, \&
  {Willner}}]{rosolowsky2007}
{Rosolowsky}, E., {Keto}, E., {Matsushita}, S., \& {Willner}, S.~P. 2007, \apj,
  661, 830, \dodoi{10.1086/516621}

\bibitem[{{Rosolowsky} \& {Simon}(2008)}]{rosolowsky2008}
{Rosolowsky}, E., \& {Simon}, J.~D. 2008, \apj, 675, 1213,
  \dodoi{10.1086/527407}

\bibitem[{{Sarajedini} {et~al.}(2000){Sarajedini}, {Geisler}, {Schommer}, \&
  {Harding}}]{sarajedini2000}
{Sarajedini}, A., {Geisler}, D., {Schommer}, R., \& {Harding}, P. 2000, \aj,
  120, 2437, \dodoi{10.1086/316807}

\bibitem[{{Schlafly} \& {Finkbeiner}(2011)}]{schlafly2011}
{Schlafly}, E.~F., \& {Finkbeiner}, D.~P. 2011, \apj, 737, 103,
  \dodoi{10.1088/0004-637X/737/2/103}

\bibitem[{Smith {et~al.}(2020)Smith, E Andrews, Moe, Milne, Bilinski,
  Kilpatrick, Fong, Badenes, Filippenko, Kasliwal, \& et~al.}]{smith2020}
Smith, N., E Andrews, J., Moe, M., {et~al.} 2020, Monthly Notices of the Royal
  Astronomical Society, 492, 5897–5915, \dodoi{10.1093/mnras/staa061}

\bibitem[{{Stephens} \& {Frogel}(2002)}]{stephens2002}
{Stephens}, A.~W., \& {Frogel}, J.~A. 2002, \aj, 124, 2023,
  \dodoi{10.1086/342538}

\bibitem[{{STSCI Development Team}(2012)}]{2012ascl.soft12011S}
{STSCI Development Team}. 2012, {{DrizzlePac}}: {{HST}} Image Software, 2.2.6.
\newblock \doeprint{1212.011}

\bibitem[{{Telford} {et~al.}(2020){Telford}, {Dalcanton}, {Williams}, {Bell},
  {Dolphin}, {Durbin}, \& {Choi}}]{telford2020}
{Telford}, O.~G., {Dalcanton}, J.~J., {Williams}, B.~F., {et~al.} 2020, \apj,
  891, 32, \dodoi{10.3847/1538-4357/ab701c}

\bibitem[{{Telford} {et~al.}(2019){Telford}, {Werk}, {Dalcanton}, \&
  {Williams}}]{telford2019}
{Telford}, O.~G., {Werk}, J.~K., {Dalcanton}, J.~J., \& {Williams}, B.~F. 2019,
  \apj, 877, 120, \dodoi{10.3847/1538-4357/ab1b3f}

\bibitem[{{Thilker} {et~al.}(2005){Thilker}, {Hoopes}, {Bianchi}, {Boissier},
  {Rich}, {Seibert}, {Friedman}, {Rey}, {Buat}, {Barlow}, {Byun}, {Donas},
  {Forster}, {Heckman}, {Jelinsky}, {Lee}, {Madore}, {Malina}, {Martin},
  {Milliard}, {Morrissey}, {Neff}, {Schiminovich}, {Siegmund}, {Small},
  {Szalay}, {Welsh}, \& {Wyder}}]{thilker2005}
{Thilker}, D.~A., {Hoopes}, C.~G., {Bianchi}, L., {et~al.} 2005, \apjl, 619,
  L67, \dodoi{10.1086/424816}

\bibitem[{{Tibbs} {et~al.}(2018){Tibbs}, {Israel}, {Laureijs}, {Tauber},
  {Partridge}, {Peel}, \& {Fauvet}}]{tibbs2018}
{Tibbs}, C.~T., {Israel}, F.~P., {Laureijs}, R.~J., {et~al.} 2018, \mnras, 477,
  4968, \dodoi{10.1093/mnras/sty824}

\bibitem[{{Toribio San Cipriano} {et~al.}(2016){Toribio San Cipriano},
  {Garc{\'\i}a-Rojas}, {Esteban}, {Bresolin}, \& {Peimbert}}]{toribio2016}
{Toribio San Cipriano}, L., {Garc{\'\i}a-Rojas}, J., {Esteban}, C., {Bresolin},
  F., \& {Peimbert}, M. 2016, \mnras, 458, 1866, \dodoi{10.1093/mnras/stw397}

\bibitem[{{T{\"u}llmann} {et~al.}(2011){T{\"u}llmann}, {Gaetz}, {Plucinsky},
  {Kuntz}, {Williams}, {Pietsch}, {Haberl}, {Long}, {Blair}, {Sasaki},
  {Winkler}, {Challis}, {Pannuti}, {Edgar}, {Helfand}, {Hughes}, {Kirshner},
  {Mazeh}, \& {Shporer}}]{tuellmann2011}
{T{\"u}llmann}, R., {Gaetz}, T.~J., {Plucinsky}, P.~P., {et~al.} 2011, \apjs,
  193, 31, \dodoi{10.1088/0067-0049/193/2/31}

\bibitem[{{van der Kruit} \& {Freeman}(2011)}]{vanderkruit2011}
{van der Kruit}, P.~C., \& {Freeman}, K.~C. 2011, \araa, 49, 301,
  \dodoi{10.1146/annurev-astro-083109-153241}

\bibitem[{{van der Marel} {et~al.}(2019){van der Marel}, {Fardal}, {Sohn},
  {Patel}, {Besla}, {del Pino}, {Sahlmann}, \& {Watkins}}]{2019ApJ...872...24V}
{van der Marel}, R.~P., {Fardal}, M.~A., {Sohn}, S.~T., {et~al.} 2019, \apj,
  872, 24, \dodoi{10.3847/1538-4357/ab001b}

\bibitem[{{van der Walt} {et~al.}(2011){van der Walt}, Colbert, \&
  Varoquaux}]{numpy}
{van der Walt}, S., Colbert, S.~C., \& Varoquaux, G. 2011, Computing in Science
  \& Engineering, 13, 22, \dodoi{10.1109/MCSE.2011.37}

\bibitem[{{Verley} {et~al.}(2009){Verley}, {Corbelli}, {Giovanardi}, \&
  {Hunt}}]{verley2009}
{Verley}, S., {Corbelli}, E., {Giovanardi}, C., \& {Hunt}, L.~K. 2009, \aap,
  493, 453, \dodoi{10.1051/0004-6361:200810566}

\bibitem[{{Wainer} {et~al.}(2020){Wainer}, {Johnson}, {Torres-Villanueva}, \&
  {Seth}}]{2020AAS...23530602W}
{Wainer}, T., {Johnson}, L., {Torres-Villanueva}, E., \& {Seth}, A. 2020, in
  American Astronomical Society Meeting Abstracts, Vol. 235, American
  Astronomical Society Meeting Abstracts \#235, 306.02

\bibitem[{Waskom {et~al.}(2018)Waskom, Botvinnik, O'Kane, Hobson, Ostblom,
  Lukauskas, Gemperline, Augspurger, Halchenko, Cole, Warmenhoven, Ruiter, Pye,
  Hoyer, Vanderplas, Villalba, Kunter, Quintero, Bachant, Martin, Meyer, Miles,
  Ram, Brunner, Yarkoni, Williams, Evans, Fitzgerald, Brian, \&
  Qalieh}]{seabornv090}
Waskom, M., Botvinnik, O., O'Kane, D., {et~al.} 2018, Mwaskom/Seaborn: V0.9.0
  ({{July}} 2018), Zenodo, \dodoi{10.5281/ZENODO.1313201}

\bibitem[{{Watkins} {et~al.}(2010){Watkins}, {Evans}, \& {An}}]{watkins2010}
{Watkins}, L.~L., {Evans}, N.~W., \& {An}, J.~H. 2010, \mnras, 406, 264,
  \dodoi{10.1111/j.1365-2966.2010.16708.x}

\bibitem[{{Weisz} {et~al.}(2015){Weisz}, {Johnson}, {Foreman-Mackey},
  {Dolphin}, {Beerman}, {Williams}, {Dalcanton}, {Rix}, {Hogg}, {Fouesneau},
  {Johnson}, {Bell}, {Boyer}, {Gouliermis}, {Guhathakurta}, {Kalirai}, {Lewis},
  {Seth}, \& {Skillman}}]{weisz2015}
{Weisz}, D.~R., {Johnson}, L.~C., {Foreman-Mackey}, D., {et~al.} 2015, \apj,
  806, 198, \dodoi{10.1088/0004-637X/806/2/198}

\bibitem[{{West} {et~al.}(2018){West}, {Lehmer}, {Wik}, {Yang}, {Walton},
  {Antoniou}, {Haberl}, {Hornschemeier}, {Maccarone}, {Plucinsky}, {Ptak},
  {Williams}, {Vulic}, {Yukita}, \& {Zezas}}]{west2018}
{West}, L.~A., {Lehmer}, B.~D., {Wik}, D., {et~al.} 2018, \apj, 869, 111,
  \dodoi{10.3847/1538-4357/aaec6b}

\bibitem[{{White} {et~al.}(2019){White}, {Long}, {Becker}, {Blair}, {Helfand},
  \& {Winkler}}]{white2019}
{White}, R.~L., {Long}, K.~S., {Becker}, R.~H., {et~al.} 2019, \apjs, 241, 37,
  \dodoi{10.3847/1538-4365/ab0e89}

\bibitem[{{Williams} {et~al.}(2009){Williams}, {Dalcanton}, {Dolphin},
  {Holtzman}, \& {Sarajedini}}]{williams2009}
{Williams}, B.~F., {Dalcanton}, J.~J., {Dolphin}, A.~E., {Holtzman}, J., \&
  {Sarajedini}, A. 2009, \apjl, 695, L15, \dodoi{10.1088/0004-637X/695/1/L15}

\bibitem[{{Williams} {et~al.}(2014){Williams}, {Lang}, {Dalcanton}, {Dolphin},
  {Weisz}, {Bell}, {Bianchi}, {Byler}, {Gilbert}, {Girardi}, {Gordon},
  {Gregersen}, {Johnson}, {Kalirai}, {Lauer}, {Monachesi}, {Rosenfield},
  {Seth}, \& {Skillman}}]{williams2014}
{Williams}, B.~F., {Lang}, D., {Dalcanton}, J.~J., {et~al.} 2014, \apjs, 215,
  9, \dodoi{10.1088/0067-0049/215/1/9}

\bibitem[{{Williams} {et~al.}(2015){Williams}, {Wold}, {Haberl}, {Garofali},
  {Blair}, {Gaetz}, {Kuntz}, {Long}, {Pannuti}, {Pietsch}, {Plucinsky}, \&
  {Winkler}}]{williams2015}
{Williams}, B.~F., {Wold}, B., {Haberl}, F., {et~al.} 2015, \apjs, 218, 9,
  \dodoi{10.1088/0067-0049/218/1/9}

\bibitem[{{Williams} {et~al.}(2017){Williams}, {Dolphin}, {Dalcanton}, {Weisz},
  {Bell}, {Lewis}, {Rosenfield}, {Choi}, {Skillman}, \&
  {Monachesi}}]{williams2017}
{Williams}, B.~F., {Dolphin}, A.~E., {Dalcanton}, J.~J., {et~al.} 2017, \apj,
  846, 145, \dodoi{10.3847/1538-4357/aa862a}

\bibitem[{{Wyse}(2002)}]{wyse2002}
{Wyse}, R.~F.~G. 2002, in EAS Publications Series, Vol.~2, EAS Publications
  Series, ed. O.~{Bienayme} \& C.~{Turon}, 295--304.
\newblock \doarXiv{astro-ph/0204190}

\bibitem[{{Xi} {et~al.}(2020){Xi}, {Zhang}, {Liu}, \& {Wang}}]{xi2020}
{Xi}, S.-Q., {Zhang}, H.-M., {Liu}, R.-Y., \& {Wang}, X.-Y. 2020, arXiv
  e-prints, arXiv:2003.07830.
\newblock \doarXiv{2003.07830}

\bibitem[{{Xilouris} {et~al.}(2012){Xilouris}, {Tabatabaei}, {Boquien},
  {Kramer}, {Buchbender}, {Bertoldi}, {Anderl}, {Braine}, {Verley},
  {Rela{\~n}o}, {Quintana-Lacaci}, {Akras}, {Beck}, {Calzetti}, {Combes},
  {Gonzalez}, {Gratier}, {Henkel}, {Israel}, {Koribalski}, {Lord}, {Mookerjea},
  {Rosolowsky}, {Stacey}, {Tilanus}, {van der Tak}, \& {van der
  Werf}}]{xilouris2012}
{Xilouris}, E.~M., {Tabatabaei}, F.~S., {Boquien}, M., {et~al.} 2012, \aap,
  543, A74, \dodoi{10.1051/0004-6361/201219291}

\end{thebibliography}

\clearpage

\begin{table}
\centering
\scriptsize
\caption{Sample exposure data for one field. A full machine-readable table is provided in the online journal.}
\label{table_obs}
\begin{tabular}{llllrlllr}
\hline
Target Name & R.A. (J2000) & Decl. (J2000) & Start Time & Exp. (s) & Inst. & Aperture & Filter & Orientation \\
\hline
\hline
   M33-B01-F01-IR &  $01^\mathrm{h}34^\mathrm{m}33^\mathrm{s}$ &  $+30^\circ47{}^\prime57{}^{\prime\prime}$ &  2017-12-28 06:53:23 &  399.23 &     WFC3 &       IR-FIX &  F160W &  -80.3544 \\
   M33-B01-F01-IR &  $01^\mathrm{h}34^\mathrm{m}33^\mathrm{s}$ &  $+30^\circ47{}^\prime58{}^{\prime\prime}$ &  2017-12-28 07:01:04 &  699.23 &     WFC3 &       IR-FIX &  F110W &  -80.3527 \\
   M33-B01-F01-IR &  $01^\mathrm{h}34^\mathrm{m}33^\mathrm{s}$ &  $+30^\circ47{}^\prime58{}^{\prime\prime}$ &  2017-12-28 07:13:45 &  399.23 &     WFC3 &       IR-FIX &  F160W &  -80.3530 \\
   M33-B01-F01-IR &  $01^\mathrm{h}34^\mathrm{m}33^\mathrm{s}$ &  $+30^\circ47{}^\prime58{}^{\prime\prime}$ &  2017-12-28 07:22:28 &  399.23 &     WFC3 &       IR-FIX &  F160W &  -80.3567 \\
   M33-B01-F01-IR &  $01^\mathrm{h}34^\mathrm{m}33^\mathrm{s}$ &  $+30^\circ47{}^\prime58{}^{\prime\prime}$ &  2017-12-28 07:31:11 &  399.23 &     WFC3 &       IR-FIX &  F160W &  -80.3554 \\
 M33-B01-F01-UVIS &  $01^\mathrm{h}34^\mathrm{m}33^\mathrm{s}$ &  $+30^\circ47{}^\prime59{}^{\prime\prime}$ &  2017-12-28 05:20:02 &  550.00 &     WFC3 &  UVIS-CENTER &  F336W &  -80.1842 \\
 M33-B01-F01-UVIS &  $01^\mathrm{h}34^\mathrm{m}33^\mathrm{s}$ &  $+30^\circ47{}^\prime59{}^{\prime\prime}$ &  2017-12-28 05:31:49 &  350.00 &     WFC3 &  UVIS-CENTER &  F275W &  -80.1837 \\
 M33-B01-F01-UVIS &  $01^\mathrm{h}34^\mathrm{m}34^\mathrm{s}$ &  $+30^\circ47{}^\prime59{}^{\prime\prime}$ &  2017-12-28 05:40:19 &  700.00 &     WFC3 &  UVIS-CENTER &  F336W &  -80.1838 \\
 M33-B01-F01-UVIS &  $01^\mathrm{h}34^\mathrm{m}34^\mathrm{s}$ &  $+30^\circ47{}^\prime58{}^{\prime\prime}$ &  2017-12-28 05:54:37 &  540.00 &     WFC3 &  UVIS-CENTER &  F275W &  -80.1831 \\
  M33-B01-F01-WFC &  $01^\mathrm{h}34^\mathrm{m}34^\mathrm{s}$ &  $+30^\circ47{}^\prime51{}^{\prime\prime}$ &  2017-07-27 22:08:21 &   15.00 &      ACS &          WFC &  F814W & -127.6122 \\
  M33-B01-F01-WFC &  $01^\mathrm{h}34^\mathrm{m}34^\mathrm{s}$ &  $+30^\circ47{}^\prime51{}^{\prime\prime}$ &  2017-07-27 22:18:26 &  350.00 &      ACS &          WFC &  F814W & -127.6120 \\
  M33-B01-F01-WFC &  $01^\mathrm{h}34^\mathrm{m}33^\mathrm{s}$ &  $+30^\circ47{}^\prime51{}^{\prime\prime}$ &  2017-07-27 22:26:56 &  700.00 &      ACS &          WFC &  F814W & -127.6124 \\
  M33-B01-F01-WFC &  $01^\mathrm{h}34^\mathrm{m}33^\mathrm{s}$ &  $+30^\circ47{}^\prime51{}^{\prime\prime}$ &  2017-07-27 22:41:14 &  430.00 &      ACS &          WFC &  F814W & -127.6123 \\
  M33-B01-F01-WFC &  $01^\mathrm{h}34^\mathrm{m}34^\mathrm{s}$ &  $+30^\circ47{}^\prime51{}^{\prime\prime}$ &  2017-07-27 23:33:11 &   10.00 &      ACS &          WFC &  F475W & -127.6115 \\
  M33-B01-F01-WFC &  $01^\mathrm{h}34^\mathrm{m}34^\mathrm{s}$ &  $+30^\circ47{}^\prime51{}^{\prime\prime}$ &  2017-07-27 23:40:10 &  600.00 &      ACS &          WFC &  F475W & -127.6114 \\
  M33-B01-F01-WFC &  $01^\mathrm{h}34^\mathrm{m}34^\mathrm{s}$ &  $+30^\circ47{}^\prime51{}^{\prime\prime}$ &  2017-07-27 23:52:51 &  370.00 &      ACS &          WFC &  F475W & -127.6116 \\
  M33-B01-F01-WFC &  $01^\mathrm{h}34^\mathrm{m}34^\mathrm{s}$ &  $+30^\circ47{}^\prime51{}^{\prime\prime}$ &  2017-07-28 00:01:39 &  360.00 &      ACS &          WFC &  F475W & -127.6116 \\
  M33-B01-F01-WFC &  $01^\mathrm{h}34^\mathrm{m}34^\mathrm{s}$ &  $+30^\circ47{}^\prime51{}^{\prime\prime}$ &  2017-07-28 00:10:17 &  360.00 &      ACS &          WFC &  F475W & -127.6114 \\
\hline
\end{tabular}
\end{table}

\begin{table}
\centering
\footnotesize
\caption{DOLPHOT parameters used for all photometry.}
\label{table_par}
\begin{tabular}{lll}
\hline
Detector &        Parameter &  Value \\
\hline
\hline
      IR &            raper &      2 \\
      IR &             rchi &    1.5 \\
      IR &            rsky0 &      8 \\
      IR &            rsky1 &     20 \\
      IR &             rpsf &     10 \\
    UVIS &            raper &      3 \\
    UVIS &             rchi &    2.0 \\
    UVIS &            rsky0 &     15 \\
    UVIS &            rsky1 &     35 \\
    UVIS &             rpsf &     10 \\
     WFC &            raper &      3 \\
     WFC &             rchi &    2.0 \\
     WFC &            rsky0 &     15 \\
     WFC &            rsky1 &     35 \\
     WFC &             rpsf &     10 \\
     All &            apsky &  15 25 \\
     All &           UseWCS &      2 \\
     All &          PSFPhot &      1 \\
     All &           FitSky &      2 \\
     All &          SkipSky &      2 \\
     All &           SkySig &   2.25 \\
     All &       SecondPass &      5 \\
     All &       SearchMode &      1 \\
     All &          SigFind &    3.0 \\
     All &      SigFindMult &   0.85 \\
     All &         SigFinal &    3.5 \\
     All &            MaxIT &     25 \\
     All &        NoiseMult &   0.10 \\
     All &             FSat &  0.999 \\
     All &         FlagMask &      4 \\
     All &            ApCor &      1 \\
     All &           Force1 &      1 \\
     All &            Align &      2 \\
     All &         aligntol &      4 \\
     All &        alignstep &      2 \\
     WFC &        ACSuseCTE &      0 \\
 UVIS/IR &       WFC3useCTE &      0 \\
     All &           Rotate &      1 \\
     All &        RCentroid &      1 \\
     All &          PosStep &    0.1 \\
     All &          dPosMax &    2.5 \\
     All &         RCombine &  1.415 \\
     All &           SigPSF &    3.0 \\
     All &           PSFres &      1 \\
     All &           psfoff &    0.0 \\
     All &     DiagPlotType &    PNG \\
     All &       CombineChi &      1 \\
     WFC &       ACSpsfType &      0 \\
      IR &    WFC3IRpsfType &      0 \\
    UVIS &  WFC3UVISpsfType &      0 \\
\hline
\end{tabular}
\end{table}
\begin{table}
\centering
\caption{50\% completeness limits by stellar density (stars/square arcsec).}
\label{table_comp}
\begin{tabular}{lrrrrrr}
\hline
Density &  F275W &  F336W &  F475W &  F814W &  F110W &  F160W \\
\hline
\hline
0 - 0.15   &  24.44 &  25.63 &  27.65 &  26.77 &  25.62 &  24.95 \\
0.15 - 0.3 &  24.43 &  25.53 &  27.20 &  26.49 &  25.09 &  24.55 \\
0.3 - 0.6  &  24.42 &  25.54 &  26.96 &  26.17 &  24.55 &  23.88 \\
0.6 - 0.9  &  24.37 &  25.43 &  26.41 &  25.65 &  23.96 &  23.27 \\
0.9+       &  24.14 &  25.10 &  25.75 &  25.23 &  23.56 &  22.91 \\
\hline
\end{tabular}
\end{table}
\begin{table}
\label{table_bias}
\centering
\caption{Sample photometric bias, AST-derived uncertainty, DOLPHOT-reported uncertainty, and AST/DOLPHOT uncertainty ratio by magnitude and stellar density (stars/square arcsec). A full machine-readable table is provided in the online journal.}
\begin{tabular}{rlrrrrr}
\hline
 Density & Filter &  Magnitude &      Bias &   Uncertainty &   DOLPHOT &     Ratio \\
\hline
\hline
 0 - 0.15 &  F275W &       17.5 & -0.003858 &     0.004518 &  0.003984 &  1.134049 \\
 0 - 0.15 &  F275W &       18.0 & -0.001987 &     0.005048 &  0.004971 &  1.015534 \\
 0 - 0.15 &  F275W &       18.5 & -0.000040 &     0.006076 &  0.005993 &  1.013889 \\
 0 - 0.15 &  F275W &       19.0 &  0.002106 &     0.007913 &  0.007987 &  0.990811 \\
 0 - 0.15 &  F275W &       19.5 &  0.006055 &     0.010892 &  0.009997 &  1.089572 \\
 0 - 0.15 &  F275W &       20.0 &  0.009879 &     0.014001 &  0.012977 &  1.078916 \\
 0 - 0.15 &  F275W &       20.5 &  0.015065 &     0.017544 &  0.015974 &  1.098268 \\
 0 - 0.15 &  F275W &       21.0 &  0.023059 &     0.023446 &  0.021009 &  1.115996 \\
 0 - 0.15 &  F275W &       21.5 &  0.034013 &     0.033485 &  0.028010 &  1.195487 \\
 0 - 0.15 &  F275W &       22.0 &  0.043005 &     0.045021 &  0.039973 &  1.126294 \\
 0 - 0.15 &  F275W &       22.5 &  0.064932 &     0.064028 &  0.055000 &  1.164144 \\
 0 - 0.15 &  F275W &       23.0 &  0.088065 &     0.092049 &  0.076996 &  1.195507 \\
 0 - 0.15 &  F275W &       23.5 &  0.137023 &     0.131494 &  0.114016 &  1.153292 \\
 0 - 0.15 &  F275W &       24.0 &  0.167010 &     0.171251 &  0.160019 &  1.070193 \\
 0 - 0.15 &  F275W &       24.5 &  0.142983 &     0.209921 &  0.207011 &  1.014055 \\
 0 - 0.15 &  F275W &       25.0 & -0.050442 &     0.239546 &  0.234015 &  1.023635 \\
\hline
\end{tabular}
\end{table}

\begin{table}
\tiny
\begin{adjustwidth}{-3.75cm}{}
\caption{Sample photometric data. A full machine-readable table is provided in the online journal.}
\label{table_phot}
\begin{tabular}{rrrrlrrlrrlrrlrrlrrl}
\hline
R.A. (J2000) &  Decl. (J2000) &   F275W &    S/N & GST &   F336W &   S/N & GST &   F475W &    S/N & GST &   F814W &    S/N & GST &   F110W &     S/N & GST &   F160W &     S/N & GST \\
\hline
\hline
 23.340786 &  30.523476 &  29.160 &   0.1 &   F &  26.679 &   2.1 &   F &  26.791 &  10.4 &   T &  26.325 &   7.2 &   T &  99.999 &  0.0 &   F &  99.999 &  0.0 &   F \\
 23.340794 &  30.523207 &  23.059 &  11.2 &   T &  23.230 &  15.4 &   T &  24.244 &  48.6 &   T &  24.062 &  38.2 &   T &  99.999 &  0.0 &   F &  99.999 &  0.0 &   F \\
 23.340808 &  30.523388 &  99.999 &  -2.2 &   F &  99.999 &  -1.5 &   F &  26.613 &  11.9 &   T &  24.796 &  26.3 &   T &  99.999 &  0.0 &   F &  99.999 &  0.0 &   F \\
 23.340814 &  30.523616 &  99.999 &  -0.5 &   F &  99.999 &  -0.0 &   F &  28.516 &   2.5 &   F &  27.409 &   3.0 &   F &  99.999 &  0.0 &   F &  99.999 &  0.0 &   F \\
 23.340824 &  30.523322 &  25.680 &   2.2 &   F &  27.131 &   1.5 &   F &  28.579 &   2.1 &   F &  27.189 &   3.4 &   F &  99.999 &  0.0 &   F &  99.999 &  0.0 &   F \\
 23.340825 &  30.523415 &  25.555 &   2.1 &   F &  28.072 &   0.7 &   F &  28.599 &   2.3 &   F &  27.096 &   3.5 &   F &  99.999 &  0.0 &   F &  99.999 &  0.0 &   F \\
 23.340839 &  30.523445 &  99.999 &  -0.4 &   F &  99.999 &  -0.6 &   F &  28.191 &   3.2 &   F &  27.260 &   3.3 &   F &  99.999 &  0.0 &   F &  99.999 &  0.0 &   F \\
 23.340845 &  30.523811 &  99.999 &  -1.7 &   F &  28.635 &   0.4 &   F &  27.973 &   3.9 &   F &  26.840 &   4.8 &   T &  99.999 &  0.0 &   F &  99.999 &  0.0 &   F \\
 23.340846 &  30.523475 &  99.999 &  -0.7 &   F &  99.999 &  -0.9 &   F &  99.999 &  -0.1 &   F &  27.709 &   2.0 &   F &  99.999 &  0.0 &   F &  99.999 &  0.0 &   F \\
\hline
\end{tabular}
\end{adjustwidth}
\end{table}

\begin{turnpage}
\begin{table}
\centering
\tiny
\begin{adjustwidth}{-5.5cm}{}
\caption{Sample artificial star data. A full machine-readable table is provided in the online journal.}
\label{table_asts}
\begin{tabular}{rrrrrlrrrlrrrlrrrlrrrlrrrl}
\hline
RA(J2000) &  Dec(J2000) &  F275W &  Out-in & S/N & GST &  F336W &  Out-in & S/N & GST &  F475W &  Out-in & S/N & GST &  F814W &  Out-in & S/N & GST &  F110W &  Out-in & S/N & GST &  F160W &  Out-in &  S/N & GST \\
\hline
\hline
 23.436616 &  30.647496 &    29.515 &    -1.793 &   0.4 &         F &    27.248 &     0.037 &   1.7 &         F &    26.114 &     0.202 &   16.0 &         T &    24.538 &     0.223 &   22.8 &         T &    24.006 &     0.577 &   3.4 &         F &    23.406 &     0.377 &  10.3 &         T \\
 23.436644 &  30.647382 &    32.387 &    99.999 &   0.0 &         F &    30.107 &    99.999 &   0.0 &         F &    29.138 &    99.999 &    0.0 &         F &    27.682 &    99.999 &    0.0 &         F &    27.229 &    99.999 &   0.0 &         F &    26.719 &    99.999 &   0.0 &         F \\
 23.436657 &  30.647784 &    31.890 &    99.999 &   0.0 &         F &    30.294 &    99.999 &   0.0 &         F &    29.845 &    99.999 &    0.0 &         F &    28.666 &    99.999 &    0.0 &         F &    28.315 &    99.999 &   0.0 &         F &    27.893 &    99.999 &   0.0 &         F \\
 23.436664 &  30.647719 &    27.948 &    99.999 &   0.0 &         F &    27.244 &    99.999 &   0.0 &         F &    27.176 &    99.999 &    0.0 &         F &    26.664 &    99.999 &    0.0 &         F &    26.540 &    99.999 &   0.0 &         F &    26.373 &    99.999 &   0.0 &         F \\
 23.436685 &  30.647769 &    24.387 &     0.413 &   4.7 &         T &    24.706 &    -0.100 &  10.9 &         T &    25.494 &     0.107 &   31.9 &         T &    25.688 &     0.362 &    8.1 &         T &    25.826 &    99.999 &  -0.9 &         F &    25.886 &    99.999 &  -0.4 &         F \\
 23.436758 &  30.647418 &    21.550 &     0.047 &  48.4 &         T &    22.045 &     0.024 &  73.0 &         T &    23.471 &     0.010 &  137.1 &         T &    23.727 &     0.027 &   53.8 &         T &    23.943 &     0.149 &   9.3 &         T &    24.036 &     1.652 &   1.8 &         F \\
 23.436774 &  30.647546 &    27.575 &    -1.026 &   1.3 &         F &    26.345 &     0.319 &   2.8 &         F &    26.029 &     0.141 &   19.3 &         T &    25.247 &    -0.004 &   17.7 &         T &    25.043 &    -0.292 &   5.9 &         T &    24.795 &     1.111 &   1.3 &         F \\
 23.436785 &  30.647818 &    32.132 &    99.999 &   0.0 &         F &    30.002 &    99.999 &   0.0 &         F &    29.102 &    99.999 &    0.0 &         F &    27.633 &    99.999 &    0.0 &         F &    27.156 &    99.999 &   0.0 &         F &    26.607 &    99.999 &   0.0 &         F \\
 23.436838 &  30.648715 &    22.383 &    -0.011 &  31.9 &         T &    22.401 &     0.031 &  58.7 &         T &    22.796 &    -0.009 &  197.2 &         T &    22.919 &    -0.023 &  114.8 &         T &    23.000 &    -0.014 &  31.8 &         T &    23.043 &    -0.143 &  22.3 &         T \\
 23.436847 &  30.648508 &    21.232 &     0.030 &  55.6 &         T &    21.679 &     0.013 &  88.2 &         T &    23.026 &    -0.008 &  192.8 &         T &    23.246 &    -0.008 &   87.6 &         T &    23.441 &     0.148 &  12.8 &         T &    23.524 &     0.334 &   6.7 &         T \\
\hline
\end{tabular}
\end{adjustwidth}
\end{table}
\end{turnpage}

\clearpage

\end{document}